\documentclass[12pt,twoside,english]{article}
\usepackage{a4wide}

\makeatletter

\def\tikzOff{}

\usepackage[utf8]{inputenc}            
\usepackage[T1]{fontenc}
\usepackage{lmodern}  
\usepackage{graphicx}
\usepackage{fancyhdr}
\usepackage{amsmath, amssymb, amsfonts, bm}
\usepackage{mathtools}
\usepackage{caption}
\usepackage[labelformat=simple]{subfig}
\usepackage{setspace}
\usepackage{verbatim} 
\usepackage[english]{babel}
\usepackage[novbox]{pdfsync} 
\usepackage{caption}
\usepackage{booktabs}
\usepackage{hyperref}
\usepackage{float}
\usepackage{relsize} 
\usepackage{amsthm}
\usepackage[numbers,sort&compress]{natbib}
\theoremstyle{remark}
\newtheorem{remark}{Remark}

\usepackage{siunitx}
\usepackage{xfrac}
\usepackage{pdfpages}
\usepackage{overpic}

\usepackage{color}

\includepdfset{turn=true,pagecommand={}}

\hypersetup{
	colorlinks = false,
	linktocpage = true,
	allbordercolors = {0.8 0.8 0.8}
}

\addto\captionsenglish{}
\addto\captionsenglish{}
\addto\extrasenglish{}
\addto\extrasenglish{}
\addto\extrasenglish{}
\addto\extrasenglish{}
\addto\extrasenglish{}
\addto\extrasenglish{}
\addto\extrasenglish{}

\usepackage{etex}
\usepackage{tikz}
\usepackage{pgfplots}
\usepackage{tikz-3dplot}
\usepackage{fp}

\pgfplotsset{compat=newest}

\ifdefined\bUseTikzExternalize
	\usetikzlibrary{external}
	\tikzsetexternalprefix{tmp/}
\else
	\usetikzlibrary{external}
	\tikzexternaldisable
\fi

\usetikzlibrary{calc,intersections}
\usetikzlibrary{shapes.geometric,shapes.arrows,decorations.pathmorphing}
\usetikzlibrary{matrix,chains,scopes,positioning,arrows,fit}
\usetikzlibrary{spy}
\usetikzlibrary{fadings}
\usetikzlibrary{shadings}
\usetikzlibrary{backgrounds}

\usepgfplotslibrary{groupplots}
\usepgfplotslibrary{patchplots}

\def\pgfplotfontsizetitle{\small}
\def\pgfplotfontsize{\small}

\def\tikzfontsizetiny{\tiny}

\pgfplotsset{
  mystyle/.style ={%
    grid = major,
    every tick label/.append style={font=\pgfplotfontsize},
    every axis label/.append style={font=\pgfplotfontsize},
    legend style={font=\scriptsize},
    label style={font=\pgfplotfontsize},
    title style={font=\pgfplotfontsizetitle},
    /pgf/number format/set thousands separator = {}, 
  }
}

\tikzset{
	>=stealth',
	axis/.style={<->},
	important line/.style={thick},
	connection/.style={thick, dotted},
	dot/.style = {
		draw,
		fill = white,
		circle,
		solid,
		thin,
		inner sep = 0pt,
		minimum size = 3pt
	},
	defP/.style = {
		inner sep = 0pt
	}	
}
\pgfarrowsdeclaredouble{doubled}{doubled}{stealth'}{stealth'}

\usepackage[outline]{contour}
\usepackage{standalone} 

\usepackage{forloop}
\usepackage{arrayjobx}

\usepackage{listings}

\usepackage{matlab-prettifier}

\usepackage{cleveref} 
\crefformat{figure}{Figure~#2#1#3} 
\Crefformat{figure}{Figure~#2#1#3} 
\crefmultiformat{figure}{Figures~#2#1#3}{ and~#2#1#3}{, #2#1#3}{ and~#2#1#3} 
\crefformat{equation}{Equation~(#2#1#3)} 
\Crefformat{equation}{Equation~(#2#1#3)} 
\labelcrefformat{equation}{(#2#1#3)}  
\crefformat{section}{Section~#2#1#3} 
\Crefformat{section}{Section~#2#1#3} 
\crefformat{listing}{Algorithm~(#2#1#3)} 
\Crefformat{listing}{Algorithm~(#2#1#3)} 

\newcommand{\vek}[1]{\mathchoice{\displaystyle\boldsymbol#1}
	{\textstyle\boldsymbol#1}{\scriptstyle\boldsymbol#1}
	{\scriptscriptstyle\boldsymbol#1}}
\newcommand{\mat}[1]{\mathchoice{\displaystyle\mathbf#1}
	{\textstyle\mathbf#1}{\scriptstyle\mathbf#1}
	{\scriptscriptstyle\mathbf#1}}

\newcommand{\ten}[1]{ \ensuremath{\bm #1} }
\renewcommand{\d}{ \ensuremath{\mathrm{d} }}
\newcommand{\p}{ \ensuremath{\partial} }
\renewcommand{\t}[1]{ \ensuremath{\text{#1}} }
\newcommand{\T}{{\ensuremath{\mathrm{T}}}}

\newcommand{\divG}[1]{\ensuremath{\text{div}_{\Gamma} #1}}

\newcommand{\gradG}[1]{\ensuremath{\nabla_\Gamma #1}}
\newcommand{\gradGD}[1]{\ensuremath{\nabla_\Gamma^\text{dir} #1}}
\newcommand{\gradGC}[1]{\ensuremath{\nabla_\Gamma^\text{cov} #1}}
\newcommand{\nG}{\ensuremath{\vek{n}_\Gamma}}
\newcommand{\nCo}{\ensuremath{\vek{n}_{\p\Gamma}}}
\newcommand{\tB}{\ensuremath{\vek{t}_{\p\Gamma}}}

\definecolor{mygreen}{rgb}{0, 0.51, 0}
\definecolor{myred}{rgb}{0.46, 0., 0.05}

\def\dirtikz{tikz} 
\def\dirdata{tikz/data} 

\newcommand{\KL}{KL} 

\newcommand{\mycomment}[1]{\textcolor{blue}{\emph{#1}}}

\ifdefined\tikzOff

\else

\fi

\ifdefined\tikzOff

\else

\fi

\newcommand\und{\,\text{and}\,}

\newcommand\with{\,\text{with}\,}

\newcommand\where{\,\text{where}\,}

\newcommand{\R}{\mathbb{R}}

\newcommand{\myMat}[1]{\mathbf{#1}}

\newcommand{\myVec}[1]{\vek{#1}}

\newcommand\operate[1]{(#1)}

\newcommand\fromto[2]{\{ #1, \dots, #2 \}}
\DeclareMathOperator{\supp}{supp}
\newcommand{\Supp}[1]{\supp\{ #1\}}

\newcommand\pt[1]{\boldsymbol{#1}}

\newcommand\indexA{i} 		
\newcommand\totalA{I} 		
\newcommand\indexB{j} 		
\newcommand\totalB{J} 		
\newcommand\indexC{k} 		
\newcommand\indexD{m} 		
\newcommand\indexG{l} 		
\newcommand\totalG{L} 		

\newcommand{\Bspline}{B} 		
\newcommand{\BsplineSeg}{\mathcal{B}} 	
\newcommand{\KV}{\varXi} 		
\newcommand\uu{\xi} 			
\newcommand\vv{\uu} 			
\newcommand\UVsurf{\myVec{\uu}} 	
\newcommand\uusurf{\uu_1} 		
\newcommand\vvsurf{\vv_2} 		
\newcommand\multi{m} 			

\newcommand\pu{p}			
\newcommand\pusurf{p_1}
\newcommand\pvsurf{p_2}	
\newcommand\x{XXXX}			
\newcommand\surface{\Gamma}
\newcommand\tangent{\myVec{t}}

\newcommand{\CP}{\pt{c}}		

\newcommand\ndofs{n}

\newcommand{\exBspline}{B^e} 		
\newcommand{\degSet}{\mathbb{J}} 	
\newcommand{\NSet}{\mathbb{T}} 	  	
\newcommand{\eweight}{e} 	  	
\newcommand{\Nbasis}{\psi} 	  	
\newcommand{\taylorpoint}{\tilde{\uu}} 
\newcommand{\poly}{{\alpha}} 		

\newcommand{\patchdomain}{\mathcal{A}^{\textnormal{v}}}
\newcommand\indexSpan{s}	
\newcommand\indexSpanTrimmed{t}	
\newcommand{\supportdomain}{\mathcal{S}^{\textnormal{v}}}
\newcommand{\supportalpha}{\alpha}
\newcommand{\BsplineMatrix}{\myMat{A}}
\newcommand{\BsplineMatrixCode}{\myMat{A}}
\newcommand{\ExMatrix}{\myMat{E}}

\newcommand{\visibledomain}{\patchdomain} 
\newcommand{\TrimCurve}{\pt{C}^{t}}     
\newcommand{\TrimCurvePS}{\tilde{\pt{C}}^{t}}     
\newcommand{\LeveSet}{\phi}

\newcommand\primary{u} 			

\pgfplotsset{compat=1.10}

\usetikzlibrary{external}
\tikzsetexternalprefix{tmp/}

\usetikzlibrary{shapes,spy,fit}

\usepgfplotslibrary{groupplots}

\def\pgfplotfontsizetitle{\small}
\def\pgfplotfontsize{\footnotesize}

\def\tikzfontsizetiny{\scriptsize}

\pgfplotsset{
  mystyle/.style ={%
    grid = major,
    every tick label/.append style={font=\pgfplotfontsize},
    every axis label/.append style={font=\pgfplotfontsize},
    legend style={font=\pgfplotfontsize},
    label style={font=\pgfplotfontsize},
    title style={font=\pgfplotfontsizetitle},
    /pgf/number format/set thousands separator = {}, 
  }
}

\colorlet{drawblue}      {blue!80!white}
\colorlet{drawred}       {red!80!white}
\colorlet{drawgray}      {gray}
\definecolor{drawgreen}  {RGB}{44,162,95}
\colorlet{drawpurple}    {purple}
\colorlet{draworange}    {orange}
\colorlet{drawlime}      {lime!80!black}
\colorlet{drawartichoke} {yellow!60!black}
\colorlet{TUGgray}{black!15}
\definecolor{TUGred}{RGB}{247,1,70}
\definecolor{IFBblue}{RGB}{51,112,169}
\definecolor{UTorange}{RGB}{191,87,0}
\definecolor{UTblue}{RGB}{0,95,134}
\definecolor{UTgreen}{RGB}{67,105,91}
\definecolor{UTgray}{RGB}{51,63,72}
\definecolor{UTyellow}{RGB}{242,169,0}

\definecolor{basisColor1}{RGB}{59,76,192}
\definecolor{basisColor2}{RGB}{87,117,225}
\definecolor{basisColor3}{RGB}{119,154,247}
\definecolor{basisColor4}{RGB}{152,185,255}
\definecolor{basisColor5}{RGB}{184,208,249}
\definecolor{basisColor6}{RGB}{195,209,230}	
\definecolor{basisColor7}{RGB}{247,200,190}	
\definecolor{basisColor8}{RGB}{247,187,160}
\definecolor{basisColor9}{RGB}{244,154,123}
\definecolor{basisColor10}{RGB}{229,112,88}
\definecolor{basisColor11}{RGB}{203,62,56}
\definecolor{basisColor12}{RGB}{180,4,38}

\definecolor{basisColor8sw}{RGB} {189,189,189}
\definecolor{basisColor9sw}{RGB} {150,150,150}
\definecolor{basisColor10sw}{RGB}{115,115,115}
\definecolor{basisColor11sw}{RGB}{91,91,91}
\definecolor{basisColor12sw}{RGB}{37,37,37}

\colorlet{myblue}    {blue}
\colorlet{myred}     {red}
\colorlet{mygreen}   {drawgreen}
\colorlet{mypurple}  {purple}
\colorlet{myorange}  {orange}

\pgfplotscreateplotcyclelist{mycyclelist}{
blue,every mark/.append style={fill=blue!80!black},mark=*\\%
red,every mark/.append style={fill=red!80!black},mark=square*\\%
brown!60!black,dashed,every mark/.append style={fill=brown!80!black},mark=otimes*\\%
black,dashed,mark=star\\%
}

\pgfplotscreateplotcyclelist{mycyclelistcompare}{
  basisColor12, semithick, loosely dashdotdotted     \\%
  basisColor11, semithick, loosely dashed      		\\%
  basisColor10, semithick, dashed \\
  basisColor9,  semithick, solid               	\\%
  basisColor12, only marks,   every mark/.append style={solid, fill=.},mark=otimes    \\%
  basisColor11, only marks,   every mark/.append style={solid, fill=.},mark=triangle*   \\%
  basisColor10, only marks,   every mark/.append style={solid, fill=.},mark=square*     \\%
  basisColor9,  only marks,   every mark/.append style={solid, fill=.},mark=*          \\%
}

\tikzset{mycyclelistcompareReferenceA/.style={basisColor12,solid}}
\tikzset{mycyclelistcompareTestA/.style={basisColor12,only marks,mark=otimes}}

\tikzset{helpline/.style={thin,dashed}}
\tikzset{labelline/.style={thin}}
\tikzset{referencePath/.style={dotted,very thick}}
\tikzset{helparrow/.style={thin,arrows={-latex}}}
\tikzset{pointer/.style={arrows={-latex}}}
\tikzset{axis/.style={thin,arrows={->}}}
\tikzset{force/.style={thick,arrows={->}}}
\tikzset{forceInverse/.style={thick,arrows={<-}}}
\tikzset{Gamma/.style={ultra thick}}
\tikzset{controlPoly/.style={draw=black}}
\tikzset{GammaFill/.style={fill=lightgray,fill opacity=0.3}}
\tikzset{colorDiri/.style={drawgreen}}
\tikzset{GammaFillDiri/.style={fill=drawgreen,fill opacity=0.5}}
\tikzset{initialgrid/.style={thin,gray}}
\tikzset{intersectioncurve/.style={dashed,thick}}
\tikzset{trimmingcurve/.style={thick}}
\tikzset{boundingbox/.style={dashed,}}
\tikzset{approx/.style={thick,line join=round,dashed}}
\tikzset{integrationRegionEdge/.style={dashed}}

\tikzstyle{anode}= [circle, inner sep=1.3pt, draw, fill=black]
\tikzstyle{gausspoint}=[shape=circle,draw=black,fill=black,inner sep=0.8pt]
\tikzstyle{controlPoint}=[shape=circle,draw=black,fill=white,thin,inner sep=0pt,minimum size=1.5mm]
\tikzstyle{abscissaPoint}=[shape=circle,draw=black,fill=white,thin,inner sep=0pt,minimum size=1.5mm]
\tikzstyle{anchorPoint}=[shape=circle,draw=black,fill=black,thin,inner sep=0pt,minimum size=1.5mm]
\tikzstyle{anchorPointDeg}=[shape=cross out,thick,draw=black,inner sep=0pt,minimum size=1.5mm]
\tikzstyle{trimmingIntersectionPoint}=[shape=cross out,thick,draw=black,inner sep=0pt,minimum size=1.5mm]
\tikzstyle{approxnode}= [circle, inner sep=1.5pt, draw, fill=black]
\tikzstyle{characterNode}= [shape=cross out,rotate=45, inner sep=1.8pt, draw,very thick,fill=white]

\tikzstyle{aPointPatchA}= [circle, inner sep=1.3pt, draw, fill=black]
\tikzstyle{aPointPatchB}= [circle, inner sep=1.3pt, draw, fill=white]
\tikzstyle{aPointModelSpace}=[shape=cross out,thick,draw=black,inner sep=0pt,minimum size=1.5mm]

\def\trianglecolor{black}
\newcommand{\upperSlopeTriangle}[4] 	
{
	\addplot[forget plot, domain=#3:#4,color=\trianglecolor,samples=2]{  #2 / (x^#1) } node (A1) [pos=1] {}; 
	\addplot[forget plot, domain=#3:#4,color=\trianglecolor,samples=2]{   #2  / (#3^#1)} node (A2) [pos=1] {} node [anchor=south,pos=0.5,black] {\tikzfontsizetiny $1$};
	\draw[color=\trianglecolor] (A1.center) -- (A2.center) node [anchor=west,pos=0.5,black] {\tikzfontsizetiny #1};
}

\newcommand{\lowerSlopeTriangle}[4] 	
{
	\addplot[forget plot, domain=#3:#4,color=\trianglecolor,samples=2]{  #2 / (x^#1) } node (A1) [pos=0] {}; 
	\addplot[forget plot, domain=#3:#4,color=\trianglecolor,samples=2]{   #2  / (#4^#1)} node (A2) [pos=0] {} node [anchor=north,pos=0.5,black] {\tikzfontsizetiny $1$};
	\draw[color=\trianglecolor] (A1.center) -- (A2.center) node [anchor=east,pos=0.5,black] {\tikzfontsizetiny #1};
}

\newcommand{\myaddgraphic}[5]
{
 \node[anchor=south west,inner sep=0] (image) {\phantom{\includegraphics[#2]{#1}}};
  \begin{scope}[x={(image.south east)},y={(image.north west)}]
      
      \begin{scope}
          
          #5
          
          \node[anchor=south west,inner sep=0] {\includegraphics[#2]{#1}};
      \end{scope} 
      
      #4
      
      \pgfmathparse{int(#3)} \let\gridIndicator\pgfmathresult
      \ifthenelse{ \gridIndicator = 1 }
      {
          \draw[help lines,xstep=.1,ystep=.1] (0,0) grid (1.001,1.001);
          \foreach \x in {1,...,9} { \node [anchor=north] at (\x/10,0) {\x};}
          \foreach \y in {1,...,9} { \node [anchor=east] at (0,\y/10) {\y};}
      }{}
      
  \end{scope}    
}

\newcounter{itR}
\newcommand{\bsplinevalue}[5] 
{                    				
	\newarray\vKnots
	\newarray\vN
	\newarray\vNumeratorL
	\newarray\vNumeratorR
	\newarray\vSave

	\readarray{vKnots}{#1}
	\readarray{vN}{1}
	\readarray{vSave}{0}
	\readarray{vNumeratorL}{0}
	\readarray{vNumeratorR}{0}

	\foreach \j in {1,...,#2}
	{        
		\pgfmathparse{ int(#4+\j+1) } \checkvKnots(\pgfmathresult)
		\pgfmathsetmacro{\numR}{\cachedata-#3}
		
		\pgfmathparse{ int(#4-\j+2) } \checkvKnots(\pgfmathresult)
		\pgfmathsetmacro{\numL}{#3-\cachedata} 					

		\expandarrayelementtrue
		\pgfmathparse{ int(\j+1) }
		\vNumeratorL(\pgfmathresult)={\numL}
		\vNumeratorR(\pgfmathresult)={\numR}
		
		\forloop[1]{itR}{0}{\value{itR} < \j }
		{
			\pgfmathparse{ int(\theitR+1) } \let\tS\pgfmathresult  	
			\checkvSave(\tS)							
			\pgfmathsetmacro{\save}{\cachedata}  
		
			\pgfmathparse{int(\j-\theitR+1)} \checkvNumeratorL(\pgfmathresult)
			\pgfmathsetmacro{\tmpL}{\cachedata} 	
		    
			\pgfmathparse{ int(\theitR+2) } \checkvNumeratorR(\pgfmathresult)
			\pgfmathsetmacro{\tmpR}{\cachedata} 	       
	
			\pgfmathparse{ int(\theitR+1) } \checkvN(\pgfmathresult)
			\pgfmathparse{ \cachedata / (\tmpL+\tmpR) } \let\tmp\pgfmathresult
			
			\pgfmathparse{ \save + \tmpR * \tmp } 
			\vN(\tS)={\pgfmathresult}

			\pgfmathparse{ \tmpL * \tmp } \let\tmpsave\pgfmathresult
			\pgfmathparse{ int(\theitR+2) } \let\tS\pgfmathresult      
			\vSave(\tS)={\tmpsave}
		}
		
		\pgfmathparse{ int(\j+1) } \checkvSave(\pgfmathresult )
		\vN(\tS)={\cachedata}
	}

	\pgfmathparse{int( #2+1) } \let\lastIndex\pgfmathresult
	\checkvN(1) \pgfmathsetmacro{\first}{\cachedata}  
	\foreach \i [remember=\a as \lasta (initially \first)] in {2,...,\lastIndex}
	{
		\checkvN(\i) \def\a{\lasta,\cachedata} 
		\ifthenelse{\i=\lastIndex}{ \xdef#5{\a} }{}
	}
    
	\foreach \i in {1,...,\lastIndex}
	{
	    \clrarray{vNumeratorR}(\i)
	    \clrarray{vNumeratorL}(\i)
	    \clrarray{vN}(\i)
	    \clrarray{vSave}(\i)
	}
	\delarray\vN
	\delarray\vNumeratorL
	\delarray\vNumeratorR
	\delarray\vSave
}

\newcounter{countvalues}
\newcounter{getBasis}
\newcommand{\bsplinebasis}[4] 
{						
    \newarray\vKnots 	
    \readarray{vKnots}{#1}

    \pgfmathparse{#3}  \let\i\pgfmathresult
    \pgfmathparse{#2}  \let\p\pgfmathresult

    \setcounter{countvalues}{0}
    \setcounter{getBasis}{\p} 
    \pgfmathparse{int(\i+\p)}
    \foreach \knotspan in {\i,...,\pgfmathresult}
    {
	\pgfmathparse{ int(\knotspan+1+1) } \checkvKnots(\pgfmathresult)
	\pgfmathsetmacro{\tmpR}{\cachedata} 						
	
	\pgfmathparse{ int(\knotspan+1) } \checkvKnots(\pgfmathresult) 
	\pgfmathsetmacro{\tmpL}{\cachedata} 						

	\pgfmathparse{ \tmpR - \tmpL } \let\spansize\pgfmathresult  			
   
        \pgfmathparse{ \spansize > 0.0 } \let\bNonZero\pgfmathresult
        \ifthenelse{ \bNonZero = 1 }
        {
            \foreach \percentU in {0,10,...,100}
            {
		\pgfmathparse{\tmpL+\spansize*\percentU/100} \let\u\pgfmathresult

		\bsplinevalue{#1}{\p}{\u}{\knotspan}{\Basis} 
		\def\basisfuncarray{{\Basis}} 				
		
		\pgfmathparse{\basisfuncarray[\thegetBasis]} 
		
		\ifthenelse{\thecountvalues=0}
		{ 
			\xdef\nodeB{"\u,\pgfmathresult"}
		}{
			\xdef\nodeB{\nodeB,"\u,\pgfmathresult"}  
		}
		\addtocounter{countvalues}{1}
            }
        }{}
        \addtocounter{getBasis}{-1}
    }
    
	\xdef#4{\nodeB}

	\delarray\vKnots
}

\newcommand{\bsplinebasisspan}[6] 
{							
    \newarray\vKnots 	
    \readarray{vKnots}{#1}

    \pgfmathparse{#5}  \let\plotknotspan\pgfmathresult
    \pgfmathparse{#4}  \let\splineknotspan\pgfmathresult
    \pgfmathparse{#3}  \let\i\pgfmathresult
    \pgfmathparse{#2}  \let\p\pgfmathresult

    \setcounter{countvalues}{0}
    \setcounter{getBasis}{\i} 
    \foreach \knotspan in {\plotknotspan}
    {
	\pgfmathparse{ int(\knotspan+1+1) } \checkvKnots(\pgfmathresult)
	\pgfmathsetmacro{\tmpR}{\cachedata} 						
	
	\pgfmathparse{ int(\knotspan+1) } \checkvKnots(\pgfmathresult) 
	\pgfmathsetmacro{\tmpL}{\cachedata} 						

	\pgfmathparse{ \tmpR - \tmpL } \let\spansize\pgfmathresult  			
   
        \pgfmathparse{ \spansize > 0.0 } \let\bNonZero\pgfmathresult
        \ifthenelse{ \bNonZero = 1 }
        {
            \foreach \percentU in {0,10,...,100}
            {
		\pgfmathparse{\tmpL+\spansize*\percentU/100} \let\u\pgfmathresult

		\bsplinevalue{#1}{\p}{\u}{\splineknotspan}{\Basis} 
		\def\basisfuncarray{{\Basis}} 				
		
		\pgfmathparse{\basisfuncarray[\thegetBasis]} 
		
		\ifthenelse{\thecountvalues=0}
		{ 
			\xdef\nodeB{"\u,\pgfmathresult"}
		}{
			\xdef\nodeB{\nodeB,"\u,\pgfmathresult"}  
		}
		\addtocounter{countvalues}{1}
            }
        }{}
        \addtocounter{getBasis}{-1}
    }
    
	\xdef#6{\nodeB}

	\delarray\vKnots
}

\newcommand{\plotbsplinebasis}[4] 	
{							
	\bsplinebasis{#1}{#2}{#3}{\nodeOut}
	\def\nodearray{{\nodeOut}}

	\xdef\name{ }
	\addtocounter{countvalues}{-1}
	\foreach \i in {0,...,\thecountvalues}
	{
		\pgfmathparse{\nodearray[\i]}
		\coordinate (point\i) at (\pgfmathresult);	  
		\xdef\name{ \name (point\i)  }
	}
	
	\draw[#4] plot coordinates{ \name };
	
	\xdef\name{ }
}

\newcommand{\plotbsplinesegment}[6] 	
{								
	\bsplinebasisspan{#1}{#2}{#3}{#4}{#5}{\nodeOut}
	\def\nodearray{{\nodeOut}}

	\xdef\name{ }
	\addtocounter{countvalues}{-1}
	\foreach \i in {0,...,\thecountvalues}
	{
		\pgfmathparse{\nodearray[\i]}
		\coordinate (point\i) at (\pgfmathresult);	  
		\xdef\name{ \name (point\i)  }
	}
	
	\draw[#6] plot coordinates{ \name };
	
	\xdef\name{ }
}

\newcommand{\plotbsplineaccumulated}[5] 	
{						
    \newarray\vKnots 	
    \readarray{vKnots}{#1}
    \newarray\vSubCoef 	
    \readarray{vSubCoef}{#3}
    
    \pgfmathparse{#4}  \let\plotknotspan\pgfmathresult
    \pgfmathparse{#4}  \let\splineknotspan\pgfmathresult
    \pgfmathparse{#2}  \let\p\pgfmathresult
    
    \setcounter{countvalues}{0}
    \pgfmathparse{int( \p+1) } \let\lastIndex\pgfmathresult
    \foreach \knotspan in {\plotknotspan}
    {
        \pgfmathparse{ int(\knotspan+1+1) } \checkvKnots(\pgfmathresult)
        \pgfmathsetmacro{\tmpR}{\cachedata} 						
        
        \pgfmathparse{ int(\knotspan+1) } \checkvKnots(\pgfmathresult) 
        \pgfmathsetmacro{\tmpL}{\cachedata} 						
        
        \pgfmathparse{ \tmpR - \tmpL } \let\spansize\pgfmathresult  			
        
        \pgfmathparse{ \spansize > 0.0 } \let\bNonZero\pgfmathresult
        \ifthenelse{ \bNonZero = 1 }
        {
            \foreach \percentU in {0,10,...,100}
            {
                \pgfmathparse{\tmpL+\spansize*\percentU/100} \let\u\pgfmathresult
                
                \bsplinevalue{#1}{\p}{\u}{\splineknotspan}{\Basis} 
                \def\basisfuncarray{{\Basis}} 				
                
                \setcounter{getBasis}{0} 
                \pgfmathparse{\basisfuncarray[\thegetBasis]} 
                \let\basisValue\pgfmathresult
                
                \checkvSubCoef(1) \pgfmathsetmacro{\coef}{\cachedata}  
                \pgfmathparse{ \basisValue * \coef } \let\first\pgfmathresult
                
                \xdef\lastx{\first}
                \foreach \i in {2,...,\lastIndex}
                { 
                    \addtocounter{getBasis}{1}       
                    \pgfmathparse{\basisfuncarray[\thegetBasis]}
                    \let\basisValue\pgfmathresult
                    
                    \checkvSubCoef(\i) \pgfmathsetmacro{\coef}{\cachedata}  
                    \pgfmathparse{ \lastx + \basisValue * \coef } \let\sum\pgfmathresult
                    
                    \xdef\lastx{\sum}
                    
                    \ifthenelse{\i=\lastIndex}
                    {
                        \ifthenelse{\thecountvalues=0}
                        { 
                            \xdef\nodeBB{"\u,\sum"}
                        }{
                            \xdef\nodeBB{\nodeBB,"\u,\sum"}  
                        }
                        \addtocounter{countvalues}{1}
                    }{}
                    
                }
            }
        }{}
    }
    
    \delarray\vKnots
    \delarray\vSubCoef
    
    \def\nodearray{{\nodeBB}}
    
    \xdef\name{ }
    \addtocounter{countvalues}{-1}
    \foreach \i in {0,...,\thecountvalues}
    {
        \pgfmathparse{\nodearray[\i]}
        \coordinate (point\i) at (\pgfmathresult);                
        \xdef\name{ \name (point\i)  }
    }
    
    \draw[#5] plot coordinates{ \name };
    
    \xdef\name{ }
}

\newcommand{\lyxaddress}[1]{
	\par {\raggedright #1
		\vspace{1.4em}
		\noindent\par}
}

\numberwithin{equation}{section}

\makeatother

\graphicspath{{Pics/}}
\def\pathToBibFile{References}

\begin{document}

\def\mytitle{A Consistent Higher-Order Isogeometric Shell Formulation}

\ifdefined\TodoListsOn
\begin{center}
    \huge Draft \today
\end{center}

\section*{CONCEPT SHEET} 

Summary of WebEx meeting from 8.April 2020:
\begin{itemize}
    \item Main aim: presenting a framework for a higher-order consistent shell formulation.
    \item The focus is NOT the simulation on engineering CAD models; trimming is applied just to realize domains without corners.
    \item Utilizing the results of Daniel's IGA2019 presentation for the first draft.
    \item If needed, comparison studies of the proposed higher-oder consistent approach and ad-hoc solutions. These studies shall focus on a specific aspect (e.g., integration accuracy, conditioning, ... ) rather than the convergence of the entire simulation. We will decide the need for these studies once the first draft is ready.
\end{itemize}

\section*{TODOs}

\begin{enumerate}
    \item Setting up the paper template \checkmark
    \item Literature review -- research gap \checkmark
    \item Method description \checkmark
    \item Adding of existing numerical results \checkmark
    \item Reading of first draft
    \item Abstract and conclusion 
    \item Fixing a title\\ (Working title ``\mytitle'')  
\end{enumerate}
\fi

\title{\mytitle}

\author{Daniel Schöllhammer$^{1}$, Benjamin Marussig$^{2}$, Thomas-Peter Fries$^{3}$}
\maketitle

\lyxaddress{
\begin{center}
    $^{1,3}$Institute of Structural Analysis\\
    Graz University of Technology\\
    Lessingstraße  25/II, 8010 Graz, Austria\\
    \texttt{www.ifb.tugraz.at}\\
    \vspace{0.75cm}
    $^{2}$Institute of Applied Mechanics\\
    Graz University of Technology\\
    Technikerstraße 4/II, 8010 Graz, Austria\\
    \texttt{www.mech.tugraz.at}\\
    \vspace{0.75cm}
    $^{1}$\texttt{schoellhammer@tugraz.at}, $^{2}$\texttt{marussig@tugraz.at}, $^{3}$\texttt{fries@tugraz.at}
\end{center}
}

\newpage
\begin{abstract}
Shell analysis is a well-established field, but achieving optimal higher-order convergence rates for such simulations is a difficult challenge.
We present an isogeometric Kirchhoff-Love shell framework that treats every numerical aspect in a consistent higher-order accurate way. 
In particular, a single trimmed B-spline surface provides a sufficiently smooth geometry, and the non-symmetric Nitsche method enforces the boundary conditions.
A higher-order accurate reparametrization of cut knot spans in the parameter space provides a robust, higher-order accurate quadrature for (multiple) trimming curves, and the extended B-spline concept controls the conditioning of the resulting system of equations. Besides these components ensuring all requirements for higher-order accuracy, the presented shell formulation is based on tangential differential calculus, and level-set functions define the trimming curves. 
Numerical experiments confirm that the approach yields higher-order convergence rates, given that the solution is sufficiently smooth.

\textbf{Keywords:  Kirchhoff-Love shells; higher-order; Trimming; Isogemetric Analysis;} 	
\end{abstract} 

\newpage\tableofcontents\newpage

\clearpage
\newpage

\section{Introduction}

\ifdefined\TodoListsOn
\begin{itemize}
    \item[\checkmark] Motivation
    \item Literature review:
    \begin{itemize}
        \item[\checkmark] Isogeometric shells -- Daniel
        \item[\checkmark] Shells within the fictitious domain context -- Daniel        
        \item[\checkmark] Shells within the trimming context -- Benjamin
    \end{itemize}
    \item[\checkmark] Research gap: consistent higher-order KL-shell model  
    \item[\checkmark] Outline    
\end{itemize}

\mycomment{
    Thomas:
    \begin{itemize}
        \item[\checkmark] **Start with outline of "IGA general", ...
        \item ...then "IGA in KL-shells", 1-2 paragraphs with 10-20 refs.
        \item[\checkmark] **Outline "trimming": Another 1-2 paragraphs with 10-20 refs.
        \item[\checkmark] **Motivation: Focus is on higher-order in this work! Therefore, we acccept being below-state-of-the-art in the sense of only enabling single-patch analysis. But multi-patch analysis is inherently low-order due to the non-smoothness of the solution (across edges between patches which is well-documented in KL-shell theory). High-order in KL-shells is an "open" topic because the requirements (also on the geometry) are high => Motivate our attempt to solve this based on an FDM-approach in the trimmed parameters space.
        \item[\checkmark] **Mention the 3 major challanges of fictitious domain methods (FDMs) and sketch our apporach:
        (1) integration
        (2) stabilization
        (3) boundary conditions
        [This will be one paragraph with ~5-10 refs. for each challenge] \\
        Benjamin: references will be discussed in the corresponding subsection
    \end{itemize}
}

\mycomment{
Benjamin's suggestion for the structure of the motivation paragraph:
\begin{enumerate}
    \item[\checkmark] Isogeometric analysis rejuvenated research on Kirchhoff-Love shell formulations because splines allow a simple realization of the $C^1$ continuity required.
    \item[\checkmark] Another advantage of the isogeometric paradigm is that application of higher-order basis function is straightforward due to the recursive structure of B-splines.
    \item[\checkmark] However, higher-order accuracy involves more than using basis functions with a high degree/order. The analysis has to consider \emph{all} aspects from $\dots$, over $\dots$ to $\dots$ in a consistent higher-order fashion.  
    \item[\checkmark] To the best of the authors' knowledge, there exists no formulation in the literature that can claim full success in this regard. + there is no higher-order accurate shell formulation in the context of trimming.
    \item[\checkmark] An aggravating factor is that the classical benchmarks tests \cite{Belytschko1985a_BM}, which mark the current gold standard for shell formulations, do not allow for higher-order convergence rates because the solutions of these benchmarks are not smooth enough (e.g., due to point forces or corners) and the reference solutions are often reported with only a few number of significant digits (see e.g., Kiendl 2009). (maybe adding an example? -> no, but references to Benzaken 2020)
\end{enumerate}
}


\mycomment{Daniel:}

\begin{itemize}
	\item[\checkmark] \mycomment{Shell in IGA or shell theory in general?}
	\item[\checkmark] \mycomment{Shell in FDM apart from trimmed patches:
	\begin{itemize}
		\item Linear Membrane with linear Cut FEM and rough embedded in continuum (kink in displacement field is not considered): \cite{Cenanovic_2016a_DS} 
		\item Non-linear membrane with h.o.~Trace FEM: \cite{Fries_2019a_DS} 
		\item Linear Reissner--Mindlin shell with h.o.~Trace FEM: \cite{Schoellhammer_2020a_DS}
		\item Linear Kirchhoff--Love shell with Trace FEM Gfrerrer (Is already available available) \cite{Gfrerer_2020a_DS}
		\item (Linear 7-parameter shell; h.o.~lift of automatic generated linear construction of zero iso-surface; Restricted to \emph{one} level-set function -> definition of curved or more complex boundaries?? and integration of boundaries) \cite{Gfrerer_2019a_DS}
	\end{itemize}
}
\end{itemize}
\fi

Shells are components of utmost importance in numerous engineering applications, and hence, the development of more accurate mechanical models and numerical analysis schemes is a long-established \cite{naghdi1973_BM,Basar_1985a_DS}, yet everlasting \cite{Bischoff_2017a_DS,wu2019a_BM}, challenge.  
The finite element method (FEM) is the most common technique for the numerical treatment of shell formulations. In this context, isogeometric analysis (IGA) \cite{Hughes2005a} has proven itself as an efficient and robust paradigm that outperforms traditional structural mechanics simulations \cite{Cottrell2006a_BM,Kiendl_2009a_DS,Lipton2010a_BM}.
The underlying idea of IGA is quite simple: using the functions of spline-based geometry representations of computer-aided design (CAD) directly for performing analysis. 
Although this concept was -- and still is -- aimed at mitigating the profound inefficiencies in the conventional interaction of CAD and simulation tools, it turned out that IGA enables several computational benefits. 
The most important advantages for shells are that it provides (i) the most accurate representation of the geometry and (ii) high continuity between elements. 
Indeed, the latter rejuvenated research on Kirchhoff-Love (KL) shell formulations, because establishing the required $C^1$ continuity is straightforward using splines. 

Since the pioneering work for KL shells from Kiendl et al.~\cite{Kiendl_2009a_DS}, IGA shell analysis gained a notable amount of research activities. The curvilinear coordinates, which are implied by the map from the parameter space to the physical space of the NURBS patch, and the high continuity are a perfect fit for the discretization of the well-known shell equations with IGA. 
It is worth noting that IGA has also been considered for Reissner-Mindlin (RM) shells, which take transverse shear deformations into account.
A successful discretization of RM shells with isogeometric analysis is given in, e.g., \cite{Benson_2010a_DS}. Furthermore, an approach with exactly calculated directors in the frame of IGA is presented in \cite{Dornisch_2013a_DS,Dornisch_2014a_DS}. Hierarchical shells that eliminate transverse shear locking are shown in \cite{Echter_2013a_DS,Oesterle_2016a_DS,Oesterle_2017a_DS}. A mixed displacement approach that also eliminates membrane locking is elaborated in \cite{Bieber_2018a_DS,Zou_2020a_DS}. Isogeometric collocation methods for KL shells and RM shells are considered in \cite{Kiendl_2017a_DS,Maurin_2018a_DS}.\par

In this paper, we employ the rotation-free Kirchhoff-Love shell 
formulated in the frame of the tangential differential calculus (TDC) 
\cite{Schoellhammer_2018a_DS,Schoellhammer_2019c_DS}.
The resulting shell formulation is independent of the concrete geometry definition and enables shell analysis on implicitly defined surfaces where a parametrization is not available nor needed, see \cite{Schoellhammer_2020a_DS}. However, herein, the midsurface is explicitly specified by a single NURBS patch. 
The benefits of the TDC-based shell formulation in our context are that
the obtained shell equations are formulated in the global Cartesian coordinate system and may be presented in a more compact and intuitive fashion employing a symbolic notation, while the classical formulation is generally expressed in curvilinear coordinates using index notation. As a result, the typically rather cumbersome effective boundary forces, see, e.g., \cite{Benzaken_2020a_DS}, needed later on for Nitsche's method, are more compact, which simplifies the implementation.\par

The desire to perform shell analysis on practical engineering CAD models leads to the challenge of trimmed geometries. Trimming refers to a procedure that is used to specify visible areas of an object independent from the object's parameter space, and it is paramount for fundamental geometrical operations denoted as Boolean operations (union, intersection, and differencing). This concept provides great flexibility in defining arbitrary shapes over spline surfaces but introduces several computational issues as detailed in \cite{marussig2017a_BM}.
In essence, trimming results in cut parameter spaces, and thus, the analysis of trimmed geometries faces the same challenges as fictitious domain methods (FDMs) \cite{Burman2012a_BM,Hoellig2003b_BM,Schillinger2015a_BM}.
That is, the presence of \emph{cut elements} complicates
\begin{enumerate}
    \item the application of boundary conditions,
    \item the conditioning of system matrices, and
    \item accurate integration.
\end{enumerate}

In the context of IGA with trimmed geometries, 
the RM shell theory was adopted in \cite{Wang2013a_BM,Kang2016a_BM,Kang2015a_BM,Kang2016aa_BM}.
The first KL shell formulation was presented in \cite{Schmidt2012a_BM}.
The work of Breitenberger et al.~\cite{Breitenberger2016phd_BM,Breitenberger2015a_BM} develops the concepts further, enabling direct use of CAD data for the simulation.
While these works rely on a penalty approach for the application of boundary conditions and a boundary-fitted quadrature scheme, the contributions of Guo et al.~\cite{Guo2015a_BM,Guo2018a_BM,Guo_2017a_DS,Guo_2019a_DS} employ the symmetric or non-symmetric Nitsche's method and a non-boundary-fitted sub-cell quadrature; except for \cite{Guo2018a_BM} where this quadrature is used only for cut elements that cannot be handled by the blending function method presented in \cite{Kudela2015a}. 
The sub-cell quadrature originates from a FDM and includes the application of a fictitious stiffness in the exterior domain to mitigate the ill-conditioning issues due to cut elements.
The literature on isogeometric formulations for trimmed shells barely covers this conditioning aspect.
Besides the fictitious stiffness, Coradello et al.~\cite{Coradello2019a_BM} use diagonal scaling for preconditioning.

As mentioned above, trimming and FDMs are closely related from a computational point of view. For instance, membranes have been introduced for Cut FEM \cite{Cenanovic_2016a_DS} and for Trace FEM \cite{Fries_2019a_DS}. 
Furthermore, linear RM shells with Trace FEM have been presented in \cite{Schoellhammer_2020a_DS}. 
A recent pre-print where a $C^1$-continuous Trace FEM approach for KL shells with boundary conforming background meshes is presented in \cite{Gfrerer_2020a_DS}.

The motivation of the present work is the realization of a \emph{higher-order KL shell} analysis, 
which is an ``open'' topic because the requirements (also on the geometry) are high.
While setting up higher-order discretizations is straightforward using IGA, it is important to note that achieving higher-order accuracy involves more than the utilization of basis functions with a high polynomial degree. The analysis has to consider \emph{all} numerical aspects in a consistent higher-order fashion: starting from the discretization  over the application of boundary conditions to the construction of the system matrices.
To the best of the authors' knowledge, there exists no finite element shell model in the literature that can claim complete success in this regard.
In fact, there is no higher-order accurate shell formulation in the context of trimming.
It is, however, noted that the already mentioned works from Fries et al.~\cite{Fries_2019a_DS} on membranes and Sch\"{o}llhammer et al.~\cite{Schoellhammer_2020a_DS} on linear RM shells are presented in the framework of higher-order Trace FEM.

An aggravating factor for the development of higher-order shell formulations is that the traditional benchmarks tests \cite{Belytschko_1985a_DS}, which mark the current gold standard for validating shell models, do not allow for higher-order convergence rates.
First of all, the solutions of these benchmarks are not smooth enough (e.g., due to point forces or geometric corners), and second, the reference solutions are often reported with only a few numbers of significant digits (see e.g., \cite{Kiendl_2009a_DS}). 
A recent work \cite{Benzaken_2020a_DS} demonstrates that these benchmarks are ``unable to assess the order of accuracy'' and maybe passed by even variationally inconsistent formulations.

In this paper, we present a higher-order accurate KL shell formulation based on the isogeometric paradigm with trimmed geometries.
Trimming allows the definition of
arbitrary shapes 
on single-patch representations. 
We do not consider multi-patch models because they are inherently low-order due to the non-smoothness of the solution (across edges between patches, which is well-documented in KL-shell theory).
In general, any corner in the shell's boundary -- even for single-patch geometries -- may typically result in singularities, immediately hindering higher-order convergence rates in the analysis.
Using trimming also provides a means to prevent such discretizations.
As listed above, 
simulations with trimmed geometries involve 
three key challenges, which we address in a consistent higher-order accurate form.
In particular, the following methods are utilized:
\begin{enumerate}
    \item The non-symmetric Nitsche method enforces the boundary conditions.
    \item The extended B-spline concept controls the conditioning.
    \item Higher-order Lagrange elements provide a reparameterization of cut elements which allows for proper distribution of integration points.
\end{enumerate}
It is worth noting that different approaches for dealing with each of these problems exist. They may be used instead of the proposed ones,  as long as they can establish higher-order accuracy; if not,
the overall convergence will be sub-optimal. 
We are using level-set functions for the definition of the trimmed domain and formulate the KL shell in the frame of tangential differential calculus (TDC) \cite{Schoellhammer_2018a_DS,Schoellhammer_2019c_DS}. 
These two particular choices are, however, no necessary ingredients for the realization of higher-order accuracy and may be substituted by alternative concepts preferred by the reader.

Three numerical examples validate the performance of the proposed KL shell model: (i) a traditional shell benchmark, (ii) the flat shell embedded in $\mathbb{R}^3$ from \cite{Schoellhammer_2018a_DS}, and (iii) a curved, clamped circular shell.
The latter addresses the three trimming challenges, as well as the representation of the geometry and the boundary conditions. Hence, it allows an overall assessment of the order of accuracy.
In particular, we measure the error of this problem in the relative $L_2$-norm of the equilibrium in strong form, which requires the evaluation of fourth-order surface derivatives of the midsurface displacement. Consequently, the theoretical optimal order of convergence is $\mathcal{O}(p-3)$. The implementation of this residual error is not for free. Yet, it is a suitable measure for the verification of the higher-order accuracy. 
To be precise, it does not account for solution jumps between elements. Those errors, however, do not occur in the context of IGA due to the high inner-element continuity.

The paper is structured as follows:
\Cref{sec:GeometryAndTrimming} details the representation of trimmed geometries as well as the proposed concept for dealing with their conditioning and integration.
\Cref{sec:GoverningEquations} introduces the actual KL shell model and the non-symmetric Nitsche's method employed. The numerical results are given in \cref{sec:numres}.
We close with concluding remarks in \cref{sec:conclusion} and provide further details on implementational aspects in \cref{app:extrapolationWeights}.

\section{Geometry representation and treatment of trimmed domains}
\label{sec:GeometryAndTrimming}
\subsection{Geometry representation and discretization}
\label{sec:disc}

\ifdefined\TodoListsOn
\begin{itemize}
    \item[\checkmark] Continuity requirements (splines/isogeometric) -- Benjamin
    \item[\checkmark] Corner-free geometries (trimming by level-sets) -- Benjamin
    \item[\checkmark] Normal vector, local triad, tangent along boundary evaluation (representation of trim curve in $\R^3$)
\end{itemize}
\fi

We are aiming for a surface discretization that allows for (i) at least $C^1$ continuity, (ii) arbitrary degree, and (iii) arbitrary boundaries.
The continuity requirement stems from the Kirchhoff-Love shell theory (see \cref{sec:shelltheory}), and high-degree basis functions are a precondition for higher-order approximation power.
Splines are a natural choice to fulfill these two properties, but regular spline surfaces are defined by a tensor product structure, which results in a four-sided topology.
Besides the related geometric limitation, it is also noted that any corner in the shell boundary typically results in singularities, which may hinder the optimal convergence.
Hence, the concept of trimming is used to specify arbitrary shapes over spline surfaces.
To be precise, the discretization employs trimmed B-spline tensor product splines.   
In the following, we recall the basic properties of these functions.
The interested reader is referred to \cite{boor2001b_BM,Piegl1997b_BM,marussig2017a_BM} for a detailed discussion. 

\begin{figure}[t]
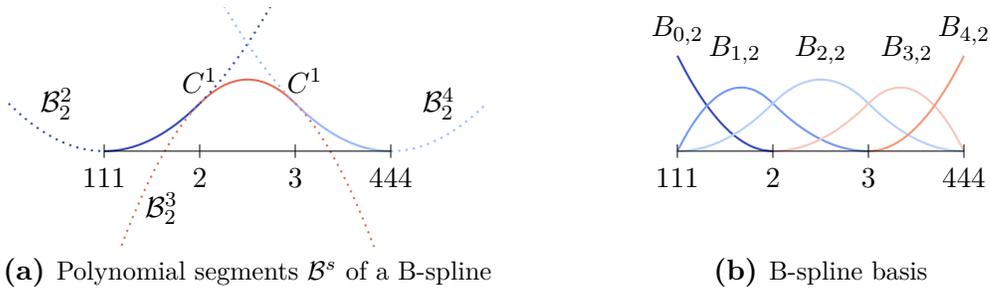

    
    \def\tkzscale{1.27}
    \def\dirtikz{tikz} 
    \def\dirdata{data} 
    \centering
    \tikzsetnextfilename{FigBsplineSegments}
    \subfloat[Polynomial segments $\BsplineSeg^{\indexSpan}$ of a B-spline]{\label{fig:BsplineExtension}\input{tikz/bsplineSegments.tex}}

    
    \def\tkzscale{1.27}
    \def\dirtikz{tikz} 
    \def\dirdata{data} 
    \centering
    \tikzsetnextfilename{FigBsplineBasis}
    \subfloat[B-spline basis]{\label{fig:BsplineBasis}\input{tikz/bsplineBasis.tex}}

    \caption{Quadratic B-splines defined by $\KV=\left\{1,1,1,2,3,4,4,4\right\}$ : (a) polynomial segments of the function $\Bspline_{2,2}$, which is part of (b) the entire B-spline basis. In (a), the resulting B-spline is illustrated by solid lines, whereas dotted lines indicate extensions of its segments.}
\end{figure}
A B-spline $\Bspline_{\indexA,\pu}$ is described by piecewise polynomial segments $\BsplineSeg^{\indexSpan}_{\indexA}$ of degree $\pu$, as shown in \cref{fig:BsplineExtension}. 
These $\BsplineSeg^{\indexSpan}_{\indexA}$ and the continuity between them are specified by the \emph{knot vector}~$\KV$, which is a non-decreasing sequence of parametric coordinates~{$\uu_\indexB \leqslant \uu_{\indexB+1}$}.  
The values of these \emph{knots} $\uu_\indexB$ define the location where adjacent $\BsplineSeg^{\indexSpan}_{\indexA}$ join. 
The continuity at these \emph{breakpoints} is $C^{\pu-\multi}$, with $\multi$ denoting the multiplicity of the corresponding knot value, i.e., $\uu_{\indexB} = \uu_{\indexB+1} = \dots = \uu_{\indexB+\multi-1}$. 
The \emph{knot span} $\indexSpan$ refers to the half-open interval $\left[\uu_{\indexSpan}, \uu_{\indexSpan+1}\right)$, and 
if its size is non-zero, 
it marks the valid region of $\BsplineSeg^{\indexSpan}_{\indexA}$. 
These regions also impose the \emph{elements} for the analysis.
The knot vector $\KV$ defines not only a single function, but an entire set of linearly independent B-splines~$\{\Bspline_{\indexA,\pu}\}_{\indexA=0}^{\ndofs}$, as illustrated in \cref{fig:BsplineBasis}. 
Each $\Bspline_{\indexA,\pu}$ has local support, $\supp{ \{\Bspline_{\indexA,\pu} \} }$, specified by the knots $\fromto{\uu_{\indexA}}{\uu_{\indexA+\pu+1}}$, and each knot span $\indexSpan$ contains $\pu+1$ non-zero B-splines represented by $\BsplineSeg^{\indexSpan}_\indexA$ with $\indexA=s-\pu,\dots,s$.

Basis functions of B-spline surfaces are obtained by computing the tensor product of univariate B-splines $\Bspline_{\indexA,\pusurf}$ and $\Bspline_{\indexB,\pvsurf}$ of  degrees $\pusurf$ and $\pvsurf$, which are defined by separate knot vectors $\KV_1$ and $\KV_2$  for the parametric directions $\uusurf$ and $\vvsurf$, respectively.
A corresponding surface is given by
\begin{align}
    \label{eq:BsplinePatch}
    \surface\operate{\UVsurf} =
    \sum_{\indexA=0}^{\totalA-1} \sum_{\indexB=0}^{\totalB-1} \Bspline_{\indexA,\pusurf} \operate{\uusurf}  \Bspline_{\indexB,\pvsurf}\operate{\vvsurf}    \: \CP_{\indexA,\indexB}
\end{align}
where $\UVsurf$ is a vector containing all parametric coordinates $\uu_\indexC$ with $\indexC=\{1,2\}$, and $\CP_{\indexA,\indexB}$ are control points in physical space $\R^3$. Following the isogeometric paradigm, we employ the same basis also for the test and trial functions.
Later, the surface's unit normal vector $\nG$ plays an essential role. It is determined by the cross-product 
\begin{align}
    \label{eq:normalvec}
    \nG = \frac{ \tangent_1 \times \tangent_2 }{ \| \tangent_1 \times \tangent_2 \|}
\end{align}
of the tangent vectors 
\begin{align}
    \label{eq:tangent}
    \tangent_1\operate{\UVsurf} &=
    \sum_{\indexA=0}^{\totalA-1} \sum_{\indexB=0}^{\totalB-1} \Bspline_{\indexA,\pusurf}^{(1)} \operate{\uusurf}  \Bspline_{\indexB,\pvsurf}\operate{\vvsurf}    \: \CP_{\indexA,\indexB} \\
    \tangent_2\operate{\UVsurf} &=
    \sum_{\indexA=0}^{\totalA-1} \sum_{\indexB=0}^{\totalB-1} \Bspline_{\indexA,\pusurf} \operate{\uusurf}  \Bspline_{\indexB,\pvsurf}^{(1)}\operate{\vvsurf}    \: \CP_{\indexA,\indexB}
\end{align}
where $\Bspline^{(1)}$ denotes the basis function's first derivative.  
Furthermore, we introduce the co-normal vector $\nCo$ along the surface's boundary $\p\surface$, which is given by
\begin{align}
    \label{eq:conormalvec}
    \nCo = \nG \times \tB
\end{align}  
with $\tB$ being the tangent vector along the boundary curve.
Note that \cref{eq:conormalvec} guarantees that $\nCo$ points out of the surface and lies in its tangent plane.
The resulting local triad $(\tB,\,\nCo,\,\nG)$ will be used
to properly represent the boundary conditions of the shell.

\begin{figure}[ht]
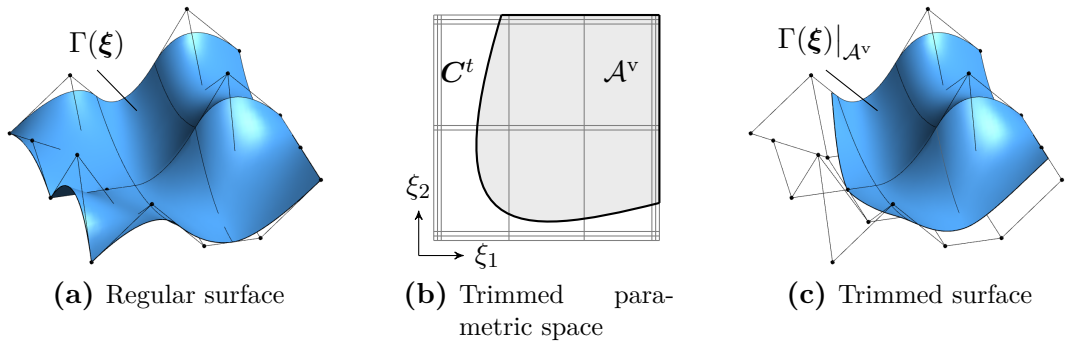

    
    \def\tkzscale{0.16}
    \def\dirtikz{tikz} 
    \def\dirdata{data} 
    \centering
    \tikzsetnextfilename{FigSplinePatch}
    \subfloat[Regular surface]{\label{fig:SplinePatch}\input{tikz/SplinePatch.tex}}

    \qquad
    
    \def\tkzscale{1.0}
    \def\dirtikz{tikz} 
    \def\dirdata{data} 
    \centering
    \tikzsetnextfilename{FigSplinePatchTrimmedBasis}
    \subfloat[Trimmed parametric space]{\label{fig:SplinePatchTrimmedBasis}\input{tikz/SplinePatchTrimmedBasis.tex}}

    \qquad
    
    \def\tkzscale{0.16}
    \def\dirtikz{tikz} 
    \def\dirdata{data} 
    \centering
    \tikzsetnextfilename{FigSplinePatchTrimmed}
    \subfloat[Trimmed surface]{\label{fig:SplinePatchTrimmed}\input{tikz/SplinePatchTrimmed.tex}}

    \caption{Trimmed surfaces: (a) regular surface as defined by \cref{eq:BsplinePatch}, (b) the related trimmed parametric space where trimming curves (thick lines) delineates the visible part $\visibledomain$ of (c) the resulting trimmed surface considered for the analysis.}
\end{figure}

Representation~\labelcref{eq:BsplinePatch} allows full control over the continuity and degree of the discretization, but it yields surfaces with an intrinsic four-sided topology due to its tensor product structure, as illustrated in \cref{fig:SplinePatch}.
Trimming provides a remedy to this restriction by defining the visible area~$\visibledomain$ of arbitrary shape over a surface $\surface\operate{\UVsurf}$. 
As indicated in \cref{fig:SplinePatchTrimmedBasis}, curves in the parametric space
specify $\visibledomain$.
Usually, these \emph{trimming curves} $\TrimCurve$ are also represented by B-splines \cite{marussig2017a_BM}, where the curve's orientation determines which regions are visible.
In this paper, we follow a different approach and define a $\TrimCurve$ as the zero-isoline of a level-set function 
$\LeveSet(\UVsurf)$. 
This choice is mainly motivated by the simplified detection of $\visibledomain$ during the analysis, i.e.,
\begin{align}
    \visibledomain \coloneqq \left\{ (\UVsurf) |  \LeveSet(\UVsurf) \geq 0\right\}.
\end{align}
We want to emphasize that the chosen representation of $\TrimCurve$ has no impact on the proposed methodology for achieving higher-order accuracy.

It remains to define the tangent vector along the boundary $\tB$ when $\p\surface$ is given by a trimming curve $\TrimCurvePS$. 
In this case, $\tB$ is no longer aligned with a parametric direction of the surface. 
Hence, we need a curve representation of $\TrimCurve$ in physical space, $\TrimCurvePS \in \R^3$, to compute $\tB$.
It is common practice to represent $\TrimCurvePS$ by an approximation because the exact description by the expansion of $\surface \circ \TrimCurve$ is of degree $\pu_t\left(\pvsurf+\pusurf\right)$ where $\pu_t$ refers to the degree of $\TrimCurve$, which is usually un-practically high.
As shall be detailed in \cref{sec:int}, knot spans cut by a trimming curve are decomposed into higher-order Lagrange elements. These elements are used to define quadrature rules in the trimmed domain and, furthermore, their edges provide higher-order accurate approximations of $\TrimCurvePS$, which we use to evaluate the tangent $\tB$. 
The normal vector $\nG$ and the co-normal vector $\nCo$, on the other hand, are computed using \cref{eq:normalvec} and \cref{eq:conormalvec}, respectively.

\subsection{Conditioning considerations}

\ifdefined\TodoListsOn
\begin{itemize}
    \item[\checkmark] Extended B-spline basics and application -- Benjamin
    \item[\checkmark] \mycomment{There is another stability issue related to the evaluation of the normal derivative, see \cite{buffa2019preprint_BM}. I do not think that extended B-splines resolve this issue, although the comments in \cite{Prenter2018a_BM} indicate this assumption. In any case, the non-symmetric Nitsche is considered to be weakly stable only \cite{arnold2002a_BM,Schillinger2016a_BM}, so I am not sure if this should be discussed here, at the Nitsche section, or even at all. $\to$ Daniel: Is discussed in \cref{sec:bc}}  
\end{itemize}
\fi

Analysis schemes that employ \emph{cut elements} may suffer from sever conditioning problems. 
In particular, ill-conditioning occurs when only a small portion of a basis function's support contributes to the overall system, which leads to small eigenvalues of the system matrix. For a detailed analysis of the conditioning issue, the interested reader is referred to \cite{Prenter2017a_BM}.

There are several remedies to this problem proposed in the literature.
The perhaps most simplistic approaches are diagonal scaling, e.g., \cite{Antolin2019a_BM,Coradello2019a_BM}, and the use of a fictitious domain stiffness, e.g., \cite{Dauge2015a_BM,Schillinger2015a_BM}. 
They are appealing due to their ease of implementation, but they do not address the actual source of the conditioning problems.
More adequate strategies are tailored preconditioning, e.g., \cite{Prenter2017a_BM,Prenter2020a_BM}, the ghost penalty, e.g., \cite{Burman2010a_BM,Burman2012a_BM,Burman2015a_BM}, and the modification of cut basis functions, e.g., \cite{Hoellig2003b_BM,Hoellig2003a_BM,marussig2016a_BM,marussig2018a_BM,Moessner2008a_BM}.   
Each of these different methods has its advantages, but a detailed comparison is beyond the scope of this paper. 
In the following, we focus on the latter approach as detailed in \cite{marussig2016a_BM}, as it is particularly well-suited for spline bases.
This so-called \emph{extended B-spline} concept removes basis functions with small support and replaces them by extensions of neighboring basis functions.
This is obviously a change of the approximation space, however, optimal approximation power is maintained, see \cite{Hoellig2003b_BM}.
As will be shown shortly, the extension procedure is done algebraically and is hence independent of the problem formulation and the structure of the resulting system matrix. 
The subsequent sections provide an introduction to extended B-splines and their application to a system of equations, while \cref{app:extrapolationWeights} gives details on the implementation.

\subsubsection{Definition of extended B-splines}
\label{ExtendedB-splines}

First, we classify the basis functions of a trimmed basis as either \emph{stable}, \emph{exterior}, or \emph{degenerate}, where the latter is referring to those that may cause conditioning problems.
The related identification utilizes the Greville abscissa (see \cite{marussig2016a_BM,marussig2018a_BM}) or the intersection of the support with the domain, i.e., $\supportdomain_\indexA \coloneqq  \Supp{ \Bspline_{\indexA,\pu} }  \cap \overline{\patchdomain}$. 
Here, we employ the latter strategy and 
label a B-spline $\Bspline_{\indexA,\pu}$ as
\begin{itemize}
	\item \emph{Stable} if $\supportdomain_\indexA \geq \supportalpha  \Supp{ \Bspline_{\indexA,\pu} }$, 
	\item \emph{Degenerate} if  $\supportdomain_\indexA < \supportalpha  \Supp{ \Bspline_{\indexA,\pu}  }$,
	\item \emph{Exterior} if $\supportdomain_\indexA = \emptyset$,
\end{itemize}
where $\supportalpha \in (0,1]$ is a user-defined threshold for the minimal relative support size considered to be stable.
Usually, we choose $\supportalpha$ between $0.4$ and $0.6$. 

\begin{figure}[thb]
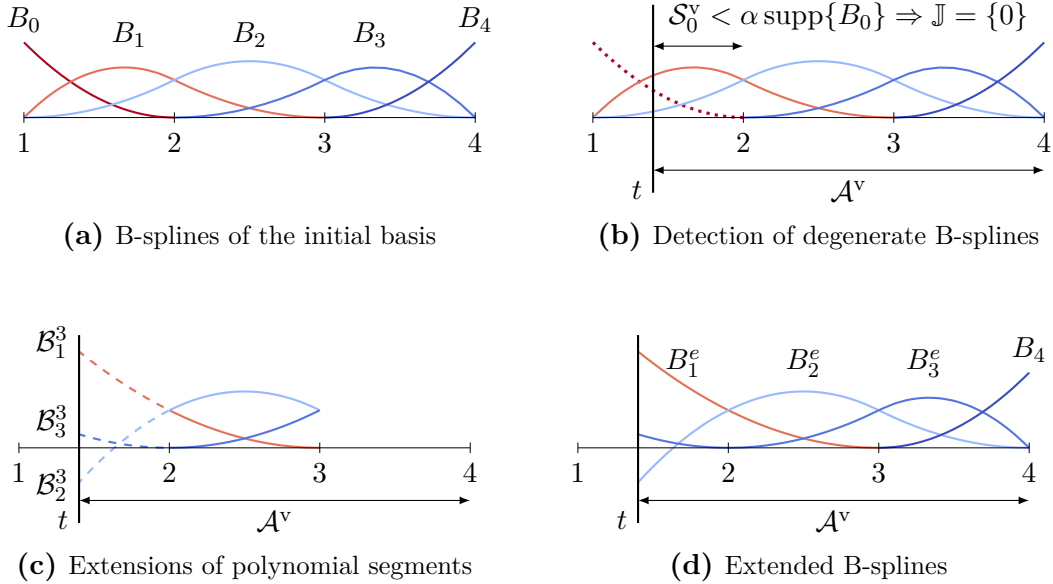

	\centering
	
    \def\tkzscale{1.0}
    \def\dirtikz{tikz} 
    \def\dirdata{data} 
    \centering
    \tikzsetnextfilename{FigExBsplineConceptA}
    \subfloat[B-splines of the initial basis]{\label{fig:extendedBsplineConcept_A}\input{tikz/conceptExBsplineA.tex}}

	\qquad
	
    \def\tkzscale{1.0}
    \def\dirtikz{tikz} 
    \def\dirdata{data} 
    \centering
    \tikzsetnextfilename{FigExBsplineConceptB}
    \subfloat[Detection of degenerate B-splines]{\label{fig:extendedBsplineConcept_B}\input{tikz/conceptExBsplineB.tex}}

    \def\tkzscale{1.0}
    \def\dirtikz{tikz} 
    \def\dirdata{data} 
    \centering
    \tikzsetnextfilename{FigExBsplineConceptC}
    \subfloat[Extensions of polynomial segments]{\label{fig:extendedBsplineConcept_C}\input{tikz/conceptExBsplineBB.tex}}

	\qquad
	
    \def\tkzscale{1.0}
    \def\dirtikz{tikz} 
    \def\dirdata{data} 
    \centering
    \tikzsetnextfilename{FigExBsplineConceptD}
    \subfloat[Extended B-splines]{\label{fig:extendedBsplineConcept_D}\input{tikz/conceptExBsplineC.tex}}
  
	\caption{
	Basic procedure to get from (a) conventional to (d) extended B-splines: (b)~determination of degenerate B-splines and substitution of trimmed polynomial segments by (c) extensions of non-trimmed ones. In (b), $\degSet$ is the index set of all degenerate B-splines.
	}
	\label{fig:extendedBsplineConcept}
  \end{figure}

Based on the resulting sets of stable and degenerate B-splines, we can now define extended B-splines. 
\Cref{fig:extendedBsplineConcept} illustrates the underlying idea: Polynomial segments of elements that contain degenerate B-splines are substituted by \emph{extensions} of stable ones.
These extensions are provided by the polynomial segments $\BsplineSeg_{\indexA}$ of the closest element that contains only stable B-splines.
The accumulation with all other stable functions yields the final extended B-spline basis. 

  %

In this construction, the extended polynomial segments $\BsplineSeg^{\indexSpan}_\indexA$ of the non-trimmed knot span~$\indexSpan$ are defined by the initial B-splines of the knot span~$\indexSpanTrimmed$ that contains degenerate functions by
\begin{align}
	\label{eq:extendedSplineInterpoalation}
	\BsplineSeg^{\indexSpan}_\indexA \operate{\uu} &= \sum_{\indexB=\indexSpanTrimmed-\pu}^{\indexSpanTrimmed}  { \Bspline_{\indexB,\pu}\operate{\uu} \: \eweight_{\indexA,\indexB} }  & \uu \in \left[ \uu_{\indexSpanTrimmed},\uu_{\indexSpanTrimmed+1} \right).
\end{align}
Hence, the definition of extended B-splines boils down to the determination of the \emph{extrapolation weights} $\eweight_{\indexA,\indexB}$.
If the index $\indexB$ corresponds to a stable B-spline, the weights are trivial, i.e.,  
\begin{align}
    \label{eq:exWeightsOne}
    \eweight_{\indexA,\indexA} = 1 && \Leftrightarrow && & \BsplineSeg^{\indexSpan}_\indexA\operate{\uu} \equiv \Bspline_{\indexA,\pu}\operate{\uu}, \quad \uu \in \left[ \uu_{\indexSpan},\uu_{\indexSpan+1} \right), \quad \forall \indexA \in \fromto{\indexSpan-\pu}{\indexSpan}, \\
    \label{eq:exWeightsZero}
    \eweight_{\indexA,\indexB} = 0 && \Leftrightarrow && & \BsplineSeg^{\indexSpan}_\indexA\operate{\uu} \neq \Bspline_{\indexB,\pu}\operate{\uu}, \quad \uu \in \left[ \uu_{\indexSpan},\uu_{\indexSpan+1} \right), \quad \forall \indexB \in \fromto{\indexSpan-\pu}{\indexSpan} \backslash \indexA.
\end{align}
The remaining extrapolation weights related to degenerate B-splines can be determined by
\begin{align}
	\label{eq:deBoorFix_explicit}
        \eweight_{{\indexA},{\indexB}} &= 
        \frac{1}{\pu!} \sum^\pu_{\indexC=0} (-1)^\indexC 
        \left( \pu - \indexC \right)! \: \beta_{\pu-\indexC} \:
        \indexC ! \:  \poly_\indexC
\end{align}
with the coefficients $\poly$ and $\beta$ denoting the constants of the polynomials
\begin{align}
	\BsplineSeg^{\indexSpan}_{\indexA}(\uu) = \sum^{\pu}_{\indexC=0} \poly_\indexC \: \uu^\indexC 
        &&\und&&
        \label{eq:deBoorFixNewtonBasis}
	\Nbasis_{\indexB,\pu}(\uu) = \prod_{\indexD=1}^\pu \left( \uu - \uu_{\indexB+\indexD}  \right) = \sum^{\pu}_{\indexC=0} \beta_\indexC \: \uu^{\indexC}.
\end{align}
The former is the power basis form of the extended polynomial segment, and the Newton polynomials $\Nbasis_{\indexB,\pu}$ result from a quasi interpolation procedure, i.e., the \emph{de Boor--Fix} functional \cite{boor2001b_BM,deBoor1973aa_BM}.
For more details, we refer to \cite{marussig2016a_BM,Hoellig2003b_BM} and \cref{app:extrapolationWeights} which provides further information on the determination of the extrapolation weights \labelcref{eq:deBoorFix_explicit}.

By taking all extrapolations weights, see Eqs.~\labelcref{eq:exWeightsOne}--\labelcref{eq:deBoorFix_explicit}, into account, an extended B-spline is defined as
\begin{align}
	\label{eq:extendedBspline}
	\exBspline_{\indexA,\pu} = \Bspline_{\indexA,\pu} + \sum_{\indexB \in \degSet_{\indexA}} \eweight_{\indexA,\indexB} \Bspline_{\indexB,\pu} 
\end{align}
where $\Bspline_{\indexA,\pu}$ is the stable B-spline from which the extension originates, and $\degSet_{\indexA}$ is the index set of all degenerate B-splines related to $\exBspline_{\indexA,\pu}$.
Definition \labelcref{eq:extendedBspline} applies to multivariate basis functions as well.
The corresponding extrapolation weights are obtained by the tensor product of the univariate components. 

\begin{remark}
	The degree of the basis affects the extended B-splines regarding the balance of improved conditioning and quality of the solution along the trimming curve.
	The parameter $\supportalpha$ provides a certain level of control, but the combination with local refinement, e.g., as suggested in \cite{marussig2018a_BM}, is a more comprehensive solution which is, however, not utilized in this paper.
\end{remark}

\subsubsection{Application of extended B-splines}
\label{sec:extensionMatrix}

A very convenient feature of extended B-splines is that they do \emph{not} need to be explicitly evaluated during an analysis. 
Suppose we set up the following linear system of equations using the initial -- potentially ill-conditioned -- B-splines basis 
\begin{align}
    \myMat{K}   \myVec{\primary} & = \myVec{f} &&  \where && 
    \myVec{\primary} \in \R^{m},\myVec{f} \in \R^{n} &&\und&& \myMat{K} \in \R^{ n \times m } &&\with&& m>n.
\end{align}
Here, $n$ refers to the number of stable B-splines, while $m$ includes the degenerate ones as well.
In order to get a well-conditioned square system matrix, $\myMat{K}$ is multiplied by an \emph{extension matrix} $\ExMatrix \in \R^{ m \times n }$ \cite{Hoellig2003b_BM}.
The entries of $\ExMatrix$ are the extrapolation weights $\eweight_{{\indexA},{\indexB}}$ of all extended B-splines. 
The trivial weights $\eweight_{{\indexA},{\indexA}} = 1$ are stored as well, even if a stable B-spline has no related degenerate ones.
The matrix entries of $\ExMatrix$ are assembled such that columns of the \emph{i}th row of $\myMat{K}$ are distributed according to the definition~\labelcref{eq:extendedBspline} of the associated extended B-spline $\exBspline_{\indexA,\pu}$. 
The final system of equations due to the extended B-spline basis is determined by
\begin{align}	
	\label{eq:stableSystem}
        {\myMat{K}_{st}}  {\myVec{\primary}_{st}} & = \myVec{f}  && \with && 
        {\myMat{K}_{st}} = \myMat{K} \ExMatrix, &&{\myMat{K}_{st}}\in \R^{ n \times n }.
\end{align}
The solution vector ${\myVec{\primary}_{st}}\in\R^{n}$ corresponds to the extended B-spline basis. Its relation to the original basis can also be expressed by the extension matrix as ${\myVec{\primary}} = \ExMatrix {\myVec{\primary}_{st}}$.

\begin{remark}
	The example given above considers a scalar problem. The generalization to vector-valued problems merely involves an appropriate repetition of the entries of $\ExMatrix$.
\end{remark}

\subsection{Numerical integration in cut knot spans}
\label{sec:int}

In isogeometric analysis, the integration of the weak form is classically performed through
an outer loop over all knot spans with an inner loop over quadrature
points. Reduced quadrature concepts following e.g.~\cite{Hiemstra2017a_BM,Johannessen2017a_BM,Schillinger2014a_BM,Auricchio2012aa_BM,Hughes2010a_BM}
use a tailored quadrature for the tensor-product B-splines or NURBS
bases and make a profit of the increased continuity of these bases.
Thus, they have shown a significantly higher efficiency in standard Galerkin IGA analyses. However, their application in the context of trimming
is delicate as trimmed basis functions lack the required continuity over the whole support being a
fundamental requirement of the reduced quadrature concepts.

Hence, we follow the classical path and use standard Gauss quadrature
in knot spans that are completely inside the domain. Obviously, no integration points are 
needed in knot spans that are fully outside the domain. However, the accurate integration
in knot spans cut by the trimming curve $\vek C^{t}$ requires
special attention. This is closely related to the context of fictitious
domain and extended finite element methods where domain boundaries
or interfaces freely cut the elements and various alternatives for
the integration in such elements exist. They differ in their ability
to enable higher-order accuracy and capture complex geometries including
corners.

The first approach to the quadrature in cut elements relies on the
decomposition into integration cells. Those may be polygonal cells
often obtained in recursive refinements of the cut element until the
desired accuracy is obtained \cite{Abedian_2013a,Moumnassi_2011a,Dreau_2010a}.
Nevertheless, the number of integration cells is often immense, even
more so in a higher-order context \cite{Stazi_2003a,Zi_2003a,Legrain_2012a,Laborde_2005a,Dreau_2010a}
which has direct implications to IGA. It was already noted in \cite{Legay_2005a,Cheng_2009a,Fries_2015a,Sala_2012a}
that a decomposition into sub-elements with curved, higher-order edges
is a strategy that consistently enables the generation of higher-order
accurate quadrature rules with only a modest number of integration
points. This approach has been investigated thoroughly in \cite{Fries_2015a,Omerovic_2016a,Fries_2016b,Fries_2016a}
and is followed herein. Other higher-order alternatives based on decompositions
are reported in \cite{Kudela_2016a,Stavrev_2016a}. Methods that do
\emph{not} decompose the cut elements are, e.g., found in \cite{Ventura_2006a,Ventura_2009a,Mueller_2013a,Mousavi_2011a,Joulaian_2016a}.
There, the number of integration points is also small, however, the
generation of proper integration weights is often involved and requires
the solution of small systems of equations, e.g., in the context of
moment fitting and Lasserre's technique.

For the standard context where the trimming curve is defined by means
of a B-spline or NURBS in the parameter space \cite{marussig2017a_BM,Sederberg_2008a},
this explicit definition is converted to an implicit level-set function
$\phi\left(\vek\xi\right)$ by computing signed distances which is called
\emph{implicitation} \cite{Chen_2008a,Hoffmann_1989a,Hoffmann_1993a,Sederberg_1984a,Selimovic_2006a,Stanford_2019a}.
The trimming curve may also be directly given in the implicit form.
In both cases, the level-set function is positive in the domain of
interest, $\phi\left(\vek\xi\right)>0$, and negative outside, $\phi\left(\vek\xi\right)<0$.
Most importantly, the trimming curve coincides with the zero-isoline
of $\phi\left(\vek\xi\right)$ in the parameter space, i.e., $\vek C^{t}=\partial\Gamma_{0}=\left\{ \vek\xi:\phi\left(\vek\xi\right)=0\right\} $.

Here, we adopt the procedure outlined in \cite{Fries_2015a,Omerovic_2016a,Fries_2016b,Fries_2016a}.
The basic idea is to first \emph{reconstruct} the zero-isoline by
curved, higher-order, interpolatory, and one-dimensional finite elements
and then \emph{decompose} the cut knot spans into higher-order, two-dimensional
finite elements serving as integration cells in the cut knot spans.

\subsubsection{Identification of the cut toplogy}

\begin{figure}
    \centering
    
    \includegraphics[width=0.25\textwidth]{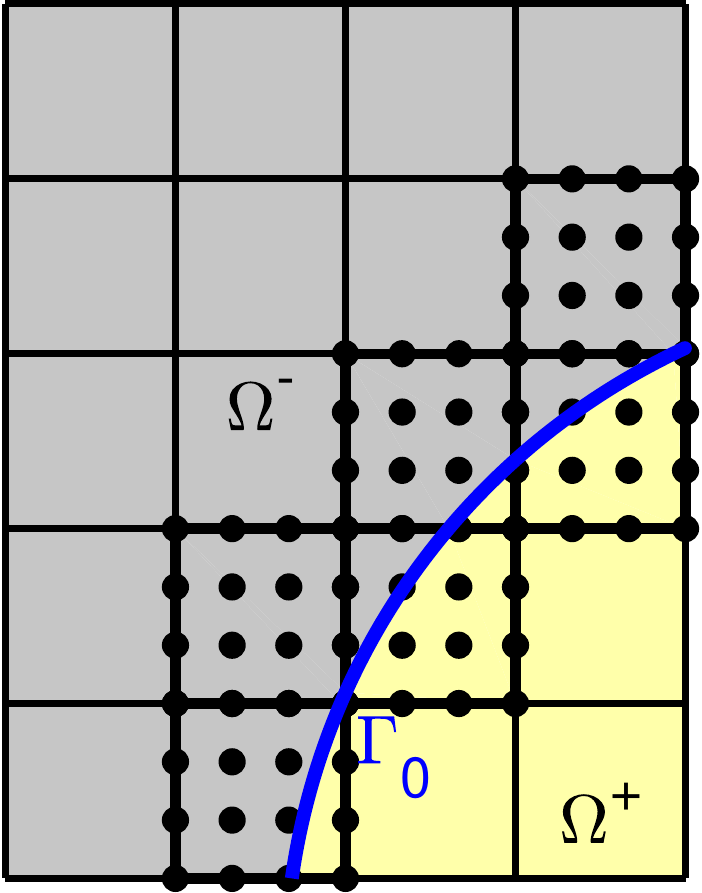}
    
    \caption{\label{fig:VisCutParamSpace}Lagrange elements are placed in cut knot
    spans of the parameter space. The level-set data is evaluated at the
    nodes in order to reconstruct integration cells.}
\end{figure}

\begin{figure}
    \centering
    
    \subfloat[level-set function $\phi^{h}$]{\includegraphics[width=3.7cm]{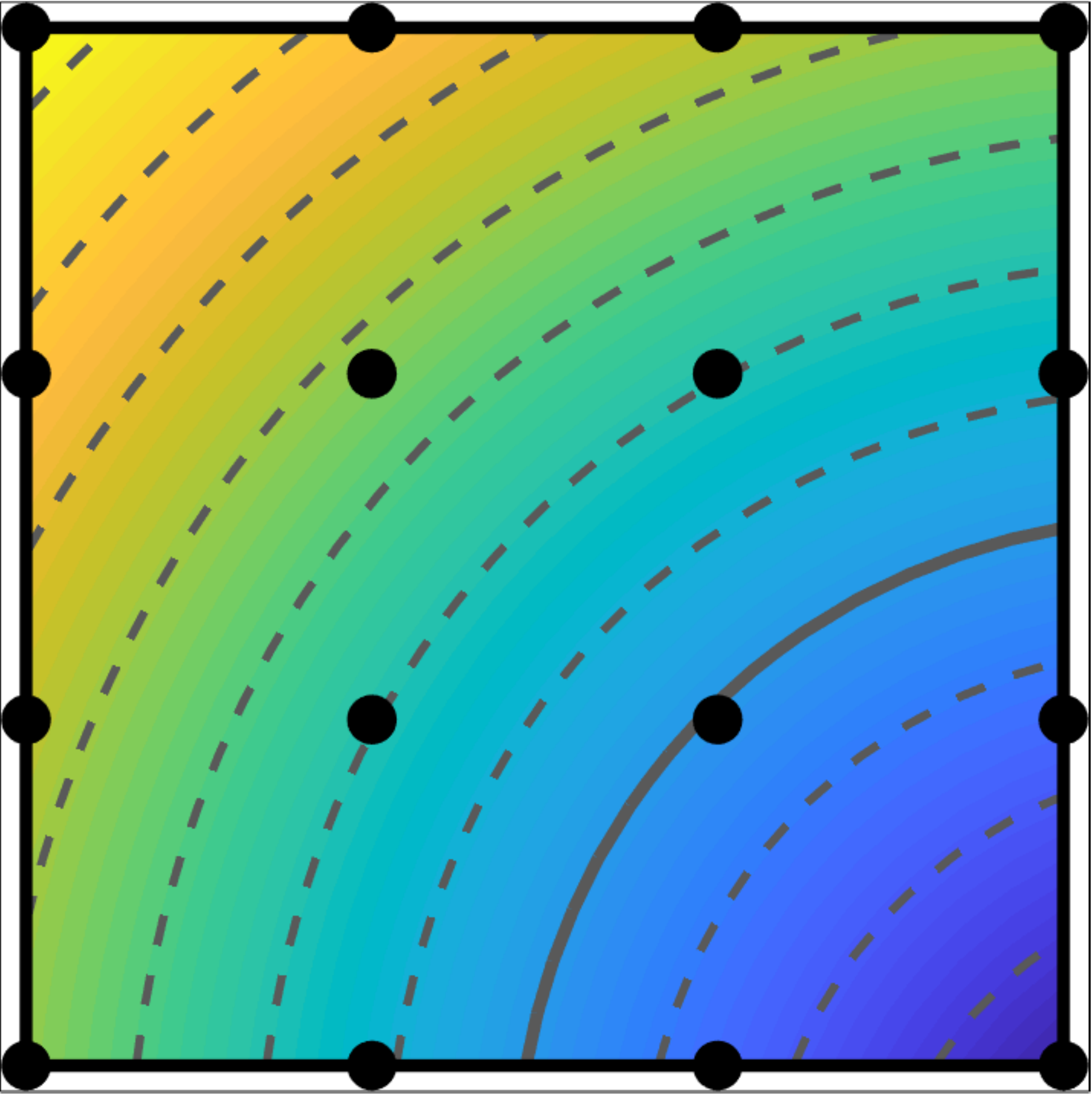}}\hfill\subfloat	
    [sample grid]{\includegraphics[width=3.7cm]{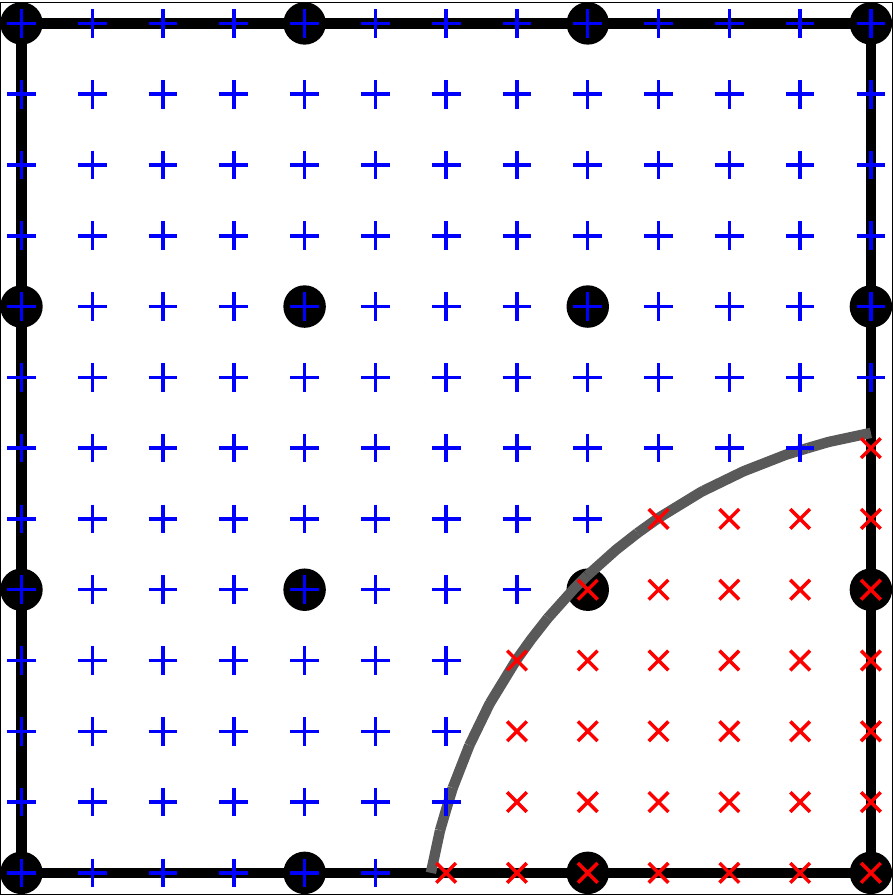}}\hfill\subfloat	
    [reconstruction]{\includegraphics[width=3.7cm]{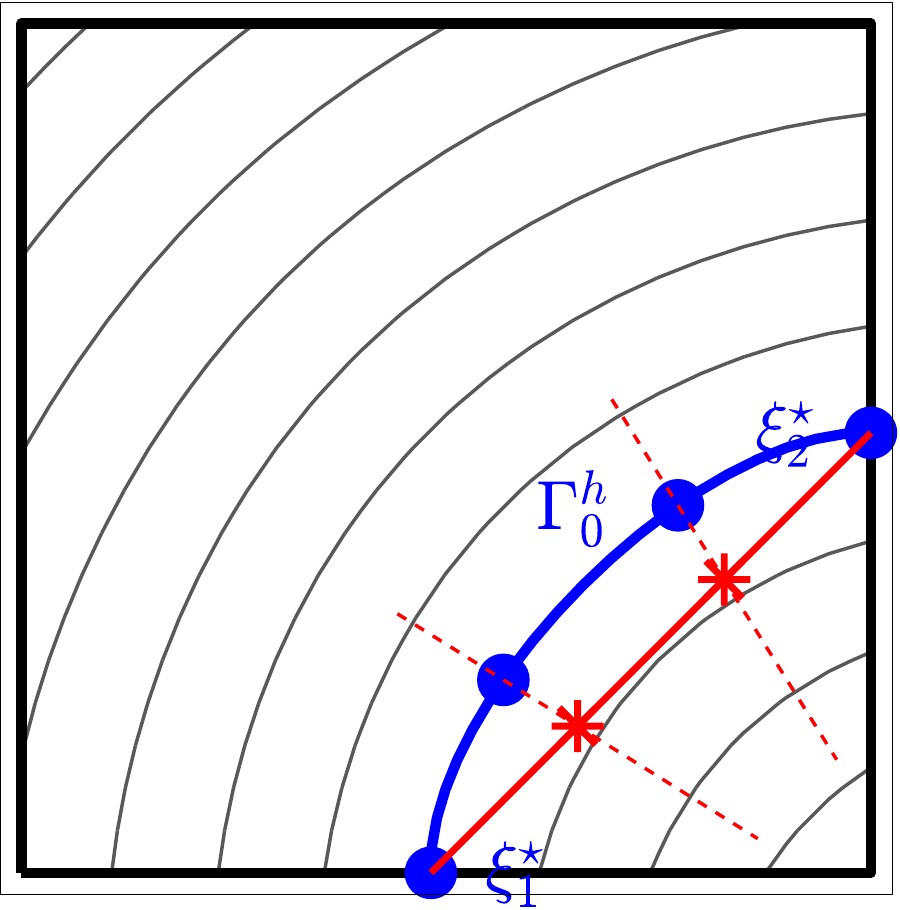}}\hfill\subfloat	
    [decomposition]{\includegraphics[width=3.7cm]{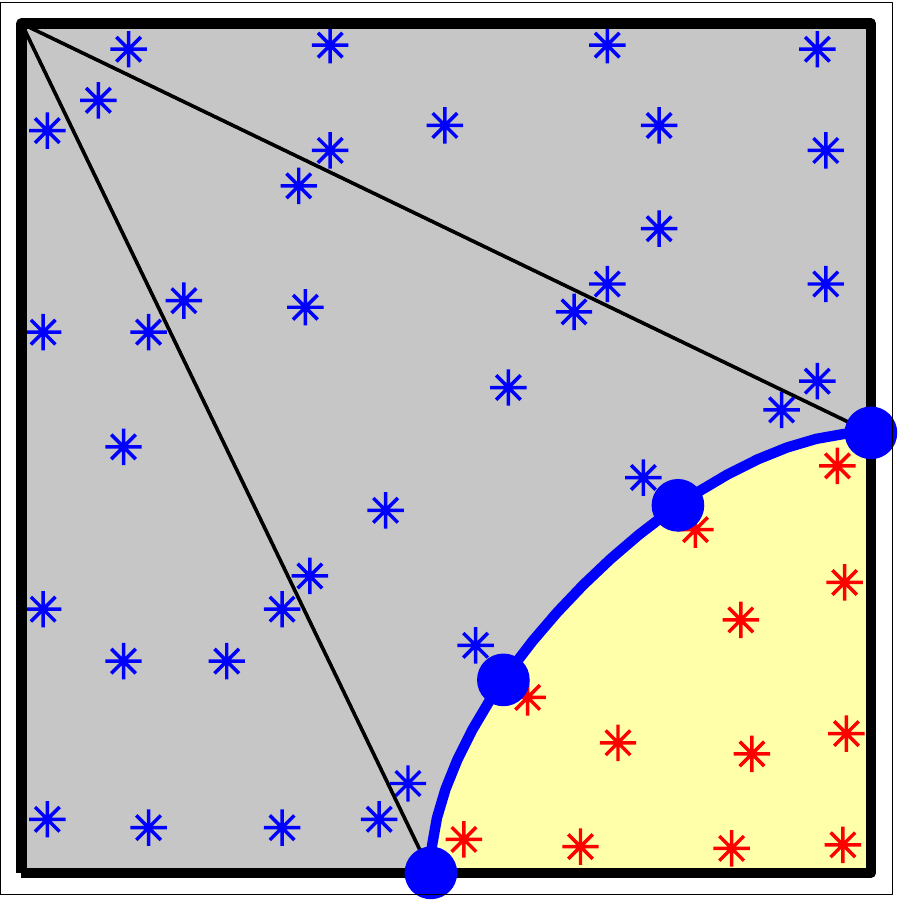}}
    
    \caption{\label{fig:VisCutType1}Two \emph{neighboring} edges of a higher-order
    finite element ($p=3$) located in a knot span are cut (topology type
    1): (a) the level-set function $\phi^{h}$ implied by the nodal level-set
    data, (b) the level-set data sampled at the grid points ($q=3$),
    (c) the reconstructed one-dimensional finite element representing
    $\partial\Gamma_{0}^{h}$, (d) the quadrature cells with resulting integration
    points.}
\end{figure}

\begin{figure}
    \centering
    
    \subfloat	
    [level-set function $\phi^{h}$]{\includegraphics[width=3.7cm]{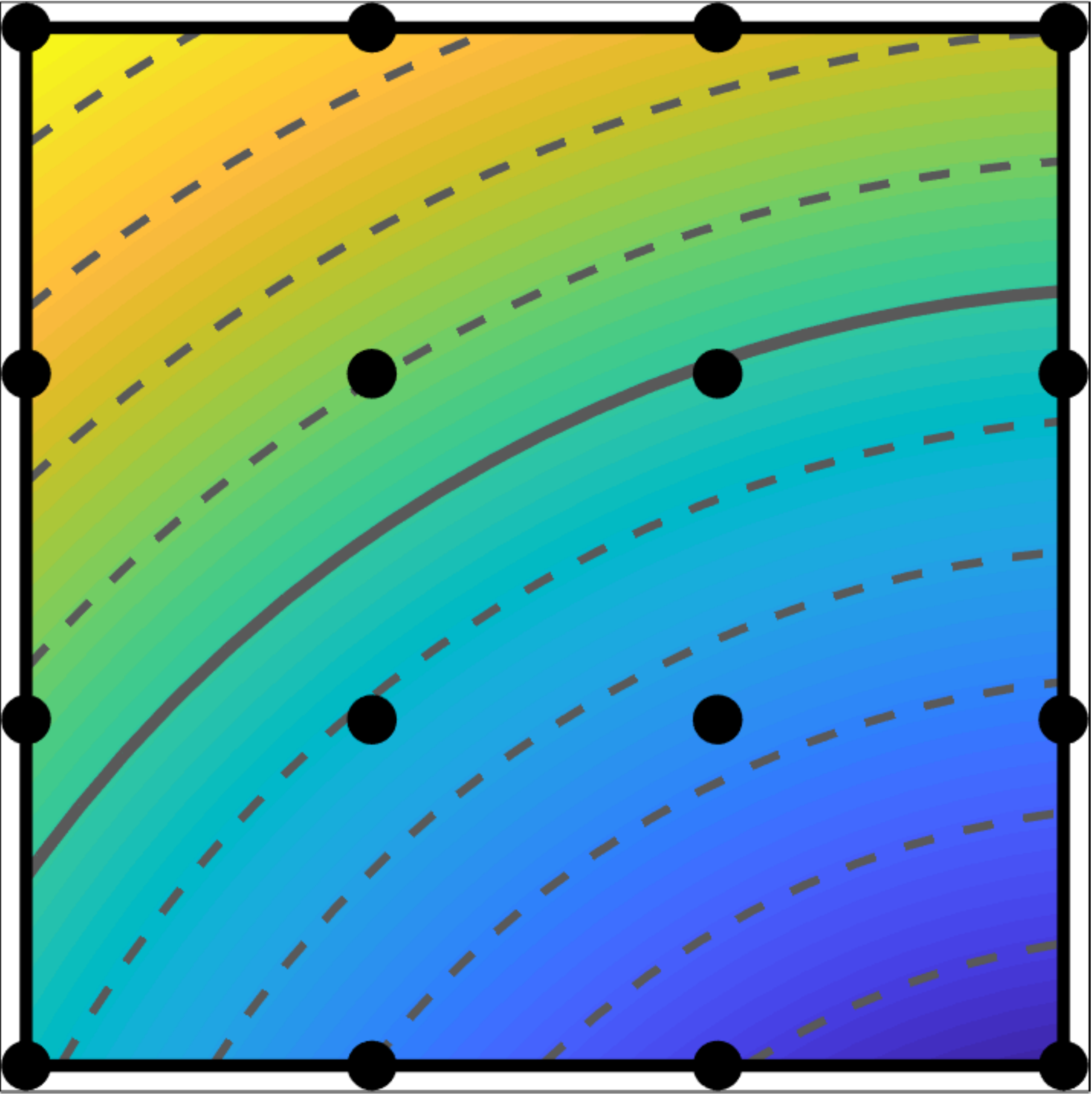}}\hfill\subfloat	
    [sample grid]{\includegraphics[width=3.7cm]{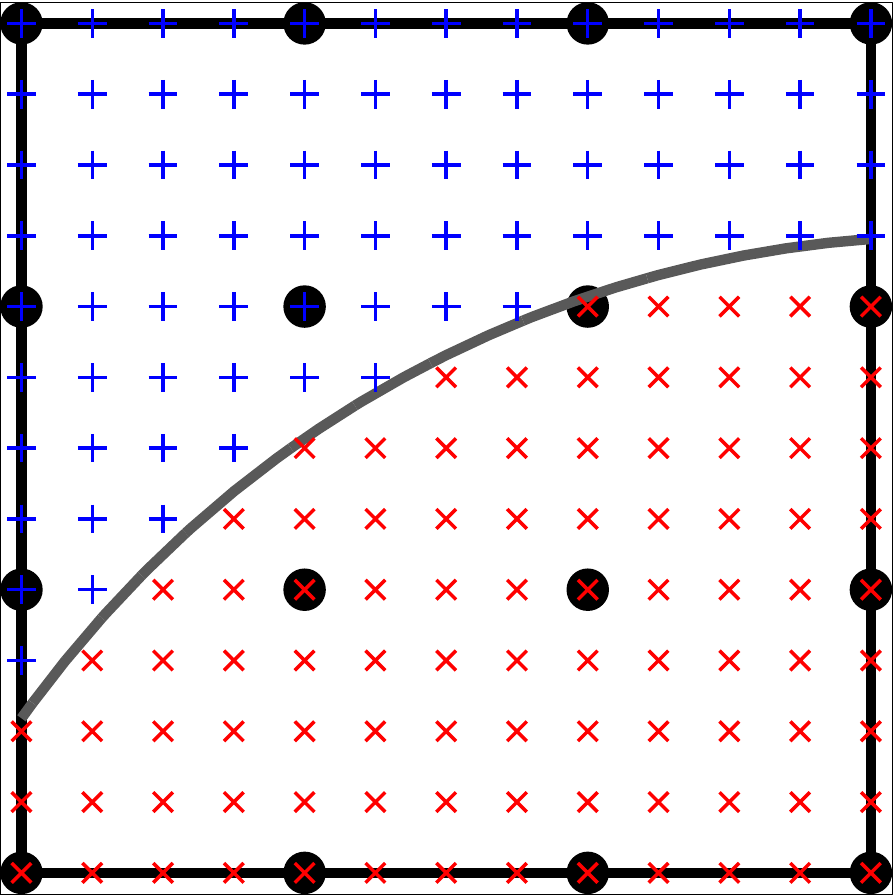}}\hfill\subfloat	
    [reconstruction]{\includegraphics[width=3.7cm]{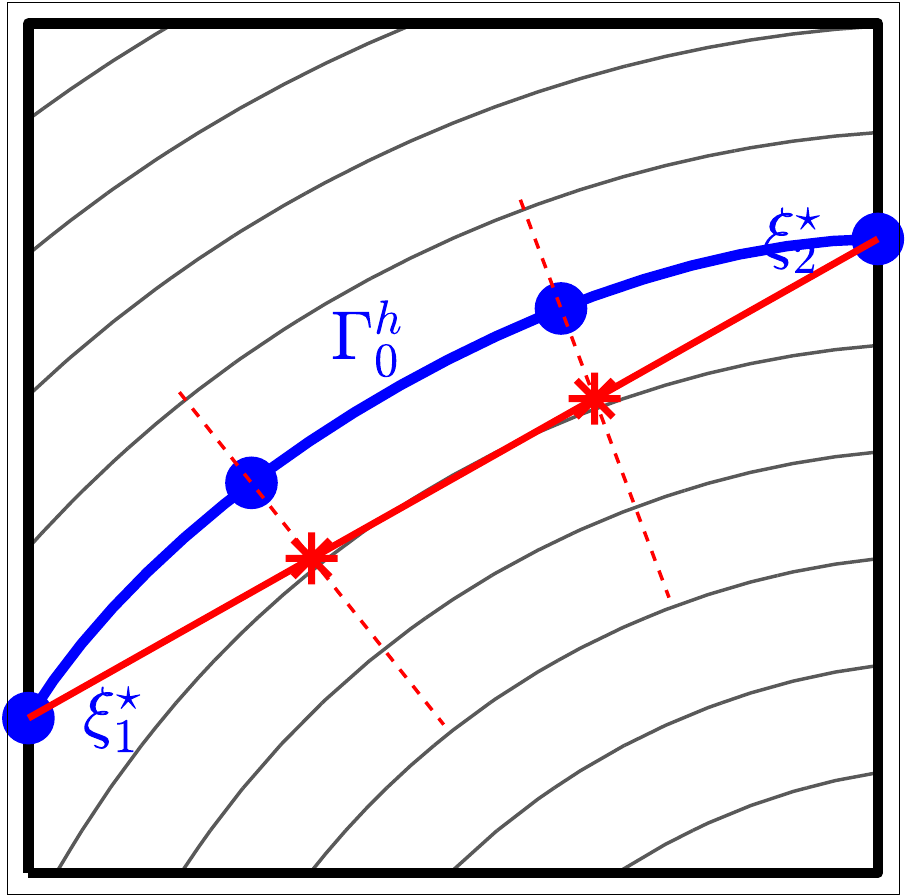}}\hfill\subfloat	
    [decomposition]{\includegraphics[width=3.7cm]{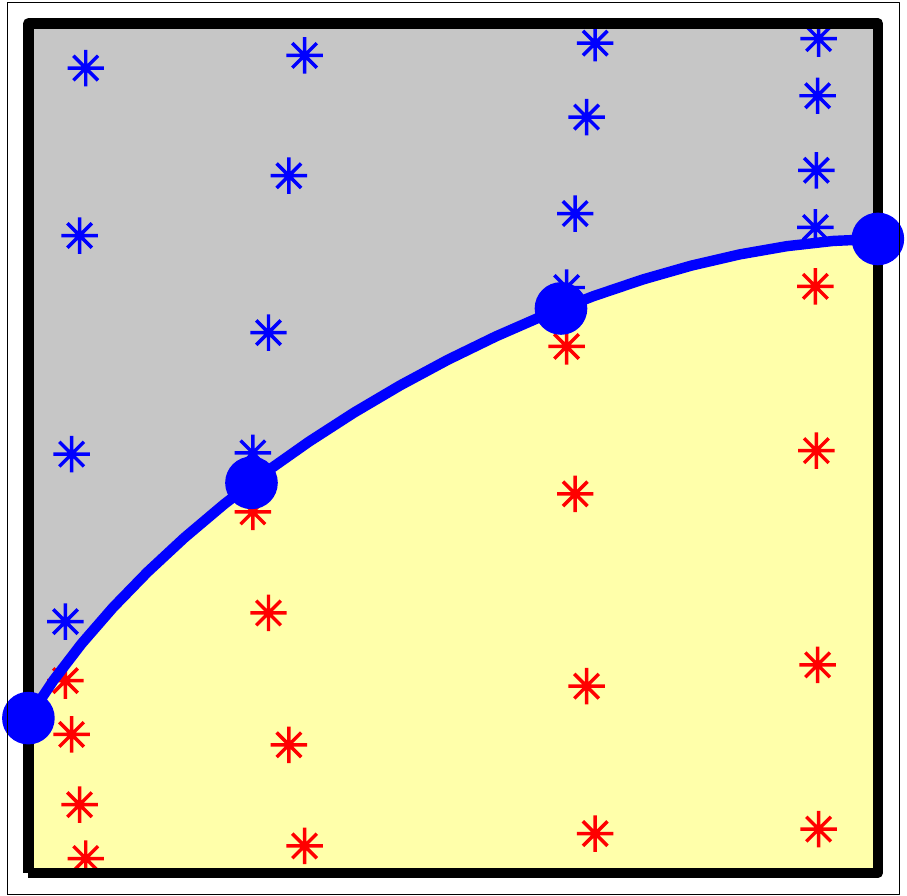}}
    
    \caption{\label{fig:VisCutType2}Two \emph{opposite} edges of a higher-order
    finite element ($p=3$) located in a knot span are cut (topology type
    2): (a) to (d) as in \cref{fig:VisCutType1}.}
\end{figure}

Starting point is a level-set function $\phi\left(\vek\xi\right)$
in the parameter space whose zero-level set $\partial\Gamma_{0}$ is the trimming
curve. Then, the following procedure is applied to each knot span:
The nodes $\vek\xi_{k}^{\mathrm{FE}}$ of a two-dimensional, quadrilateral
Lagrange finite element of order $p$ are defined in the current knot
span, see \cref{fig:VisCutParamSpace}. These are $n^{\mathrm{FE}}=(p+1)^{2}$ nodes,
hence, $k=1,\dots,n^{\mathrm{FE}}$ and $p$ matches the order
of the underlying B-spline (or NURBS) basis in the parameter space.
One may use equally spaced nodes \cite{Fries_2015a,Fries_2016b} for
simplicity or special node distributions such as Gauss-Lobatto or
Chen-Babu{\v{s}}ka nodes \cite{Fries_2016a,Chen_1995a,Chen_1996a}.
The level-set function is evaluated at the nodes of the finite element
resulting in $\phi_{k}=\phi\left(\vek\xi_{k}^{\mathrm{FE}}\right)$.
For explicit trimming curves, such function evaluations are basically
signed distance computations at each node. Through the basis functions
of the element, $N_{k}^{\mathrm{FE}}\left(\vek\xi\right)$, these
nodal values define an approximation $\phi^{h}\left(\vek\xi\right)=\sum_{k=1}^{n^{\mathrm{FE}}}N_{k}^{\mathrm{FE}}\left(\vek\xi\right)\cdot\phi_{k}$
of the exact function $\phi\left(\vek\xi\right)$, see Figs.~\ref{fig:VisCutType1}(a)
and \ref{fig:VisCutType2}(a). Also the corresponding zero-level set
$\partial\Gamma_{0}^{h}$ of $\phi^{h}\left(\vek\xi\right)$ is a higher-order
accurate approximation of $\partial\Gamma_{0}$.

Next, sample grid points $\vek\xi_{m}^{\mathrm{grid}}$ are introduced
in the knot span, typically related to the order $p$ such that there
are $n^{\mathrm{grid,1D}}=p\cdot\left(q+1\right)+1$ nodes in each
direction of the knot span resulting in a total of $n^{\mathrm{grid}}=\left(n^{\mathrm{grid,1D}}\right)^{2}$
grid points, hence, $m=1,\dots n^{\mathrm{grid}}$. $q\in\mathbb{N}_{0}^{+}$
is a user-defined number related to the complexity (e.g., curvature)
of $\phi\left(\vek\xi\right)$ and is often chosen as $q=3$. The
level-set function $\phi^{h}\left(\vek\xi\right)$ in the element
is evaluated at the grid points giving $\phi_{m}^{\mathrm{grid}}=\phi\left(\vek\xi_{m}^{\mathrm{grid}}\right)=\sum_{k=1}^{n^{\mathrm{FE}}}N_{k}^{\mathrm{FE}}\left(\vek\xi_{m}^{\mathrm{grid}}\right)\cdot\phi\left(\vek\xi_{k}^{\mathrm{FE}}\right)$,
see Figs.~\ref{fig:VisCutType1}(b) and \ref{fig:VisCutType2}(b).

Then, the cut situation in the element is determined based on the
level-set data at the sample grid points $\phi_{m}^{\mathrm{grid}}$.
The situation is called \emph{valid} if (i) each element edge is only
cut once, (ii) the overall number of cut edges is two, and (iii) if
no edge is cut then the element is completely uncut. Otherwise, the
situation is \emph{invalid}, however, a valid situation may then always
be recovered by recursive (quad-tree like) refinements of the element.
For further details, we refer to \cite{Fries_2015a,Fries_2016b}.
For valid situations, each (possibly refined) element falls into two
topologically different cases: When two neighboring edges are cut
(case 1), there results a triangle and a pentagon, and for two opposite
edges being cut (case 2), there result two quadrilaterals on each
side, see Figs.~\ref{fig:VisCutType1}(d) and \ref{fig:VisCutType2}(d).

\begin{remark}
It is noted that finite element operations such as the evaluation
of shape functions are evaluated in the (quadrilateral) reference
element. However, the map to each knot span is affine,
even linear in each direction $\xi_{i}$. Therefore, we avoid to complicate
the situation notationally by distinguishing finite element operations
in the reference element versus the knot span.
\end{remark}

\begin{remark}
The described procedure may be generalized in various
ways. Instead of approximating a given level-set function $\phi$
using interpolatory finite elements, one could also use the B-spline
(or NURBS) basis in the parameter space to determine $\phi^{h}$ based
on an $L_{2}$-projection. The boundary of the trimmed shell geometry
would then be implied by level-set values at the control points. 
\end{remark}

\subsubsection{Reconstruction of the trimming curve by a 1D finite element}

Let us assume that the nodal level-set data in a cut knot span is
valid in the above sense. First, the two intersections $\vek\xi_{1}^{\star}$
and $\vek\xi_{2}^{\star}$ of the level-set function $\phi^{h}\left(\vek\xi\right)$
with the two relevant element edges are determined for which a Newton-Raphson
procedure is employed, see Figs.~\ref{fig:VisCutType1}(c) and \ref{fig:VisCutType2}(c).
Next, $p+1$ nodes are distributed along the straight line going through
$\vek\xi_{1}^{\star}$ and $\vek\xi_{2}^{\star}$ yielding starting
points for another Newton-Raphson loop which, for each inner node
on the line, finds the intersection with the zero-level set of $\phi^{h}\left(\vek\xi\right)$
using the gradient $\nabla\phi^{h}$ at the starting points as the
search direction, see Figs.~\ref{fig:VisCutType1}(c) and \ref{fig:VisCutType2}(c).
Hence, the $p+1$ nodes are placed on $\partial\Gamma_{0}^{h}$ up to some
user-define tolerance, usually in the range of $10^{-12}\dots10^{-10}$.
These nodes imply a curved, one-dimensional element being a higher-order
approximation of $\partial\Gamma_{0}^{h}$ (and $\partial\Gamma_{0}$). Concerning the reconstruction, 
many other choices of the start values and search directions are
possible, see \cite{Fries_2015a,Fries_2016b} for details.

\begin{figure}
\centering

\subfloat	
[situation in cut knot span]{\includegraphics[width=0.25\textwidth]{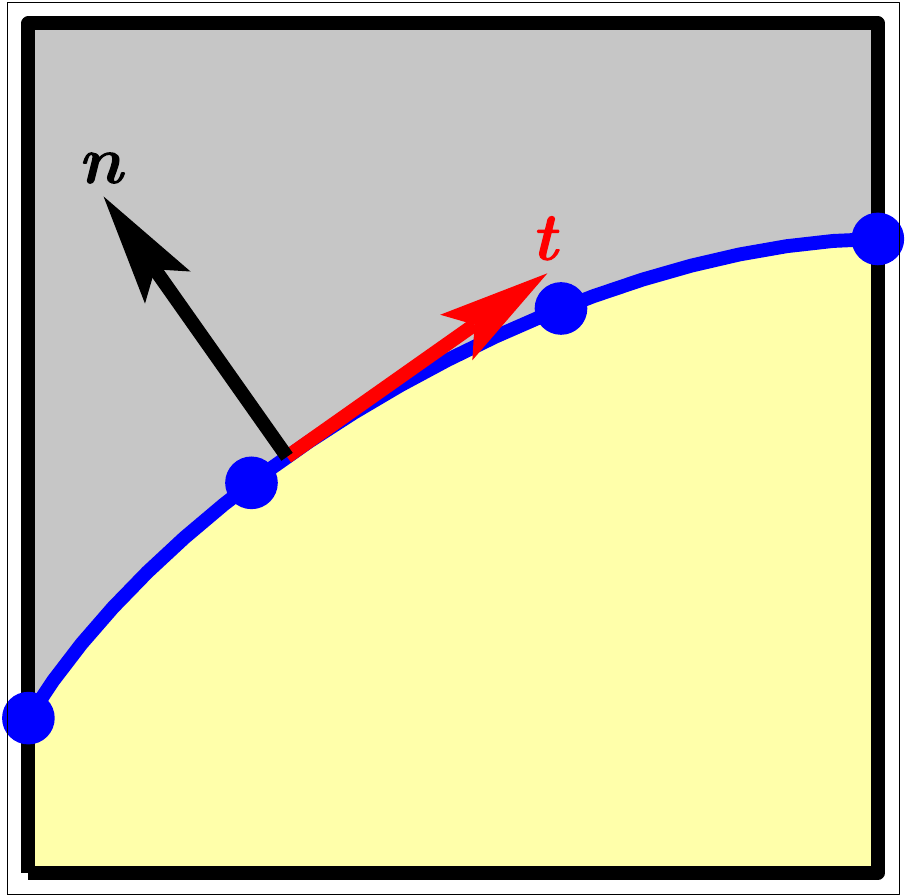}}\qquad\qquad\subfloat	
[situation in $\mathbb{R}^{3}$]{\includegraphics[width=0.4\textwidth]{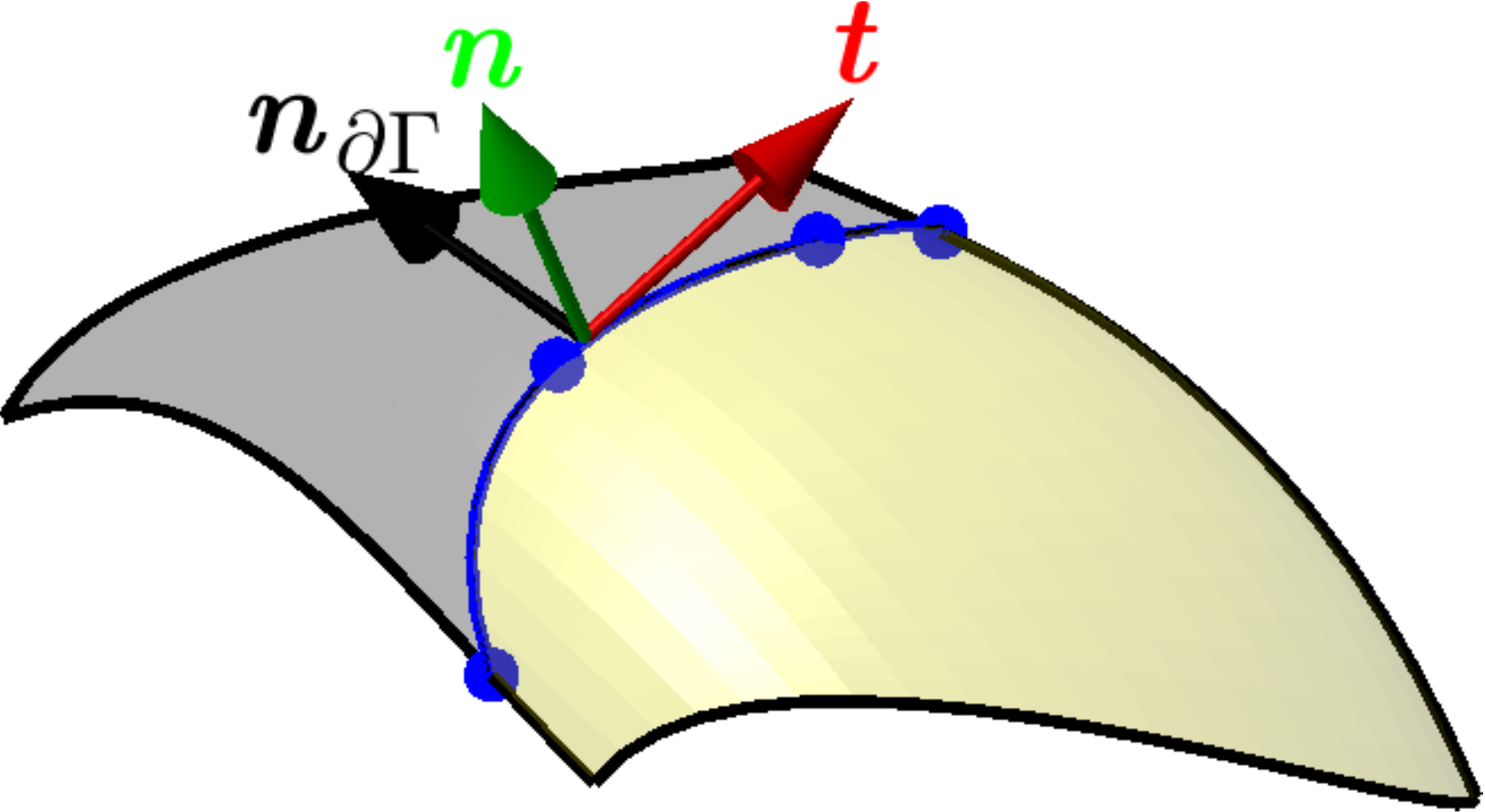}}

\caption{\label{fig:VisNormalAndTangVecs}Normal and tangential vectors at
the trimming curves in (a) parameter space and (b) the shell geometry
in $\mathbb{R}^{3}$.}
\end{figure}

The reconstructed element representing the zero-isoline and, hence,
the trimming curve, is also used to determine normal and tangent
vectors of the boundary in the parameter space. After the isogeometric
map to the real shell geometry in $\mathbb{R}^{3}$, this supports
the definition of the full triad $\left(\vek t_{\partial\Gamma},\vek n_{\partial\Gamma},\vek n_{\Gamma}\right)$
on the shell boundary, see \cref{fig:VisNormalAndTangVecs}.

\subsubsection{Decomposition into integration sub-cells}

The final task is to decompose the element in the knot span based on the
reconstructed one-dimensional element representing the zero-isoline.
The resulting two-dimensional elements depend on the topology of the
cut. For case 1, where two neighboring edges are cut, the two sides
are decomposed into four triangles featuring straight edges except
for the one edge coinciding with the reconstructed zero-isoline, see
\cref{fig:VisCutType1}(d). For case 2, where two opposite edges
are cut, the two sides fall into two quadrilateral elements, each
with one curved edge being the reconstructed zero-isoline, see \cref{fig:VisCutType2}(d).
Standard Gauss rules in triangles and quadrilaterals are mapped to
the decomposed elements that are on the positive side of $\phi^{h}$.
Those on the negative side are neglected. Concerning the mapping of
integration points to the decomposed elements (with one curved, higher-order
side representing $\partial\Gamma_{0}^{h}$), several approaches may be used
\cite{Gordon_1973a,Gordon_1973b,Szabo_2004a,Solin_2003a,Scott_1973a,Gockenbach_2006a}.
Herein, we use the transfinite maps suggested by {\v{S}}ol{\'{i}}n in \cite{Solin_2003a}
as they apply to all finite element shapes. See Figs.~\ref{fig:VisCutType1}(d)
and \ref{fig:VisCutType2}(d) for the resulting integration points,
only the red ones have positive level-set data and are considered;
the blue ones are neglected as they are outside the domain of interest.

It is noted that the reconstruction and decomposition may also fail
for valid level-set data at the sample grid points. For example, the
Newton-Raphson loops for placing nodes on the zero-isoline may not
converge or result in points outside the knot span or the Jacobians
in the decomposed elements may be negative. Such cases are then also
treated by recursive refinements until the reconstruction and decomposition
are successful, see \cite{Fries_2015a,Fries_2016b}. This typically
only applies to complex trimming curves and only in very few knot
spans.

\section{Governing equations}
\label{sec:GoverningEquations}

\subsection{Kirchhoff-Love shell based on the TDC}
\label{sec:shelltheory}

\ifdefined\TodoListsOn
\begin{itemize}
    \item[\checkmark] Short introduction to KL-shell model based on TDC -- Daniel 
    \item[\checkmark] Note that ``classical'' KL-shell formulation would also be applicable
    \item[\checkmark] Equilibrium in weak form
\end{itemize}
\fi

In the following, the fundamentals of the employed shell model are presented. In particular, the rotation-free \KL~shell \cite{Schoellhammer_2018a_DS,Schoellhammer_2019c_DS} formulated in the frame of the tangential differential calculus (TDC) is used herein. It has been shown in \cite{Schoellhammer_2020a_DS} that the TDC-based formulation is more general than the classical outline based on curvilinear coordinates as it also applies to implicitly defined shell geometries. Anyhow, herein, the midsurface is explicitly defined by a single NURBS patch, see \cref{sec:disc}, and the classical shell formulations as presented in, e.g., \cite{Basar_1985a_DS,Bischoff_2017a_DS,Kiendl_2009a_DS}, are also applicable and lead to equivalent results. Nevertheless, also in the present context, one may find the shell formulation based on the TDC also beneficial because the obtained shell equations are formulated in the global Cartesian coordinate system and may be presented in a more compact and intuitive fashion employing symbolic notation whereas the classical formulation is typically formulated in curvilinear coordinates using index notation.\par

\begin{figure}[ht]
	\centering
	\includegraphics{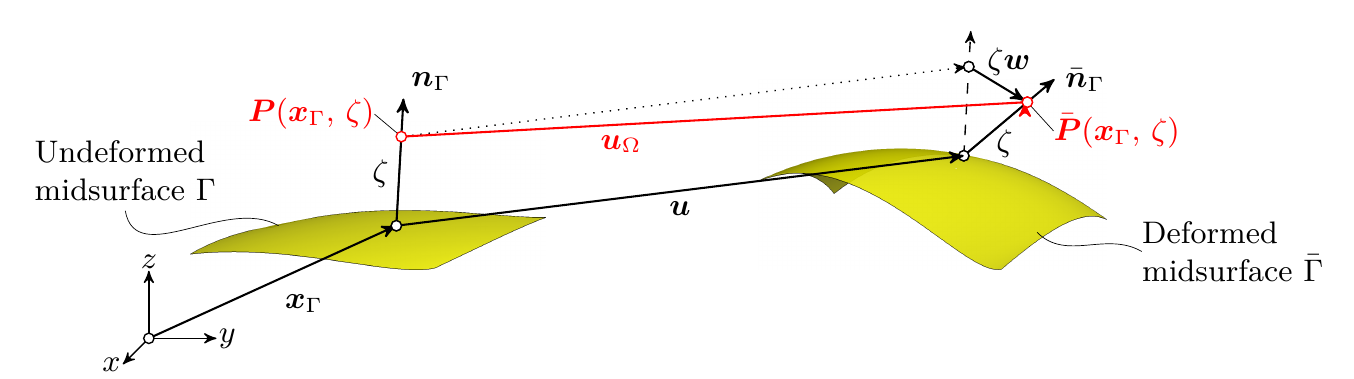}
	\caption{Displacement field $\vek{u}_\Omega$ of the \KL~shell.}
	\label{fig:klshelldisp}
\end{figure}

The whole displacement field $\vek{u}_{\Omega}(\vek{x})$ of a material point within the shell of thickness $t$ is decomposed into the displacement of the midsurface $\vek{u}$ and the rotation of the normal vector which is modelled with a difference vector approach $\vek{w}$, see \cref{fig:klshelldisp}.
With the absence of transverse shear deformations, the rotation of the normal vector is
only a function of the midsurface displacement
\begin{align} \label{eq:diffvec}
\begin{split}
\vek{w} &= - \left[\gradGD{\vek{u}} + (\gradGD{\vek{u}})^\T\right]\cdot \nG  
=\mat{H}\cdot\vek{u} - \gradG{(\vek{u}\cdot\nG)} = -\begin{bmatrix}
\vek{u}_{,x}^\t{dir} \cdot \nG \\[.0cm]
\vek{u}_{,y}^\t{dir} \cdot \nG \\[.0cm]
\vek{u}_{,z}^\t{dir} \cdot \nG
\end{bmatrix}\ ,
\end{split}
\end{align}
where $\gradGD{\vek{u}}$ is the directional surface gradient, $\vek{u}_{,i}^\t{dir}$ are partial directional surface derivatives w.r.t.~$i$ with $i = \lbrace x, y, z \rbrace$ and $\mat{H} = \gradGC{\nG}$ is the Weingarten map. Based on that, the displacement field $\vek{u}_{\Omega}(\vek{x})$ takes the form
\begin{align}
\vek{u}_{\Omega} = \vek{u} + \zeta \vek{w} = \vek{u} - \zeta \begin{bmatrix}
\vek{u}_{,x}^\t{dir} \cdot \nG \\[.0cm]
\vek{u}_{,y}^\t{dir} \cdot \nG \\[.0cm]
\vek{u}_{,z}^\t{dir} \cdot \nG
\end{bmatrix}\ ,
\end{align}
and is only a function of the midsurface displacement. The parameter $\zeta$ is the coordinate in thickness direction $\vert \zeta \vert \le \sfrac{t}{2}$. The linear strain tensor $\ten{\varepsilon}_\Gamma$ is defined by the symmetric part of the surface gradient of the displacement field, i.e.,  $\ten{\varepsilon}_\Gamma = \t{sym}\left[\gradGD{\vek{u}_\Omega}\right]$. The in-plane strains $\ten{\varepsilon}_\Gamma^{\t{P}}$ are obtained with a double projection of $\ten{\varepsilon}_\Gamma$ onto the tangent space of the midsurface $\Gamma$
\begin{align}
\ten{\varepsilon}_\Gamma^{\t{P}} &= \mat{P} \cdot \ten{\varepsilon}_\Gamma \cdot \mat{P} = \ten{\varepsilon}_{\Gamma,\t{M}}^{\t{P}} + \zeta \ten{\varepsilon}_{\Gamma,\t{B}}^{\t{P}}\ ,
\intertext{with}
\ten{\varepsilon}_{\Gamma,\t{M}}^{\t{P}} &= \t{sym}\left[\gradGC{\vek{u}}\right]\ ,\\
\ten{\varepsilon}_{\Gamma,\t{B}}^{\t{P}} &= -\left[\vek{u}_{,ij}^{\t{cov}}\cdot \nG\right]\ ,
\end{align}
where $\ten{\varepsilon}_{\Gamma,\t{M}}^{\t{P}}$ are membrane and $\ten{\varepsilon}_{\Gamma,\t{B}}^{\t{P}}$ are bending strains, respectively. The superscript ``cov'' refers to covariant derivatives. The covariant gradient of $\vek{u}$ is obtained with an additional projection of the directional gradient onto the tangent plane, i.e., $\gradGC{\vek{u}} = \mat{P} \cdot \gradGD{\vek{u}}$. The transverse shear strains $\ten{\varepsilon}_\Gamma^\t{S}(\vek{x})$ are vanishing due to the kinematic assumption from above
\begin{align}
\ten{\varepsilon}_\Gamma^\t{S} = \mat{Q} \cdot \ten{\varepsilon}_\Gamma + \ten{\varepsilon}_\Gamma \cdot \mat{Q} = \mat{0}\ .
\end{align}
A linear elastic material governed by Hooke's law with the modified Lam\'e constants $\mu = \frac{E}{2(1+\nu)}$, $\lambda = \frac{E\nu}{1-\nu^2}$ in order to eliminate the normal stress in thickness direction shall be employed herein. Based on that, the linear in-plane stress tensor yields
\begin{align}
\ten{\sigma}_\Gamma(\vek{x}) = 2\mu\ten{\varepsilon}_\Gamma^{\t{P}}(\vek{x}) + \lambda\t{tr}[\ten{\varepsilon}_\Gamma^{\t{P}}(\vek{x})]\mathbb{I}  \ .
\end{align}
With the assumption of a constant shifter in thickness direction, an analytical pre-integration w.r.t.~the thickness is possible and the stress resultants, such as moment tensor $\mat{m}_\Gamma$ and effective normal force tensor $\tilde{\mat{n}}_\Gamma$ are identified as
\begin{align}
\mat{m}_\Gamma &= \int_{-\sfrac{t}{2}}^{\sfrac{t}{2}} \zeta\, \mat{P} \cdot \ten{\sigma}_\Gamma \cdot \mat{P}\ \d\zeta = \dfrac{t^3}{12} \ten{\sigma}_\Gamma(\ten{\varepsilon}^\t{P}_{\Gamma,\t{B}}) = \mat{P} \cdot \mat{m}_\Gamma^{\t{dir}} \cdot \mat{P} \ , \\
\tilde{\mat{n}}_\Gamma &=\int_{-\sfrac{t}{2}}^{\sfrac{t}{2}}  \mat{P} \cdot \ten{\sigma}_\Gamma \cdot \mat{P}\ \d\zeta = t \ten{\sigma}_\Gamma(\ten{\varepsilon}^\t{P}_{\Gamma,\t{M}}) = \mat{P} \cdot \tilde{\mat{n}}_\Gamma^{\t{dir}} \cdot \mat{P}\ ,
\end{align}
where the superscript ``dir'' indicates that only directional derivatives of $\vek{u}$ are used and, therefore, the physical stress resultants $(\mat{m}_\Gamma, \tilde{\mat{n}}_\Gamma)$ can be computed without the need of covariant derivatives which yields a significant simplification in the implementation \cite{Schoellhammer_2018a_DS}.
The moment and effective normal force tensors are symmetric, in-plane tensors and their two non-zero eigenvalues are the principal moments or forces, respectively. Note that in the case of curved shells, the physical normal force tensor is $\mat{n}^{\t{real}}_\Gamma = \tilde{\mat{n}}_\Gamma + \mat{H}\cdot\mat{m}_\Gamma$ and is, in general, not symmetric, but features one zero eigenvalue just as $\tilde{\mat{n}}_\Gamma$.\par

Based on the stress resultants, the force equilibrium for the \KL~shell in strong form becomes
\begin{align}
\divG{\mat{n}^{\t{real}}_\Gamma} + \nG \divG{\vek{q}_\Gamma} + \mat{H}\cdot\divG{\mat{m}_\Gamma} &= - \vek{f}\ , \label{eq:sff}
\end{align} 
where$\vek{f}$ is the load vector per area and $\vek{q}_\Gamma = \mat{P}\cdot \divG{\mat{m}_\Gamma}$ is the transverse shear force vector obtained by equilibrium considerations. The equilibrium in strong form is a fourth-order surface PDE for the unknown field $\vek{u}$. With suitable boundary conditions, the complete boundary value problem (BVP) for \KL~shells in the frame of the TDC is defined.\par

\begin{figure}[ht]\centering
	\subfloat[displacements and rotation]{
			\ifdefined\tikzOff
				\includegraphics{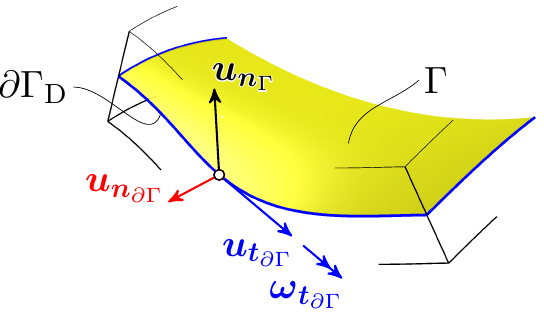}
			\else
				\def\tkzscale{0.15}
				\centering
				\tikzsetnextfilename{DirBC_KL}  
				\input{tikz/DirBC_KL}
			\fi			
\label{fig:dirbc}}
	\hfil
	\subfloat[effective forces and bending moment]{
			\ifdefined\tikzOff
				\includegraphics{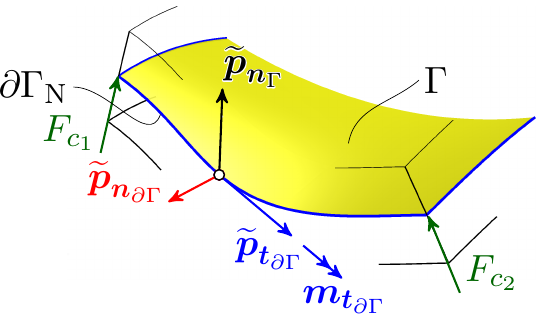}
			\else
				\def\tkzscale{0.15}
				\centering
				\tikzsetnextfilename{NeumBC_KL}  
				\input{tikz/NeumBC_KL}
			\fi			
\label{fig:neumbc}}	
	\caption{Boundary conditions for the \KL~shell in terms of the local triad $(\tB,\,\nCo,\nG)$: (a) decomposition of the midsurface displacement $\vek{u}$ and the rotation $\omega_{\tB}$ along the boundary, (b) conjugated, effective forces, bending moment and corner forces at the boundary.}	
	\label{fig:bcdecomp}
\end{figure}

For the unknown displacement field $\vek{u}$ and the rotation $\omega_{\tB}$ along the boundary, there exist two non-overlapping parts at the boundary of the shell $\p\Gamma$. In particular, the Dirichlet boundary $\p\Gamma_{\t{D},i}$ and the Neumann boundary $\p\Gamma_{\t{N},i}$ with $i = \lbrace \vek{u}, \omega \rbrace$. The corresponding boundary conditions are
\begin{align}
\begin{alignedat}{2}\label{eq:pdebcs}
\vek{u} &= \hat{\vek{g}}_{\vek{u}} &&\t{ on } \p\Gamma_{\t{D},\vek{u}}\ , \\
\widetilde{\vek{p}} &= \hat{\vek{p}} &&\t{ on } \p\Gamma_{\t{N},\vek{u}}\ ,\\
\omega_{\tB} = -[\gradGD{\vek{u}}^\T\cdot\nG]\cdot\nCo &= \hat{g}_{\omega} &&\t{ on } \p\Gamma_{\t{D},\omega}\ , \\
m_{\tB} = \nCo \cdot \mat{m}_\Gamma \cdot \nCo &= \hat{m} &&\t{ on } \p\Gamma_{\t{N},\omega}\ ,
\end{alignedat}
\end{align}
where $\hat{\vek{g}}_{\vek{u}}$ are the displacements, $\hat{\vek{p}}$ are the conjugated, effective forces to $\vek{u}$, $\hat{g}_{\omega}$ is the rotation along the boundaries and $\hat{m}$ is the conjugated bending moment to $\omega_{\tB}$ at their corresponding part of the boundary. In \cref{fig:bcdecomp}, the possible boundary conditions expressed in terms of the local triad $(\tB,\,\nCo,\,\nG)$ and their conjugated, effective forces $(\widetilde{\vek{p}}_{\tB},\,\widetilde{\vek{p}}_{\nCo},\,\widetilde{\vek{p}}_{\nG})$ and bending moment $\vek{m}_{\tB}$ are visualized. Furthermore, in \cref{fig:neumbc}, the Kirchhoff forces at corners are shown in green.

The effective boundary forces $\widetilde{\vek{p}}$ are
\begin{align}
\widetilde{\vek{p}} &= \widetilde{p}_{\tB}\tB + \widetilde{p}_{\nCo}\nCo + \widetilde{p}_{\nG}\nG
\intertext{with}
\widetilde{p}_{\tB} &=  \underset{{p}_{\tB}}{\underbrace{(\mat{n}^{\t{real}}_\Gamma \cdot \nCo)\cdot\tB}} + \left(\mat{H}\cdot\tB\right)\cdot\tB\,m_{\nCo} \ ,\\
\widetilde{p}_{\nCo} &=  \underset{{p}_{\nCo}}{\underbrace{(\mat{n}^{\t{real}}_\Gamma \cdot \nCo)\cdot\nCo}} + \left(\mat{H}\cdot\tB\right)\cdot\nCo\,m_{\nCo} \ ,\\
\widetilde{p}_{\nG} &=  \underset{{p}_{\nG}}{\underbrace{\vek{q}_\Gamma \cdot \nCo}} + \gradG{m_{\nCo}}\cdot\tB\ , \label{eq:effpn}
\end{align}
where $m_{\nCo} = (\mat{m}_\Gamma \cdot \nCo)\cdot \tB$. The Kirchhoff forces are defined as
\begin{align}
\vek{F}_c = \nG m_{\nCo} \big|_{+C}^{-C}\ ,
\end{align}
where ${+C}$ and $-C$ are points close at a corner $C$. Further details regarding the derivation of the effective boundary forces and the Kirchhoff forces are presented in \cite{Schoellhammer_2018a_DS}.
\subsection{Enforcement of boundary conditions and weak form}
\label{sec:bc}

\ifdefined\TodoListsOn
\begin{itemize}
    \item[\checkmark] Non-symmetric Nitsche's Method -- Daniel
    \item[\checkmark]  \mycomment{TODO: Mention disadvantages of non-sym. Nitsche method: possible non optimal convergence in L2, weakly stable, non-sym system of equations}
\end{itemize}
\fi

The enforcement of boundary conditions on trimmed surfaces is a non-trivial task, in particular if higher-order accuracy is desired. The treatment of Neumann boundary condition is, compared to Dirichlet (essential) boundary conditions, less challenging because the involved terms are already incorporated in the weak form and ``only'' a higher-order accurate integration, see \cref{sec:int}, along the corresponding Neumann boundary remains. The situation for Dirichlet boundary conditions is significantly different and there are various approaches for their enforcement.\par

In principal, essential boundary conditions may be enforced in a strong sense by directly prescribing nodal values or in a weak sense by modifying the weak form with additional constraints. In the frame of IGA, a strong enforcement of essential boundary conditions for the rotation-free \KL~shell is \emph{only} possible in special cases and would limit the generality of the approach significantly. Therefore, we follow the same rationale as in \cite{Ruess_2013a_DS,Guo_2017a_DS} and employ a weak enforcement of the essential boundary conditions.\par

A popular strategy for weakly enforcing essential boundary conditions is the penalty method \cite{Babuska_1973a_DS}. The main advantages of the penalty method are the built-in linear independence of the constraints which maintains the positive definiteness of the stiffness matrix and the straight forward implementation. On the other hand, the overall approach is variationally inconsistent and suffers from the interplay between the accuracy and violation of the constraint conditions, in particular, if optimal higher-order convergence rates are desired. Furthermore, the penalty parameter has direct influence on the conditioning of the stiffness matrix. Another approach is the Lagrange multiplier method \cite{Babuska_1973b_DS,Schoellhammer_2018a_DS,Burman_2010a_DS}. Although the method is variationally consistent, additional degrees of freedom are introduced and the positive definiteness of the augmented system of equations is not guaranteed. Furthermore, in the context of fictitious domain methods (FDMs) the discretization of the Lagrange multiplier fields along the inner-element boundaries is not an easy task and depends on the cut scenarios. Due to these shortcomings, Nitsche's method \cite{Nitsche_1971a_DS} has been developed to be a standard choice in FDMs because Nitsche's method is variationally consistent, suitable for higher-order and does not require the discretization of auxiliary fields. The original approach from \cite{Nitsche_1971a_DS} has been adopted to various applications for enforcing essential boundary conditions, see, e.g., \cite{Schoellhammer_2020a_DS,Fries_2019a_DS,Ruess_2013a_DS,Embar_2010a_DS,Mendez_2003a_DS,Hansbo_2002a_DS} and coupling, see, e.g., \cite{Guo_2017a_DS,Schillinger_2016a_DS,Guo_2015a_DS,Apostolatos_2014a_DS}. In principal, there are two different versions of Nitsche's method. The symmetric version of Nitsche's method requires an additional stabilization to ensure positive definiteness \cite{Mendez_2003a_DS}. The choice of the stabilization parameter is rather crucial because when the magnitude of the parameter is too large, the overall approach degenerates to a penalty method. When it is too small, the stability is lost \cite{dePrenter_2018a_DS,Guo_2017a_DS}. In contrast, the non-symmetric version of Nitsche’s method does not require an additional stabilization term for imposing boundary conditions \cite{Burman_2012a_DS,Guo_2017a_DS,Guo_2019a_DS}. However, as indicated by the name, the resulting system of equations is not symmetric and the theoretical error estimates in the $L_2$-norm of the primal variables are possibly suboptimal \cite{arnold2002a_BM,Burman_2012a_DS}. Nevertheless, we prefer the non-symmetric version of Nitsche’s method because our numerical studies showed optimal behaviour even in the $L_2$-norm (in agreement to the results in \cite{Schillinger_2016a_DS,Burman_2012a_DS}), and the absence of an additional stabilization term compared to the symmetric Nitsche method is very beneficial. In addition, a direct solver is employed so that the non-symmetric system of equations is not a problem.\par

The non-symmetric version of Nitsche's method for the \KL~shell formulated in curvilinear coordinates is presented for the linear shell in \cite{Guo_2017a_DS} and for large deformations in \cite{Guo_2019a_DS}. Herein, we propose the non-symmetric version of Nitsche's method for \KL~shell in the frame of the TDC in symbolic notation which results in a more compact formulation and the effective boundary forces are not needed explicitly.\par


The following function spaces are introduced in order to convert the equilibrium in strong form, see \cref{eq:sff}, to the weak form using the non-symmetric version of Nitsche's method for the enforcement of essential boundary condtions
\begin{align}
\mathcal{S} &:= \lbrace \vek{u} : \Gamma \to \mathbb{R}^3\ \vert\ \vek{u} \in [\mathcal{H}^1(\Gamma)]^3 : \vek{u} \cdot \nG \in \mathcal{H}^2(\Gamma) \rbrace \label{eq:funspaceV}\ ,\\
\mathcal{V} &:= \mathcal{S} \ .
\end{align}
A detailed discussion regarding the function space for the \KL~shell with Nitsche's method is given in \cite{Benzaken_2020a_DS}. 
Following the usual procedure as presented in \cite{Schoellhammer_2018a_DS}, the weak form of the equilibrium reads as follows: The task is to find $\vek{u} \in \mathcal{S}$ such that
\begin{align}
&a(\vek{u},\vek{v}) - \int_{\p\Gamma_{\t{D},\vek{u}}} \vek{v}\cdot\widetilde{\vek{p}}(\vek{u}) \ \d s + \vek{v} \cdot \nG m_{\nCo}(\vek{u}) \big|_{+D}^{-D}\\
& \quad - \int_{\p\Gamma_{\t{D},\omega}} \omega_{\tB}(\vek{v}) m_{\tB}(\vek{u})\ \d s = \langle \vek{F},\,\vek{v}\rangle \quad \forall\ \vek{v} \in \mathcal{V}\ , \label{eq:wf}
\intertext{with}
&a(\vek{u},\,\vek{v}) = \int_{\Gamma} \gradGD{\vek{v}}:\tilde{\mat{n}}_{\Gamma} -
\ten{\varepsilon}_{\Gamma,\t{B}}^\t{dir}(\vek{v}):\mat{m}_{\Gamma}\ \d A \nonumber ,\\
\begin{split}
&\langle \vek{F},\,\vek{v}\rangle = \int_{\Gamma} \vek{v} \cdot \vek{f}\ \d A + \int_{\p\Gamma_{\t{N},\vek{u}}} \vek{v} \cdot \hat{\vek{p}}\ \d s + \int_{\p\Gamma_{\t{N},\omega}} \omega_{\tB}(\vek{v}) \hat{m}\ \d s - \vek{v} \cdot \hat{\vek{F}}_c, \nonumber
\end{split}
\end{align}
where $D$ is a corner within the Dirichlet boundary for the displacements $\p\Gamma_{\t{D},\vek{u}}$ and $\hat{\vek{F}}_c$ are corner forces at the corresponding Neumann boundary $\p\Gamma_{\t{N},\vek{u}}$. The additional terms along the Dirichlet boundaries are caused by the fact that the test functions do not vanish there.\par

The Nitsche terms which have to be added to the weak form in order to enforce the displacements $\hat{\vek{g}}_{\vek{u}}$ and rotations $\hat{g}_{\omega}$ at the Dirichlet boundaries are the energy conjugated terms for the displacements and rotation at their Dirichlet boundaries
\begin{align}
&\int_{\p\Gamma_{\t{D},\vek{u}}} \left(\hat{\vek{g}}_{\vek{u}} - \vek{u}\right)\cdot\widetilde{\vek{p}}(\vek{v}) \ \d s + \left(\hat{\vek{g}}_{\vek{u}} - \vek{u}\right) \cdot \nG m_{\nCo}(\vek{v}) \big|_{+D}^{-D}\ ,\\
&\int_{\p\Gamma_{\t{D},\omega}} \left[\hat{g}_{\omega} - \omega_{\tB}(\vek{u})\right] m_{\tB}(\vek{v})\ \d s\ .
\end{align}
Adding these terms to the RHS of \cref{eq:wf} and reordering terms yields 
\begin{align}
\begin{alignedat}{3}
\label{eq:wfnitschecorner}
&a(\vek{u},\vek{v}) &&- \int_{\p\Gamma_{\t{D},\vek{u}}} \vek{v}\cdot\widetilde{\vek{p}}(\vek{u}) \ \d s + \vek{v} \cdot \nG m_{\nCo}(\vek{u}) \big|_{+D}^{-D}\ \\
& &&+ \int_{\p\Gamma_{\t{D},\vek{u}}} \vek{u}\cdot\widetilde{\vek{p}}(\vek{v}) \ \d s - \vek{u} \cdot \nG m_{\nCo}(\vek{v}) \big|_{+D}^{-D}\ \\
& &&- \int_{\p\Gamma_{\t{D},\omega}} \omega_{\tB}(\vek{v}) m_{\tB}(\vek{u})\ \d s + \int_{\p\Gamma_{\t{D},\omega}} \omega_{\tB}(\vek{u}) m_{\tB}(\vek{v})\ \d s \\
&= \langle \vek{F},\,\vek{v}\rangle && + \int_{\p\Gamma_{\t{D},\vek{u}}} \hat{\vek{g}}_{\vek{u}} \cdot\widetilde{\vek{p}}(\vek{v}) \ \d s + \hat{\vek{g}}_{\vek{u}} \cdot \nG m_{\nCo}(\vek{v}) \big|_{+D}^{-D} \\
& && \int_{\p\Gamma_{\t{D},\omega}} \hat{g}_{\omega} m_{\tB}(\vek{v})\ \d s  &&&\forall\ \vek{v} \in \mathcal{V}\ .
\end{alignedat}
\end{align}
Applying integration by parts on the effective normal force $\widetilde{p}_{\nG}$ we can get rid of the corner forces. Furthermore, a shift of the rather cumbersome derivative on the drilling moment $m_{\nCo}$ in $\widetilde{p}_{\nG}$, see \cref{eq:effpn}, simplifies the obtained terms in \cref{eq:wfnitschecorner} significantly to
\def\hOne{\t{bound.~terms due to $\vek{v} \in \mathcal{V}_{\t{N}}$}}
\def\hTwo{\t{Nitsche terms for displ.~on LHS}}
\def\hThree{\t{Nitsche term for rot.~on LHS}}
\def\hFour{\t{Nitsche terms for displ.~on RHS}}
\def\hFive{\t{Nitsche term for rot.~on RHS}}
\begin{align}
\begin{alignedat}{3}
\label{eq:wfnitsche}
&a(\vek{u},\vek{v}) && -\underset{\hOne}{\underbrace{\int_{\p\Gamma_{\t{D},\vek{u}}} \omega_{\nCo}(\vek{v}) m_{\nCo}(\vek{u}) + \vek{v}\cdot\vek{p}(\vek{u})\ \d s}} \\
& && + \underset{\hTwo}{\underbrace{\int_{\p\Gamma_{\t{D},\vek{u}}}  \omega_{\nCo}(\vek{u}) m_{\nCo}(\vek{v}) + \vek{u}\cdot\vek{p}(\vek{v})\ \d s}}\\
& &&- \underset{\hOne}{\underbrace{\int_{\p\Gamma_{\t{D},\omega}} \omega_{\tB}(\vek{v}) m_{\tB}(\vek{u})\ \d s}} + \underset{\hThree}{\underbrace{\int_{\p\Gamma_{\t{D},\omega}} \omega_{\tB}(\vek{u}) m_{\tB}(\vek{v})\ \d s}} \\
&= \langle \vek{F},\,\vek{v}\rangle && + \underset{\hFour}{\underbrace{\int_{\p\Gamma_{\t{D},\vek{u}}} \hat{\vek{g}}_{\vek{u}} \cdot\vek{p}(\vek{v}) + \omega_{\nCo}(\hat{\vek{g}}_{\vek{u}}) m_{\nCo}(\vek{v})\ \d s}}\\
& && + \underset{\hFive}{\underbrace{\int_{\p\Gamma_{\t{D},\omega}} \hat{g}_{\omega} m_{\tB}(\vek{v})\ \d s}}  &&&\forall\ \vek{v} \in \mathcal{V} \ ,
\end{alignedat}
\end{align}
with $\omega_{\nCo}(\vek{u}) = -[\gradGD{\vek{u}}^T \cdot \nG]\cdot\tB$.
Circumventing the corner forces and shifting the tangent derivative on the drilling moment has already been presented for the Kirchhoff plate in \cite{Schillinger_2016a_DS} and is, herein, straightforwardly extended to the \KL~shell. Compared to \cite{Guo_2017a_DS} this may lead to a more compact implementation of the Nitsche terms because the effective boundary forces are not needed explicitly in the implementation. Note that in case of slip supports, e.g., membrane support, or displacement constraints in a selected, arbitrary unit direction $\vek{d}$ with the magnitude $\hat{G}_d$, the involved terms for the displacement along $\p\Gamma_{\t{D},\vek{u}}$ in \cref{eq:wfnitsche} have to be converted accordingly.

\section{Numerical results}
\label{sec:numres}

\ifdefined\TodoListsOn
\begin{itemize}
    \item[\checkmark] Test setting (e.g., error in strong form) -- Daniel
    \item[\checkmark] Examples of Daniel's IGA2019 presentation -- Daniel
\end{itemize}
\fi

In this section, the proposed approach is tested on a set of benchmark examples consisting of the well-known Scordelis-Lo roof \cite{Belytschko_1985a_DS}, a flat shell embedded in $\mathbb{R}^3$ \cite{Schoellhammer_2018a_DS} and a curved, clamped circular shell loaded by gravity and inhomogeneous Dirichlet boundary conditions. All examples are computed on trimmed patches. The trimming curves are defined in the parameter space by means of level-set functions, see \cref{sec:disc}. The convergence criteria or error measurement becomes more strict from the first to last test case. This is useful in order to confirm the higher-order accuracy of the overall approach. In particular, in the Scordelis-Lo roof problem, the convergence criterion is a comparison between a given reference displacement at a certain point. In the second example, relative $L_2$-norms in the displacements and stress resultants $(\mat{n}_{\Gamma}^{\t{real}}, \mat{m}_\Gamma)$ are investigated. In the last test case, where the geometry and boundary conditions are more challenging, the residual and energy errors are computed due to the fact that an analytical solution is not available. Note that the definitions of the residual and energy errors are given below. In addition to the error plots, the estimated condition numbers for all test cases are shown.\par

The employed basis functions are NURBS as defined in \cref{sec:disc}. Following the isoparametric concept, the discrete midsurface displacement is then defined as $ \vek{u}^h =  u^{h,i}\vek{E}_i$, with $\vek{E}_i$ being Cartesian base vectors, with $i = 1,\,2,\,3$ and $u^{h,i} = \vek{\Bspline}^\T \cdot \hat{\vek{u}}^i$. The degrees of freedom (DOFs) $\hat{\vek{u}}^i$ are stored at the control points $\vek{c}$. In the presented examples, the order of the basis functions is varied between $3 \le p \le 6$. Therefore, the discrete midsurface displacement $\vek{u}^h$ is in the Sobolev space $[\mathcal{H}^p(\Gamma)]^3 \subset \mathcal{S}$ with $p \ge 3$. Consequently, the additional continuity requirements, i.e., third-derivatives in the transverse shear forces at the Dirchlet boundary $\p\Gamma_{\t{D},\vek{u}}$, resulting from the additional Nitsche terms are easily met when cubic NURBS are employed. The element scale factor $n$, which is the number of knot spans (elements) in one direction, is varied as $ 4 \le n \le 64$. The element scale factor $n$ is proportional to the element size $ h \sim \sfrac{1}{n}$. The presented numerical results are computed with a threshold for the minimal relative support size $\alpha \in [0.4, 0.6]$ in the context of the extended B-splines, see \cref{ExtendedB-splines}.

\subsection{Scordelis-Lo roof}
\label{sec:numres1}

\begin{figure}[ht]
	\begin{minipage}{.5\textwidth}\centering
		\includegraphics[width=.9\textwidth, height = .9\textwidth, keepaspectratio]{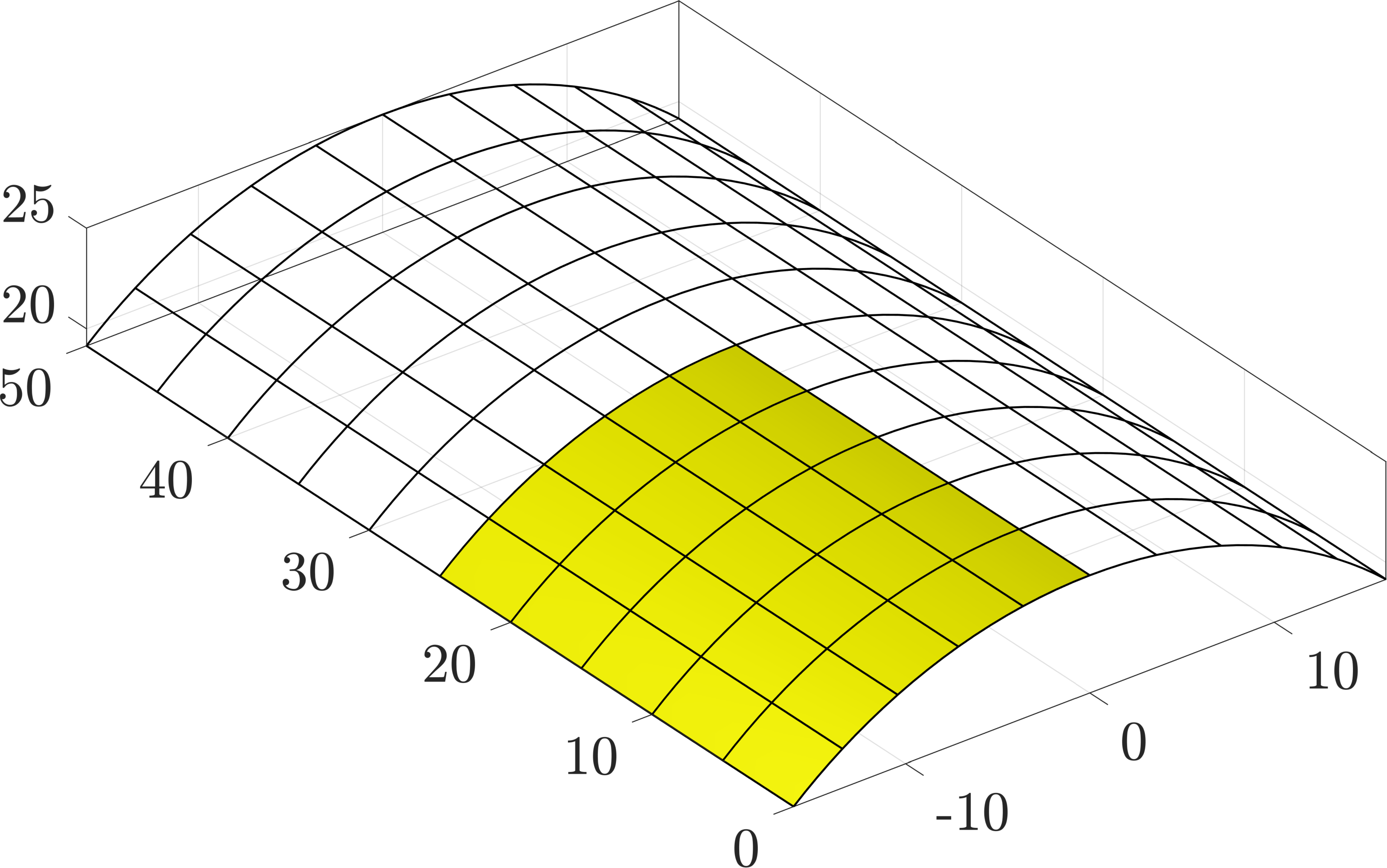}
	\end{minipage}
	\begin{minipage}{.48\textwidth}\scriptsize\flushleft
		\begin{tabular}[h]{ll}
			Geometry: & Cylindrical shell\\
			& $L = 50$ \\
			& $R = 25 $ \\
			& $\phi = \SI{80}{\degree}$ \\
			& $t = \SI{0.25}{}$\\[.25 cm]
			Parameter space: & $(r,s) \in [0,2]^2$ \\[.1 cm]
			Trimming curves: & $\phi_1(\vek{r}) = -(r+1),\ \phi_2(\vek{r}) = - (s+1)$\\[.25 cm]
			Material parameters: & $E = \SI{4.32e8}{}$ \\
			& $\nu = \SI{0.0}{}$\\[.25 cm]
			Load: & Gravity load $\vek{f} = [0,\,0,\,-90]^\T $\\[.25cm]
			Support: & Rigid diaphragms at it ends\\[.25cm]
			Ref. displacement: & $\vert u_{z,i,\t{Ref}} \vert = \SI{0.3006}{}$ from \cite{Bieber_2018a_DS}\\
			& $\vek{x}_i = (-R\cdot\sin(\sfrac{\phi}{2}),\,25,\,R\cdot\cos(\sfrac{\phi}{2}))^\T$
		\end{tabular}
	\end{minipage}
	\caption{Definition of Scordelis-Lo roof problem.}
	\label{fig:overscor}
\end{figure}

The first example is the Scordelis-Lo roof, which is part of the well-known shell obstacle course, and is taken from \cite{Belytschko_1985a_DS}. The whole problem including the definition of the parameter space and its corresponding trimming curves is defined in \cref{fig:overscor}. As a reference displacement for the \KL~shell, we choose the converged displacement presented in \cite{Bieber_2018a_DS} as reference displacement.

\begin{figure}[ht]\centering
	\subfloat[trimmed parameter space]{\includegraphics[width=.3\textwidth]{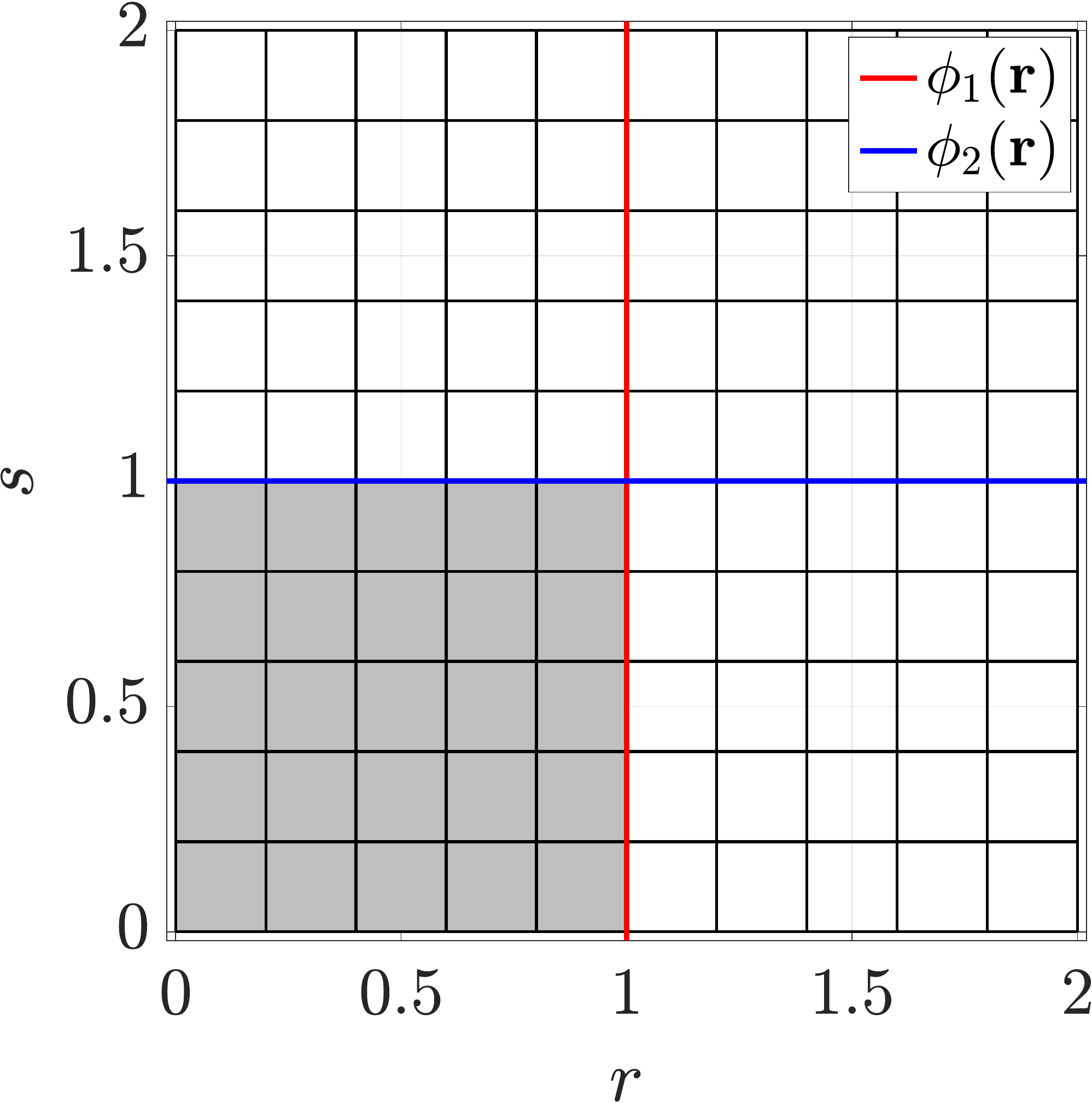}\label{fig:scorpara}}
	\hfil
	\subfloat[displacements]{\includegraphics[width=.45\textwidth]{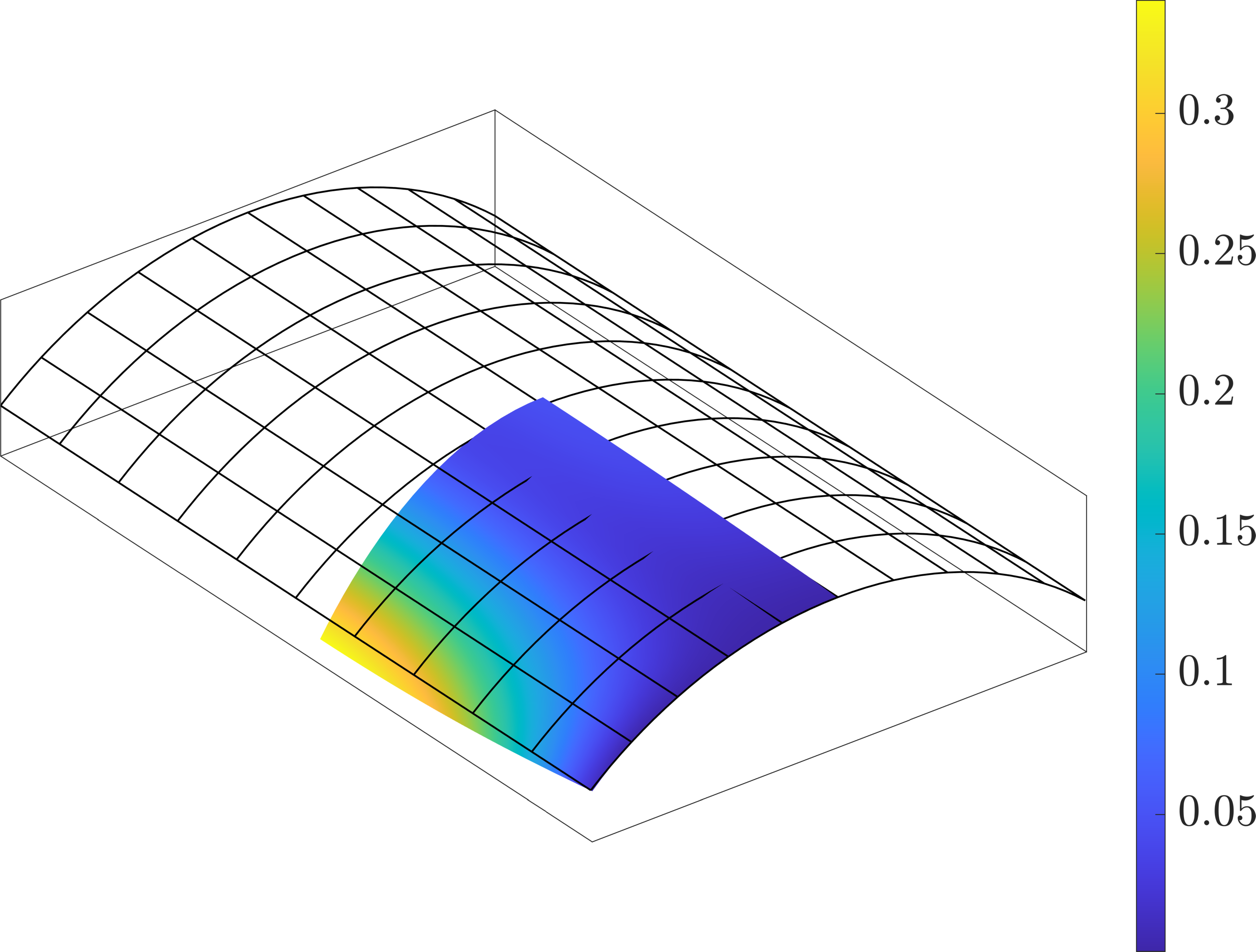}\label{fig:scordisp}}
	\caption{(a) The grey domain is the trimmed parameter space $\visibledomain$ defined by $\phi_1$ and $\phi_2$, (b) deformed domain with scaled displacement $\vek{u}$ by a factor $\SI{10}{}$.}
	\label{fig:scorpd}
\end{figure}

Employing symmetry boundary conditions only one fourth of the whole geometry is considered in the numerical simulation. The symmetry boundaries are modelled as trimming curves in the parameter space and depending on the number of knot spans $n$, the knot spans may be cut or aligned at the knot spans as visualized in \cref{fig:scorpara}. In \cref{fig:scordisp}, the whole, undeformed patch is plotted in the physical space. Furthermore, the deformed midsurface with magnified displacements by a factor 10 is added to the figure. The colors on the surface are the Euclidean norm of the displacement field $\vek{u}^h$.
In this particular example, the element scale factor $n$ in the convergence study is varied between $\lbrace 6, 10, 20, 40\rbrace$ and the threshold $\alpha = 0.6$ is constant for all orders and levels of refinement.\par

\begin{figure}[ht]\centering
	\subfloat[convergence]{\includegraphics[width=.4\textwidth]{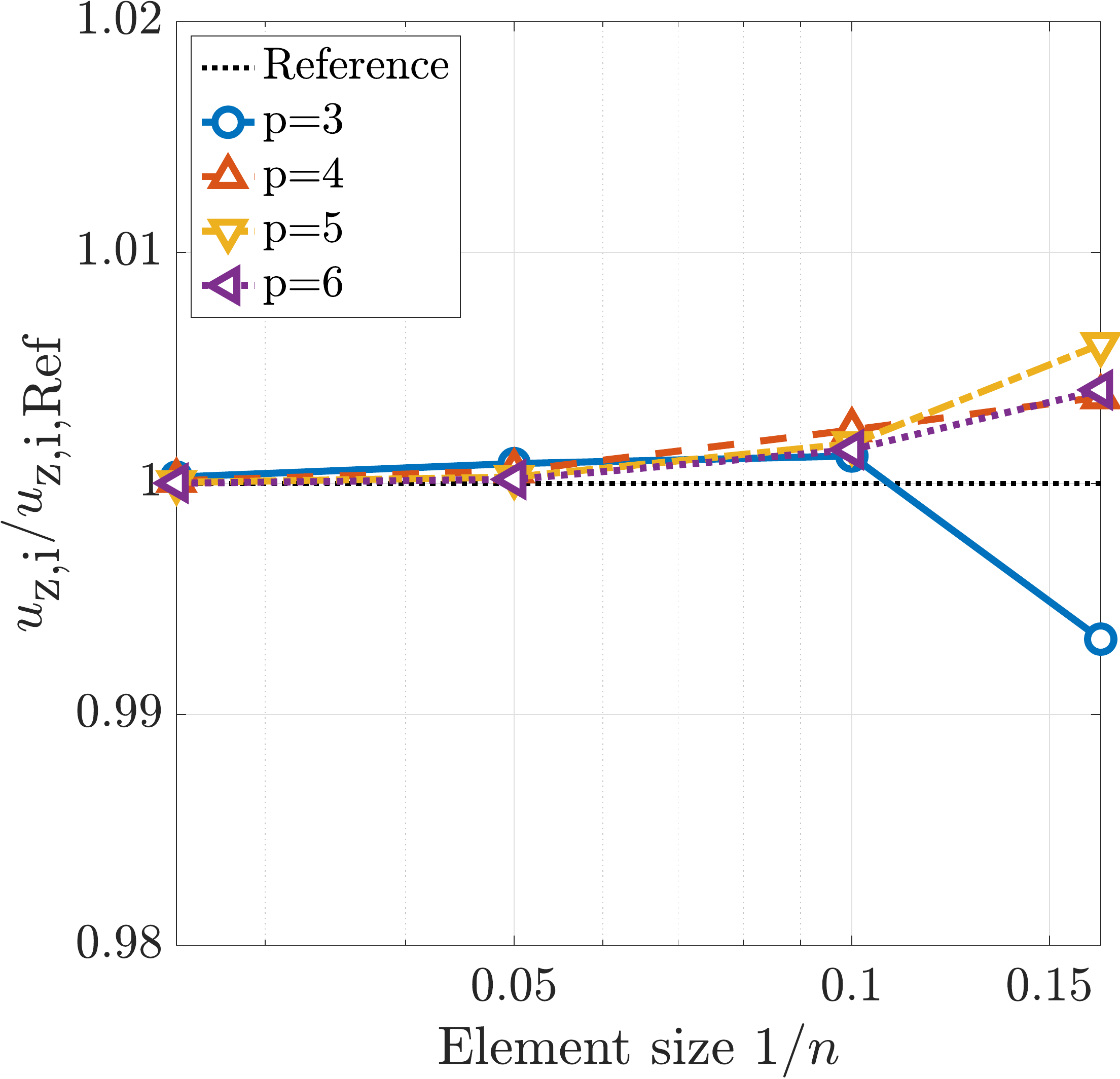}\label{fig:scorconv}}
	\hfil
	\subfloat[condition number]{\includegraphics[width=.4\textwidth]{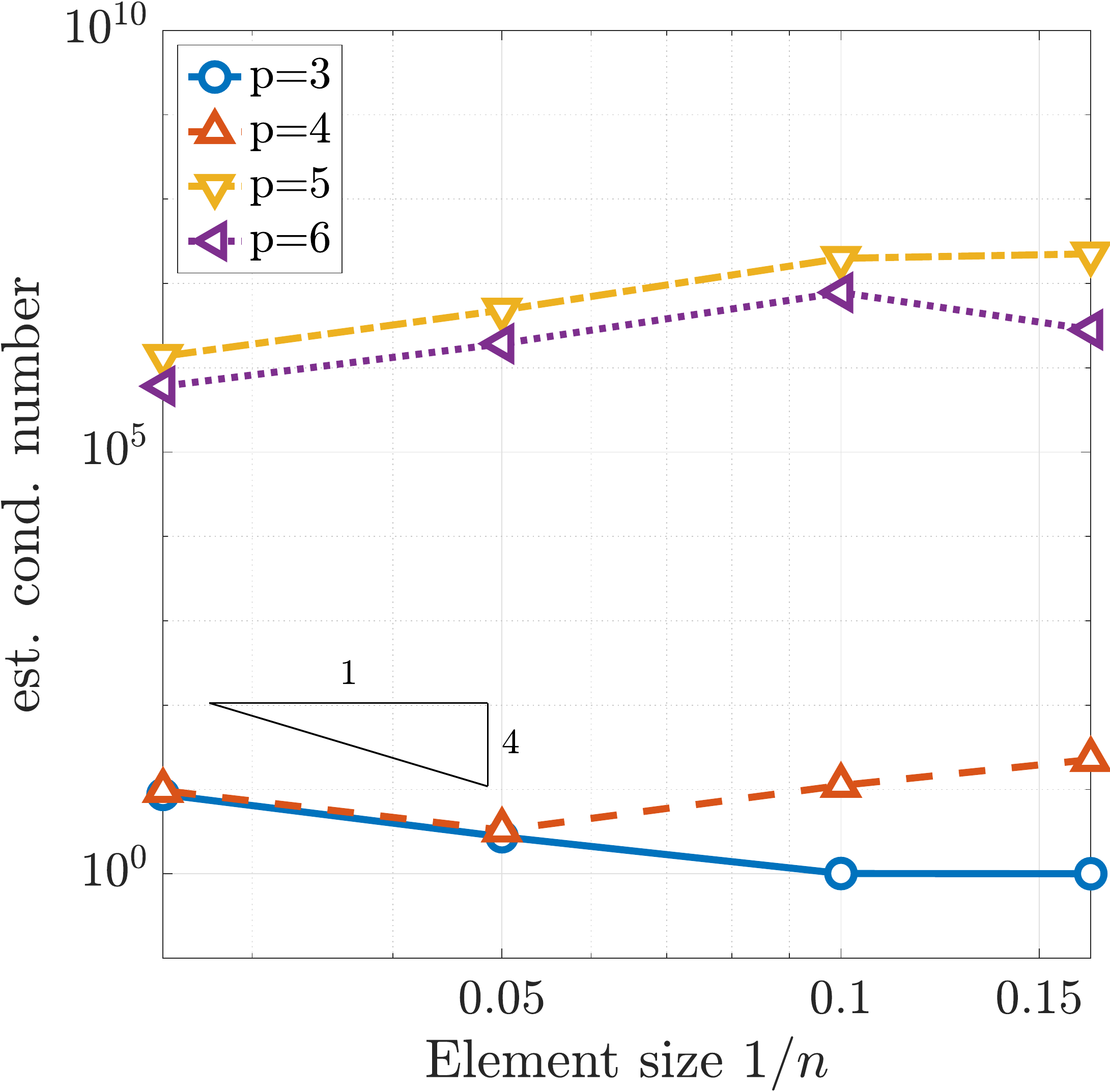}\label{fig:scorcond}}
	\caption{(a) Normalized convergence to the reference displacement $u_{z,i,\t{Ref}} = -0.3006$ at point $\vek{x}_i$, (b) normalized, estimated condition numbers, the reference value is
		$\SI{8.8419e+05}{}$ , which is the estimated condition number at $n=6,\,p=3$.}
	\label{fig:scorcd}
\end{figure}

In \cref{fig:scorcd}, the results of the convergence analyses are presented. In particular, in \cref{fig:scorconv}, the normalized convergence to the given reference displacement is plotted with respect to the element scale factor $\sfrac{1}{n}$. It can be concluded that the accuracy at each level increase for higher orders and the overall behaviour of the converge is in agreement with the results presented in
, e.g., \cite{Schoellhammer_2018a_DS,Kiendl_2009a_DS,Bieber_2018a_DS}. An interesting phenomenon occurs at the coarsest level for the orders $p\ge 4$, where the displacements are overestimated. This may be due to large errors in the boundary conditions caused by a combination of rather coarse meshes together with the elimination of unstable DoFs at the boundaries and Nitsche's method. However, this is only relevant on very coarse meshes and decreases drastically for finer levels when the resolution allows a reasonable boundary discretization. 
Due to the lack of significant digits in the given reference solution, the error plot is not in double logarithmic scale. However, the style of the presentation follows those in many other references, e.g., \cite{Belytschko_1985a_DS,Cirak_2000a_DS,Kiendl_2009a_DS,Schoellhammer_2018a_DS,Bieber_2018a_DS}. In \cref{fig:scorconv}, the normalized estimated condition numbers of the stiffness matrix are plotted. The condition numbers are computed using the ``condest'' function of MATLAB. The condition numbers are bounded for all orders and levels of refinement. The rather large jump between the orders $4$ and $5$ indicates that there might be potential for improvement in the choice of the threshold $\alpha$ for the higher orders. However, a further improvement is not mandatory since the resulting condition numbers are moderate. As a result of this, it can be concluded that the particular choice of the threshold is rather flexibel.\par

Unfortunately, this test case does not feature a sufficiently smooth solution and, therefore, optimal higher-order convergence rates cannot be achieved. 
Consequently, the higher-order accuracy of the proposed approach cannot be numerically confirmed with this test case. Nevertheless, the Scordelis-Lo roof problem is a well-known and accepted benchmark example in the field of shell analysis. 

\subsection{Simply supported flat shell in $\mathbb{R}^3$}
\label{sec:numres2}

\begin{figure}[htb]
	\begin{minipage}[h]{.5\textwidth}\centering
		\includegraphics[width=.95\textwidth, height = .7\textwidth, keepaspectratio]{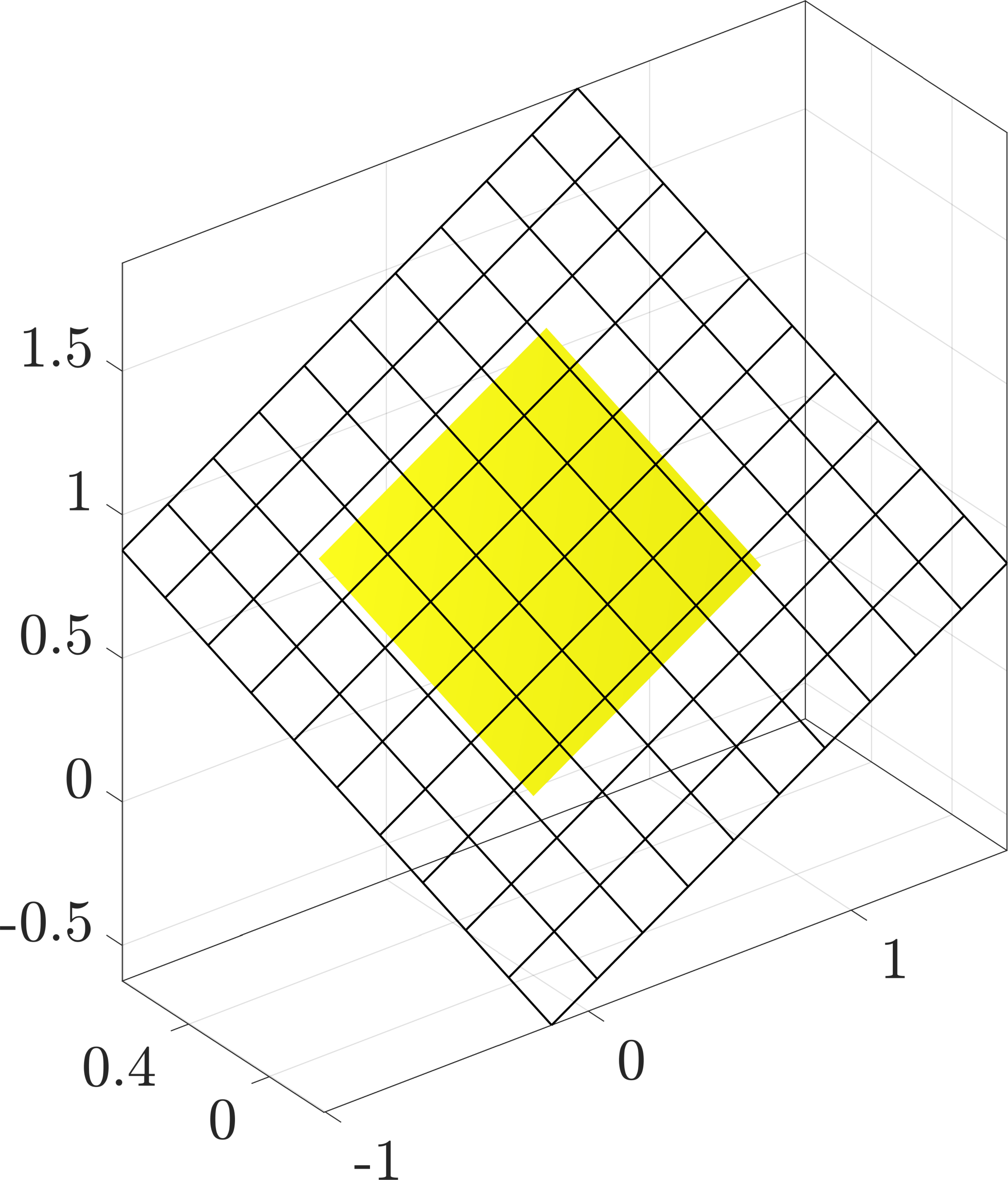}
	\end{minipage}
	\begin{minipage}[h]{.48\textwidth}\scriptsize\flushleft
		\begin{tabular}[h]{ll}
			Geometry: & Quadrilateral flat shell\\
			& $L = 1$ \\
			& $t = \SI{0.01}{}$\\
			& $\vek{n}_\Gamma~=~[-\sfrac{1}{4},\,-\sfrac{\sqrt{3}}{2},\,\sfrac{\sqrt{3}}{4}]^\T$\\[.25 cm]
			Parameter space: &$r \in [0,2]^\T- \sfrac{2}{3}+ 0.234$\\
			&$s \in [0,2]^\T- \sfrac{2}{3}+ 0.123$ \\[.1 cm]
			Trimming curves: &$\phi_1(\vek{r}) = s,\ \phi_2(\vek{r}) = -(r+1)$, \\	
			&$\phi_3(\vek{r}) = r,\ \phi_4(\vek{r}) = -(s+1)$ \\[.25cm]				
			Material parameters: & $E = \SI{10000}{}$ \\
			& $\nu = \SI{0.3}{}$\\[.25 cm]
			Load: & $\vek{f}_t $ and $f_n$  from \cite{Schoellhammer_2018a_DS}\\[.25 cm]
			Support: & Simple support on all edges
		\end{tabular}
	\end{minipage}
	\caption{Definition of flat shell problem.}
	\label{fig:overplate}
\end{figure}

\begin{figure}[ht]\centering
	\subfloat[trimmed parameter space]{\includegraphics[height=.275\textwidth]{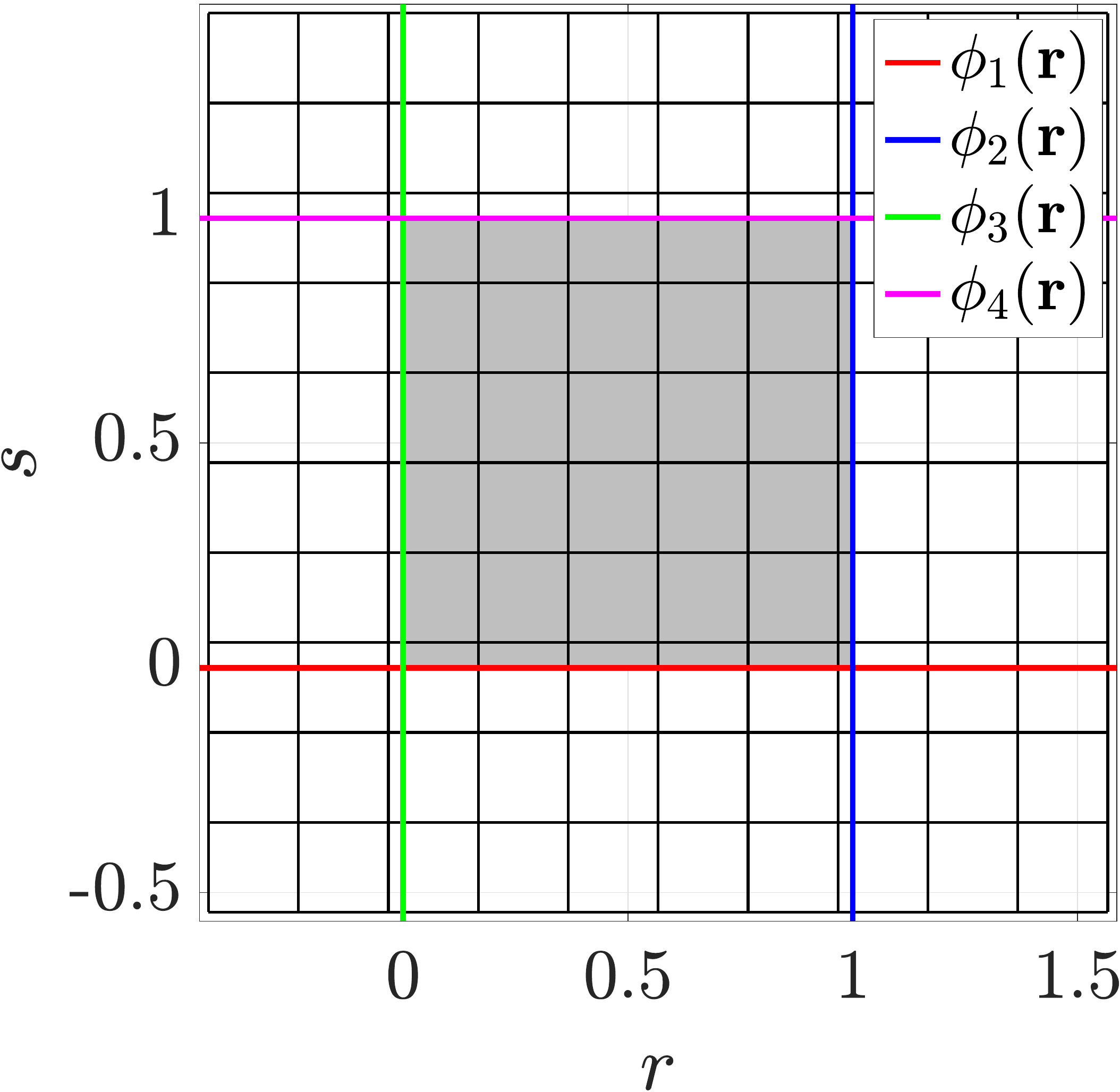}\label{fig:platepara}}
	\hfil
	\subfloat[front view]{\includegraphics[height=.3\textwidth]{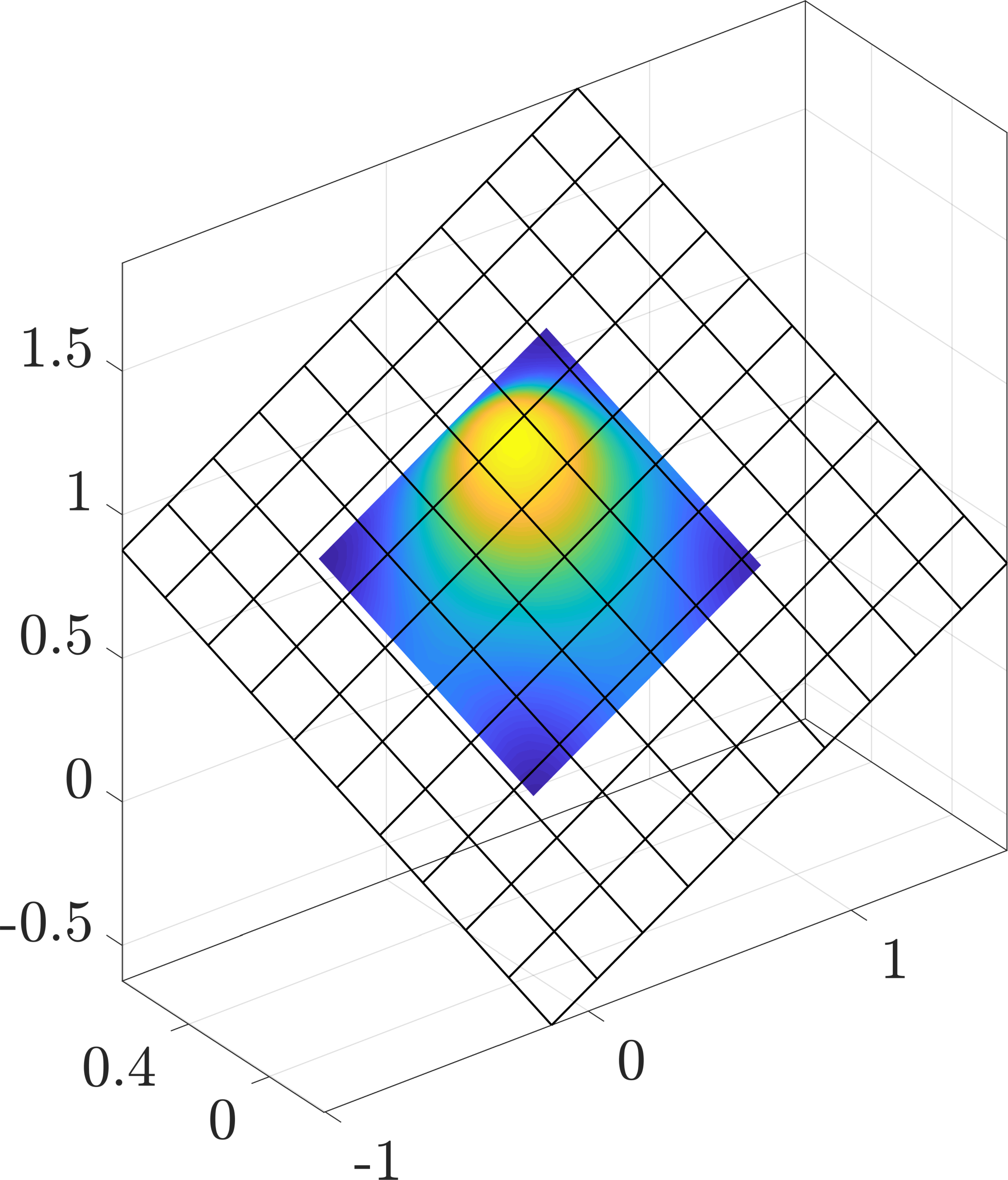}\label{fig:plated1}}
	\hfil
	\subfloat[rotated view]{\includegraphics[height=.3\textwidth]{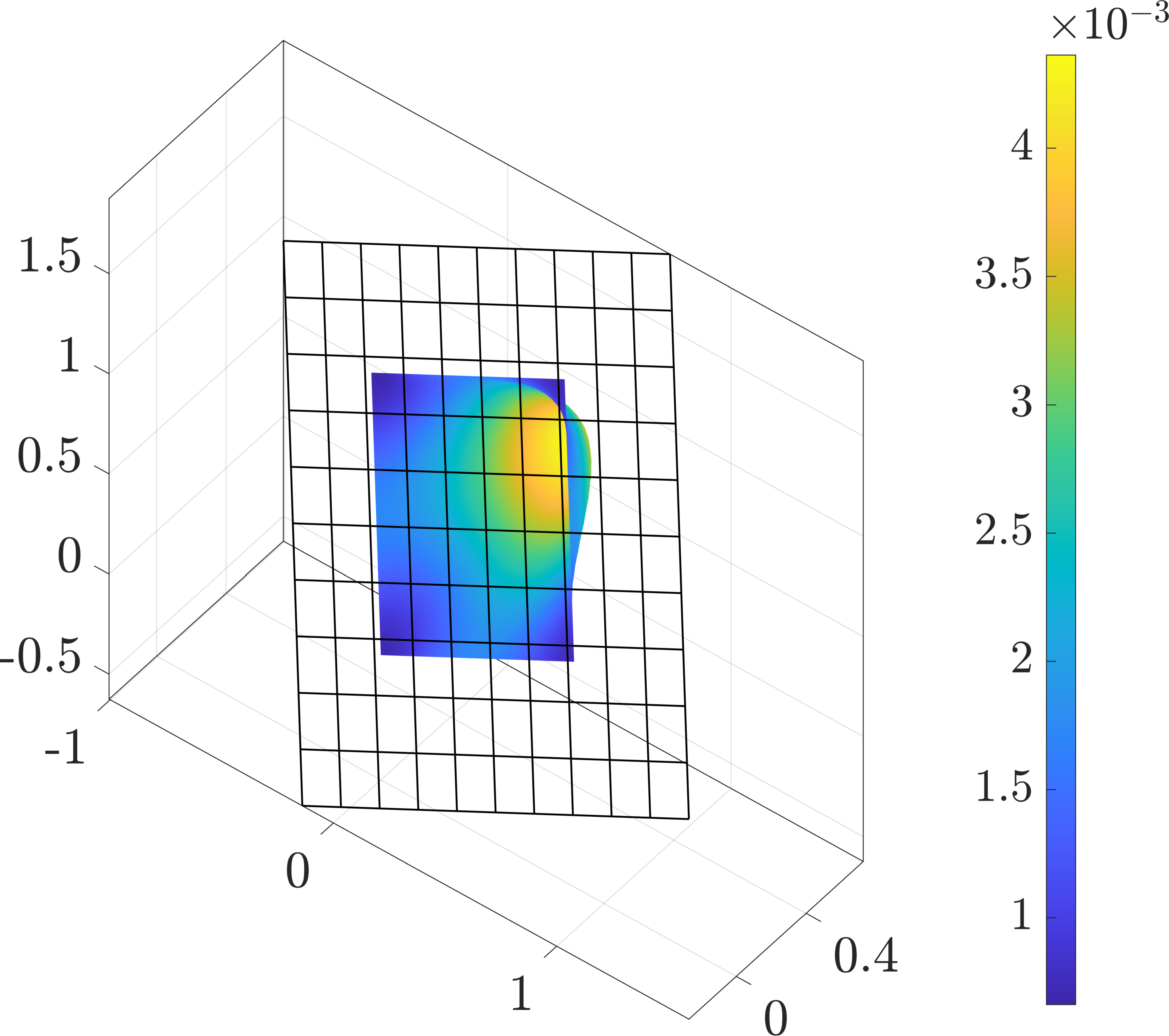}\label{fig:plated2}}
	\caption{(a) The grey domain is the trimmed parameter space $\visibledomain$ defined by $\phi_i$ with $i = \in [1,4]$, (b) deformed domain with scaled displacement $\vek{u}$ by two orders of magnitude, (c) rotated view of deformed shell}
	\label{fig:platepd}
\end{figure}

The second example is taken from \cite{Schoellhammer_2018a_DS} and is a flat shell embedded in the physical space $\mathbb{R}^3$. The problem and parameter space including trimming curves is defined in \cref{fig:overplate}.
The simply supported shell is loaded in tangential and normal direction and in this particular case, an analytical solution is available. The exact solution is given in \cite{Schoellhammer_2018a_DS}. In \cref{fig:platepara}, the trimmed parameter space $\visibledomain$ and the corresponding trimming curves are visualized.
The deformed domain is illustrated in \cref{fig:plated1} and \cref{fig:plated2} with scaled displacements by a factor 100. Analogously as above, the color on the surface indicates the Euclidean norm of the displacement field $\vek{u}^h$.\par

\begin{figure}[ht]\centering\def\mw{.4}
	\subfloat[rel.~$L_2$-norm displacements]{\includegraphics[width=\mw\textwidth]{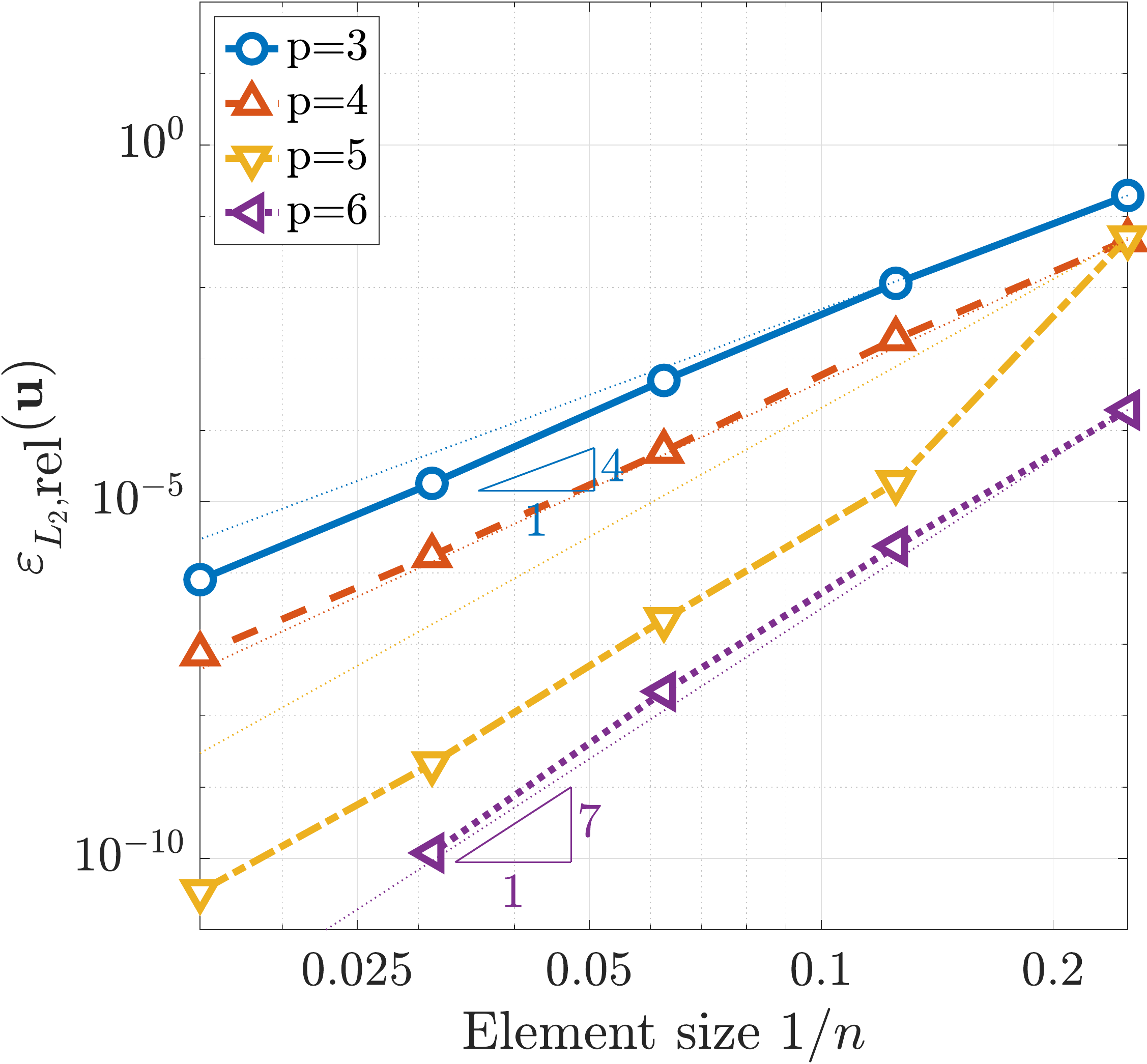} \label{fig:platecu}}
	\hfil
	\subfloat[rel.~$L_2$-norm normal forces]{\includegraphics[width=\mw\textwidth]{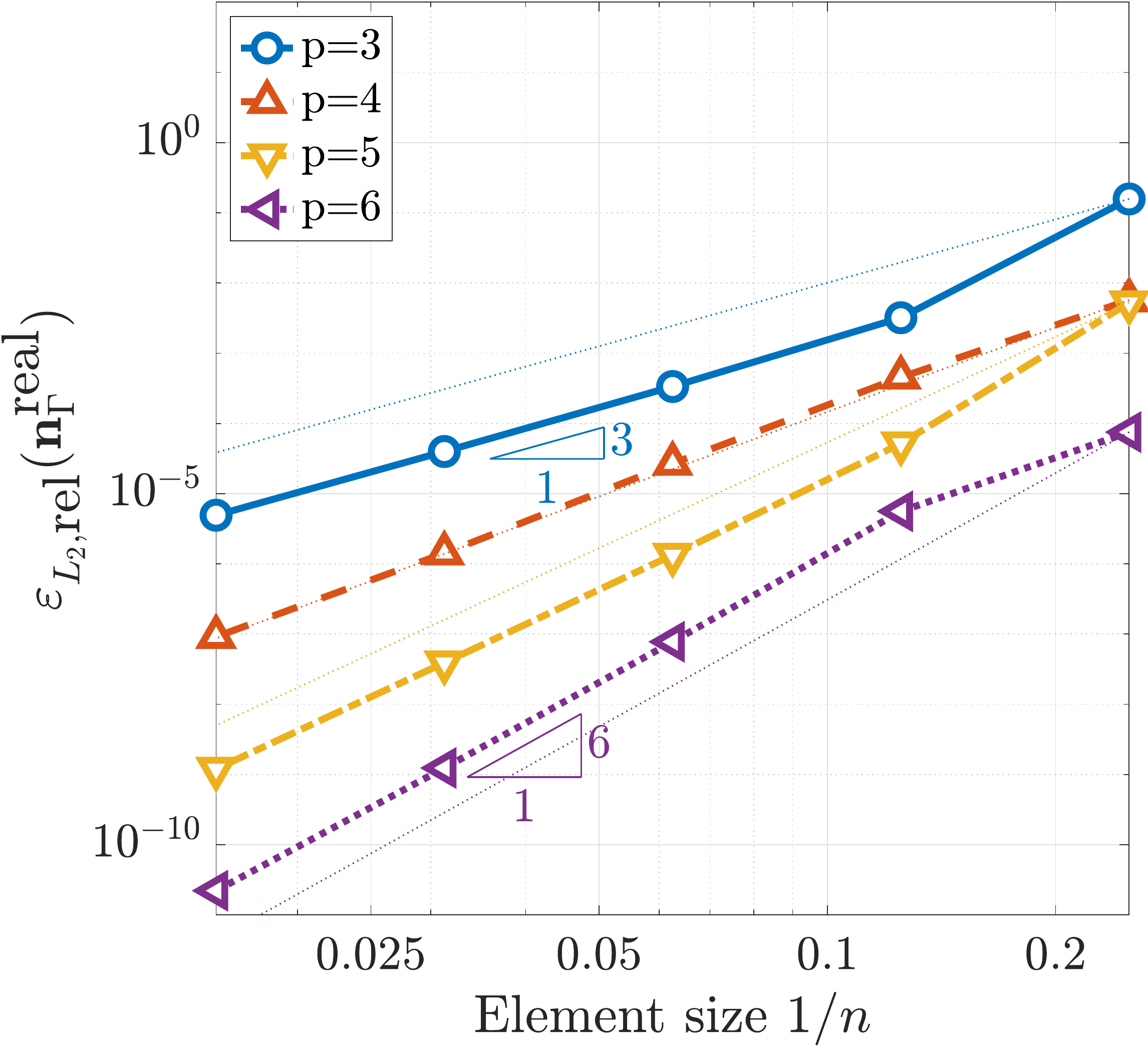}\label{fig:platecn}}
	\hfil	
	\subfloat[rel.~$L_2$-norm bending moments]{\includegraphics[width=\mw\textwidth]{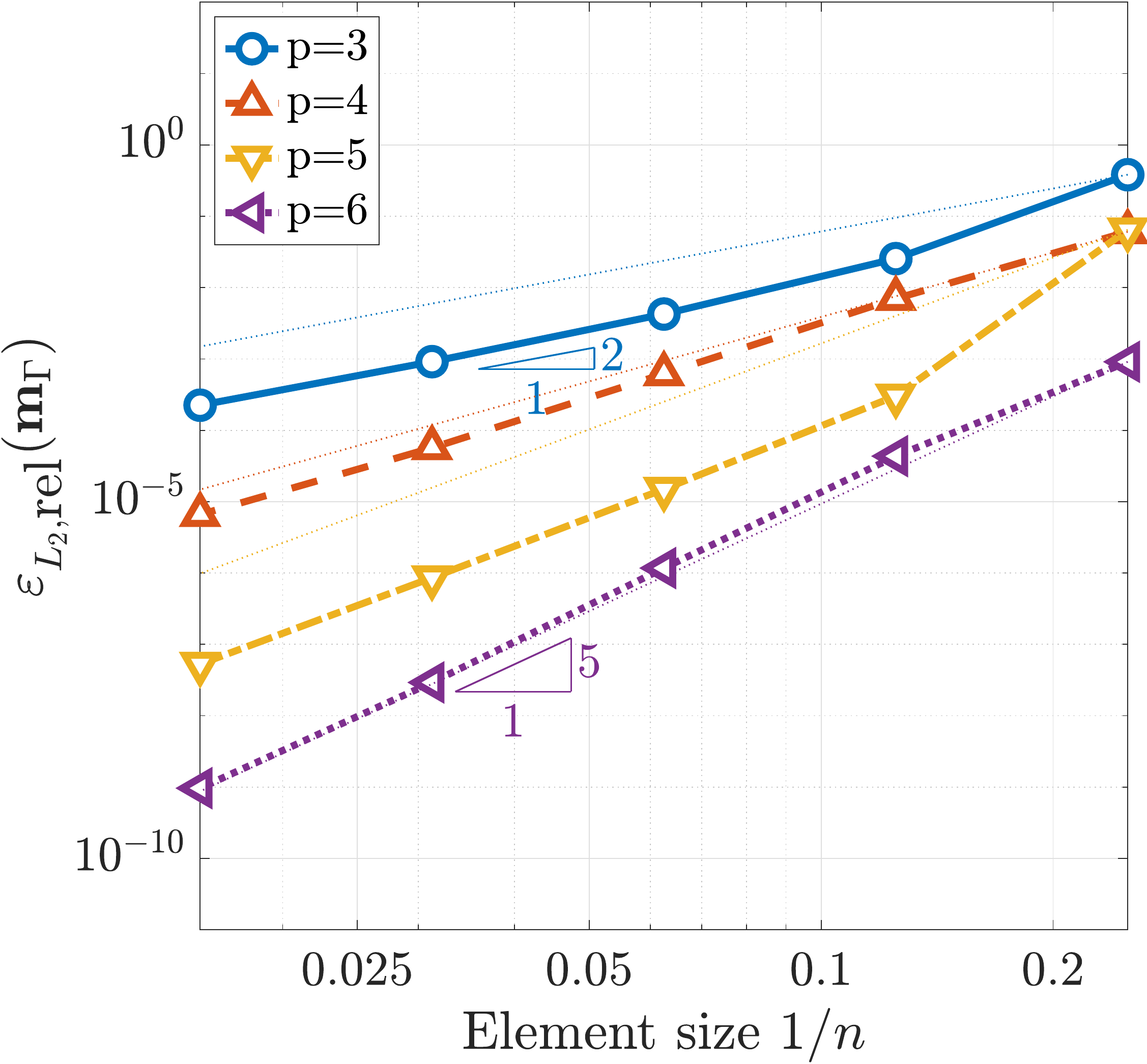}\label{fig:platecm}}
	\hfil
	\subfloat[condition number]{\includegraphics[width=\mw\textwidth]{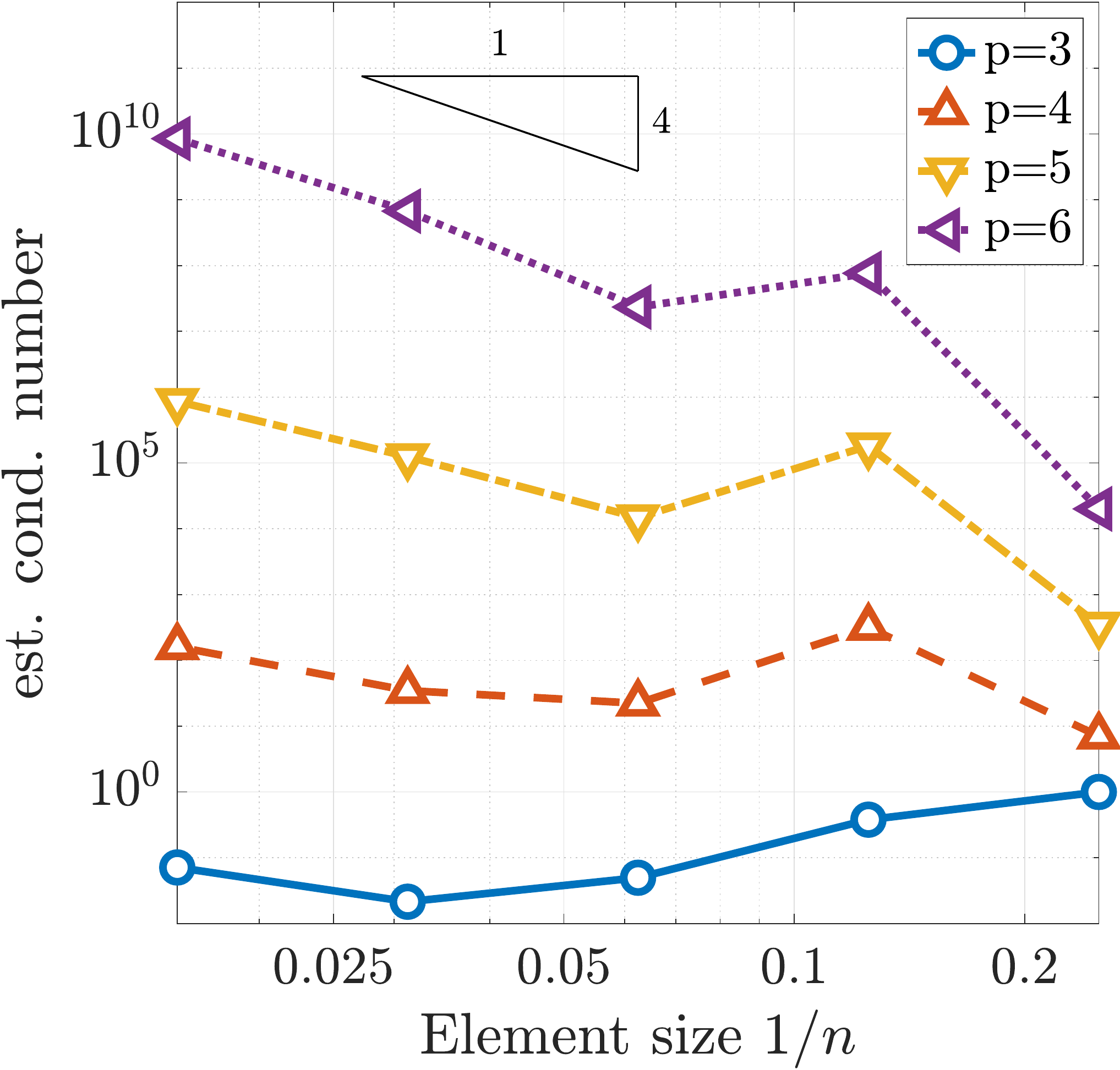}\label{fig:platecond}}	
	\caption{Results of convergence studies for simply supported flat shell: (a) Relative $L_2$-norm of displacements $\vek{u}$, (b) relative $L_2$-norm of internal normal forces $\mat{n}^{\t{real}}_\Gamma$, (c) relative $L_2$-norm of bending moments $\mat{m}_\Gamma$, (d) normalized condition numbers, the reference value is
		$\SI{1.0774e+08}{}$ , which is the condition number at $n=4,\,p=3$.}
	\label{fig:plateconv}
\end{figure}

The results of the convergence studies are presented in \cref{fig:plateconv}. The element scale factor $n$ is varied between $\lbrace 4, 8, 16, 32, 64 \rbrace$ and $\alpha = 0.4$. In \cref{fig:platecu}, the relative $L_2$-norm of the primal variable, i.e., the midsurface displacement $\vek{u}^h$ is plotted. Optimal higher-order convergence rates $\mathcal{O}(p+1)$ w.r.t.~to the element size $h$ are achieved which is an agreement to results shown in \cite{Burman_2012a_DS,Schillinger_2016a_DS}. Note that, theoretically, sub-optimal convergence rates could have been expected \cite{Burman_2012a_DS}. Optimal convergence rates are also achieved in the relative $L_2$-norms of the normal forces $\mathcal{O}(p)$, see \cref{fig:platecn}, and the bending moments $\mathcal{O}(p-1)$, see \cref{fig:platecm}. Achieving optimal convergence in all $L_2$-errors clearly emphasises the higher-order accuracy of the proposed approach for the flat shell problem.
In \cref{fig:platecond}, the estimated condition numbers are plotted. In this test case, the condition numbers increase with $\mathcal{O}(4)$ which is expected for fourth-order PDEs. The jumps in the condition numbers between each order is also well-known in the context of higher-order finite element approaches, see, e.g., \cite{Fries_2017b_DS,Schoellhammer_2020a_DS}. The overall behaviour of the condition numbers compared to the first test case is different which indicates that the choice of the threshold $\alpha$ seems to be almost optimal (with the exception for $p=3$ where it might be slightly too high). Nevertheless, a significant influence on the results in the relative $L_2$-norms has not been observed. 

\clearpage
\subsection{Clamped circular shell}
\label{sec:numres3}

\begin{figure}[htb]
	\begin{minipage}[h]{.45\textwidth}\centering
		\includegraphics[width=.9\textwidth, height = .7\textwidth, keepaspectratio]{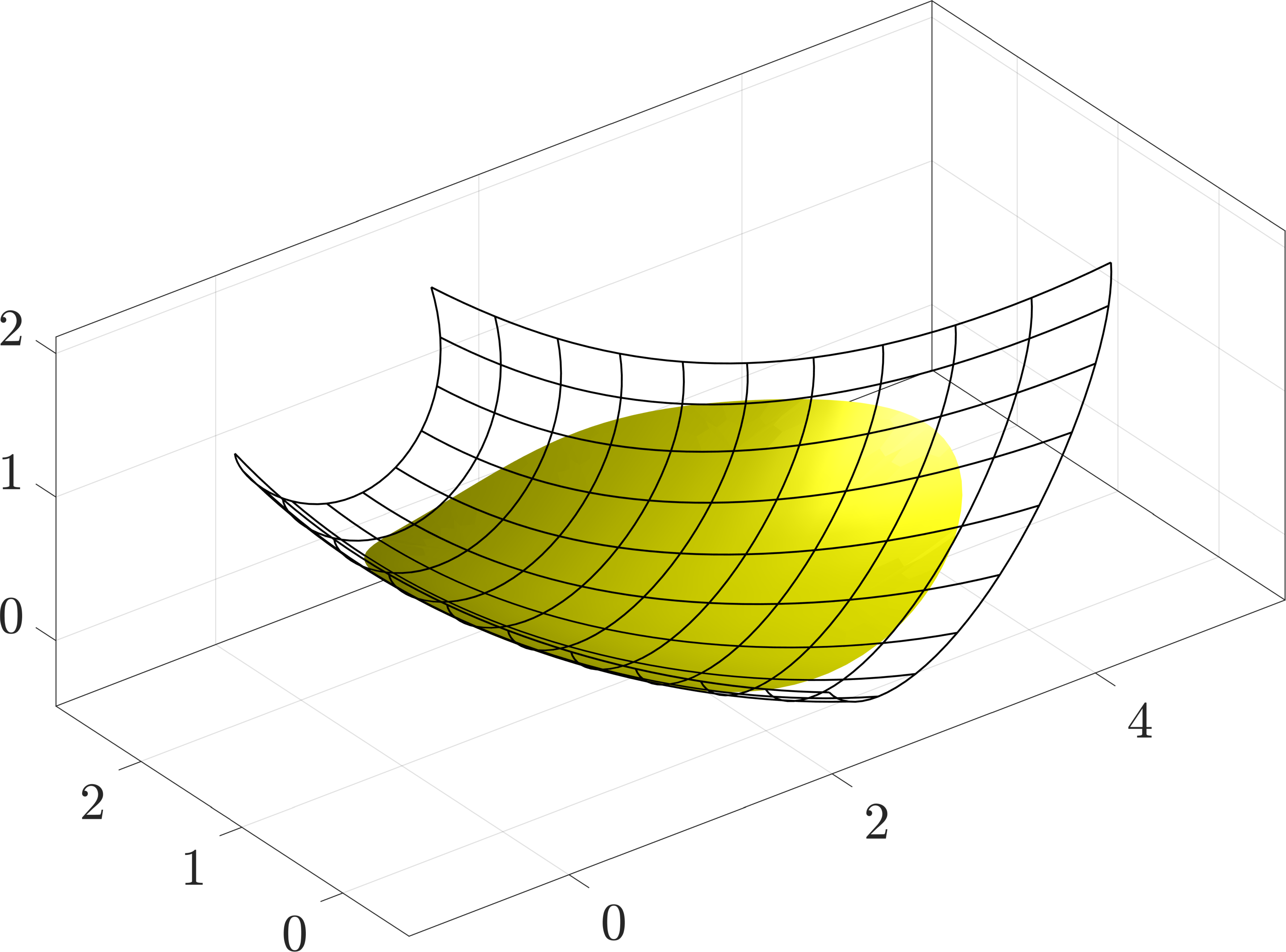}
	\end{minipage}
	\begin{minipage}[h]{.53\textwidth}\scriptsize\flushleft
		\begin{tabular}[h]{ll}
			Geometry: & Mapped circular shell\\
			& $x(\vek{r}) = 2 + r - 2s - 0.2rs + 0.75\sin(2r+0.3)$ \\
			& $y(\vek{r}) = 1 + r + s + 0.5rs + 0.5\cos(r+1.5s)$ \\
			& $z(\vek{r}) = -0.3 + 0.5r^2 + 0.75s + \sin(rs) + 0.2(x-2)^2$ \\[.25 cm]
			Parameter space: &$(r,s) \in [-0.875, 0.875]^\T$\\[.1 cm]
			Trimming curves: &$\phi_1(\vek{r}) = 0.7123 - \Vert \vek{r} \Vert $\\[.25cm]				
			Material parameters: & $E = \SI{10000}{}$ \\
			& $\nu = \SI{0.3}{}$\\[.25 cm]
			Load: & Gravity load $\vek{f} = [0,\,0,\,-100]^\T $\\[.1 cm]
			\multicolumn{2}{l}{Inhomogeneous essential boundary condition: $\hat{\vek{g}}_{\vek{u}} = -0.1\nG$}\\[.25 cm]
			Support: & Clamped edge
		\end{tabular}
	\end{minipage}
	\caption{Definition of the clamped, circular shell problem.}
	\label{fig:overcirc}
\end{figure}

\begin{figure}[ht]\centering
	\subfloat[trimmed parameter space]{\includegraphics[width=.3\textwidth]{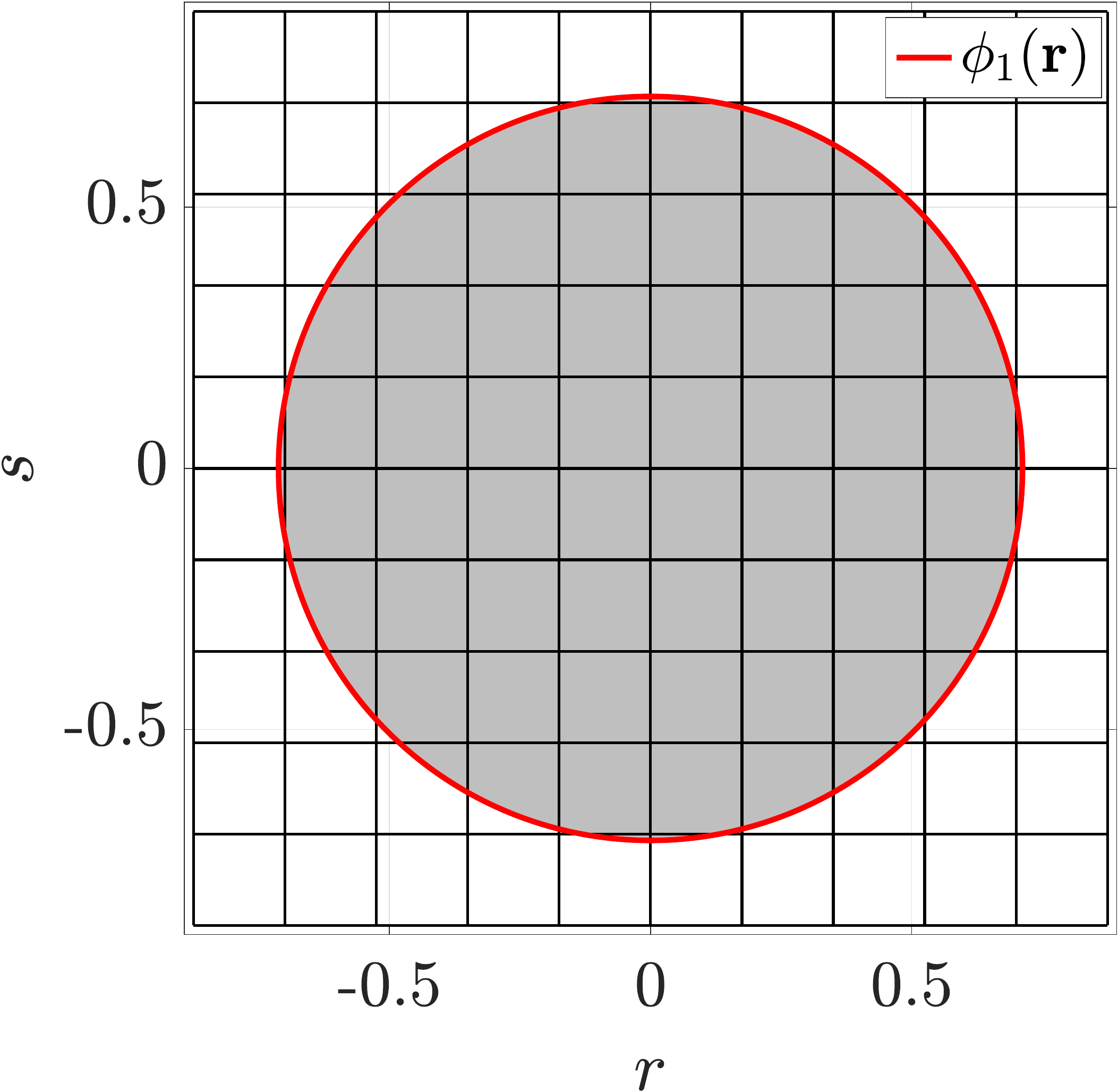} \label{fig:circp}}
	\hfil
	\subfloat[displacements]{\includegraphics[width=.45\textwidth]{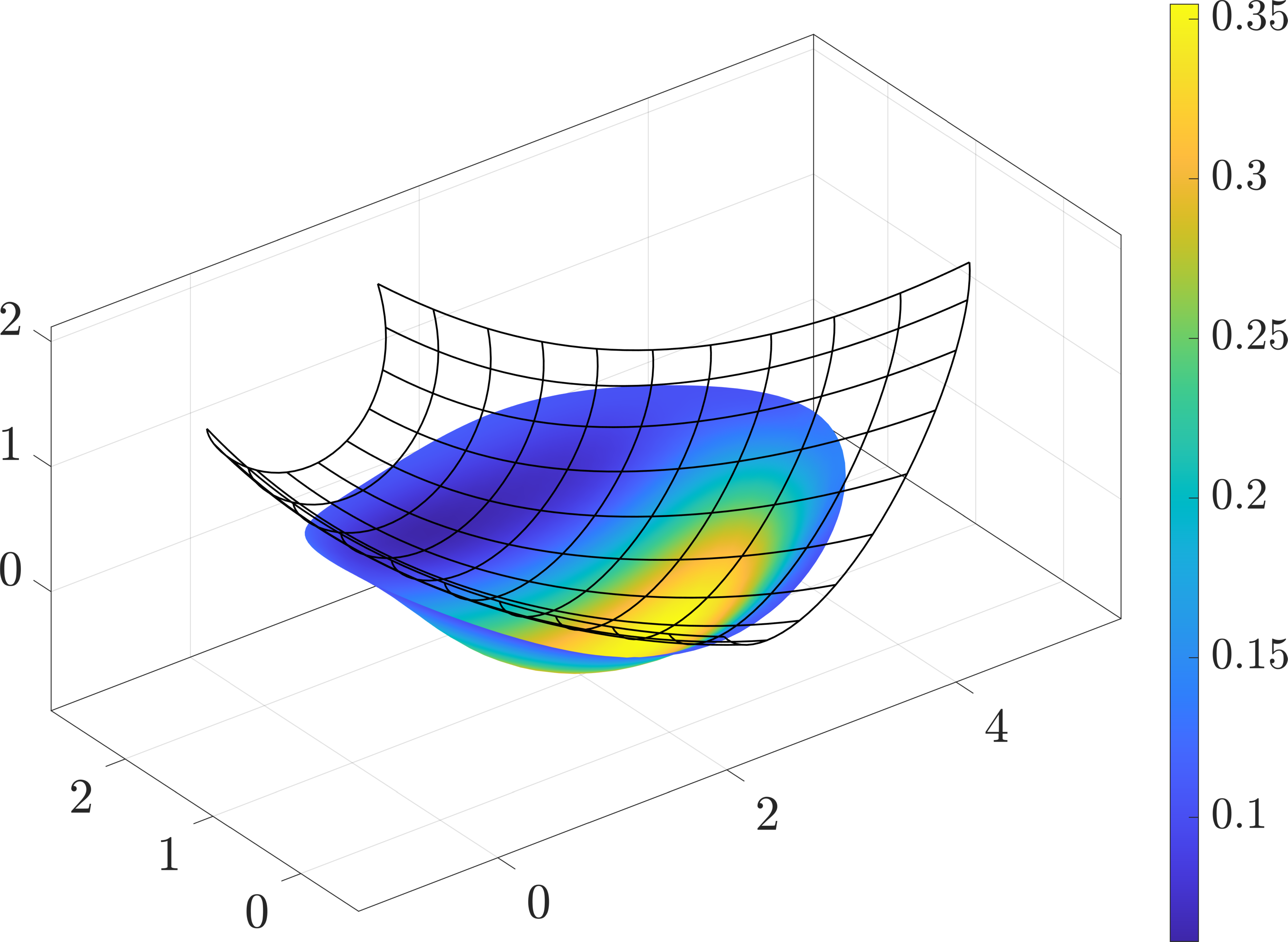} \label{fig:circd}}
	\caption{(a) The grey domain is the trimmed parameter space $\visibledomain$ defined by $\phi_1$, (b) deformed domain with scaled displacement $\vek{u}$ by a factor $\SI{1.5}{}$.}
	\label{fig:circpd}
\end{figure}

In the last example, a more general geometry and inhomogeneous boundary conditions are considered. In particular, the shape of the shell is defined through a non-linear map from the parameter space to the physical domain. The boundary is defined by a circular trimming curve in the parameter space. The details of the problem definition are given in \cref{fig:overcirc}.
For this example, an analytical solution is not available and the concept of manufactured solutions is not practical for the defined problem. As an alternative error measure, we employ the concept of residual errors \cite{Schoellhammer_2018a_DS}. The residual error is the relative $L_2$-error of the equilibrium in strong form, see \cref{eq:sff}, and is defined as
\begin{align}
	\varepsilon^2_{\t{rel,residual}} = \dfrac{\int_{\visibledomain} \left[ \divG{\mat{n}^{\t{real}}_\Gamma} + \nG \divG{\vek{q}_\Gamma} + \mat{H}\cdot\divG{\mat{m}_\Gamma} + \vek{f}\right]^2 \d A}{\int_{\visibledomain} \vek{f}^2 \d A}\ .
\end{align}
The evaluation of the strong form requires fourth-order surface derivatives of the midsurface displacement $\vek{u}^h$. Consequently, the theoretical optimal order of convergence is $\mathcal{O}(p-3)$. The implementation and the computation of such higher-order derivatives is not without effort. However, for verifying the proposed approach on a complex geometry in combination with non-trivial boundary conditions, this is a reliable measure for the verification of the higher-order accuracy. The only restriction is that the solution need to be sufficiently smooth, as always in higher-order suitable test cases. In \cref{fig:circpd}, the trimmed parameter space and deformed domain are visualized analogously to the test cases above.

\begin{figure}[ht]\centering\def\mw{.32}
	\subfloat[force equilibrium]{\includegraphics[width=\mw\textwidth]{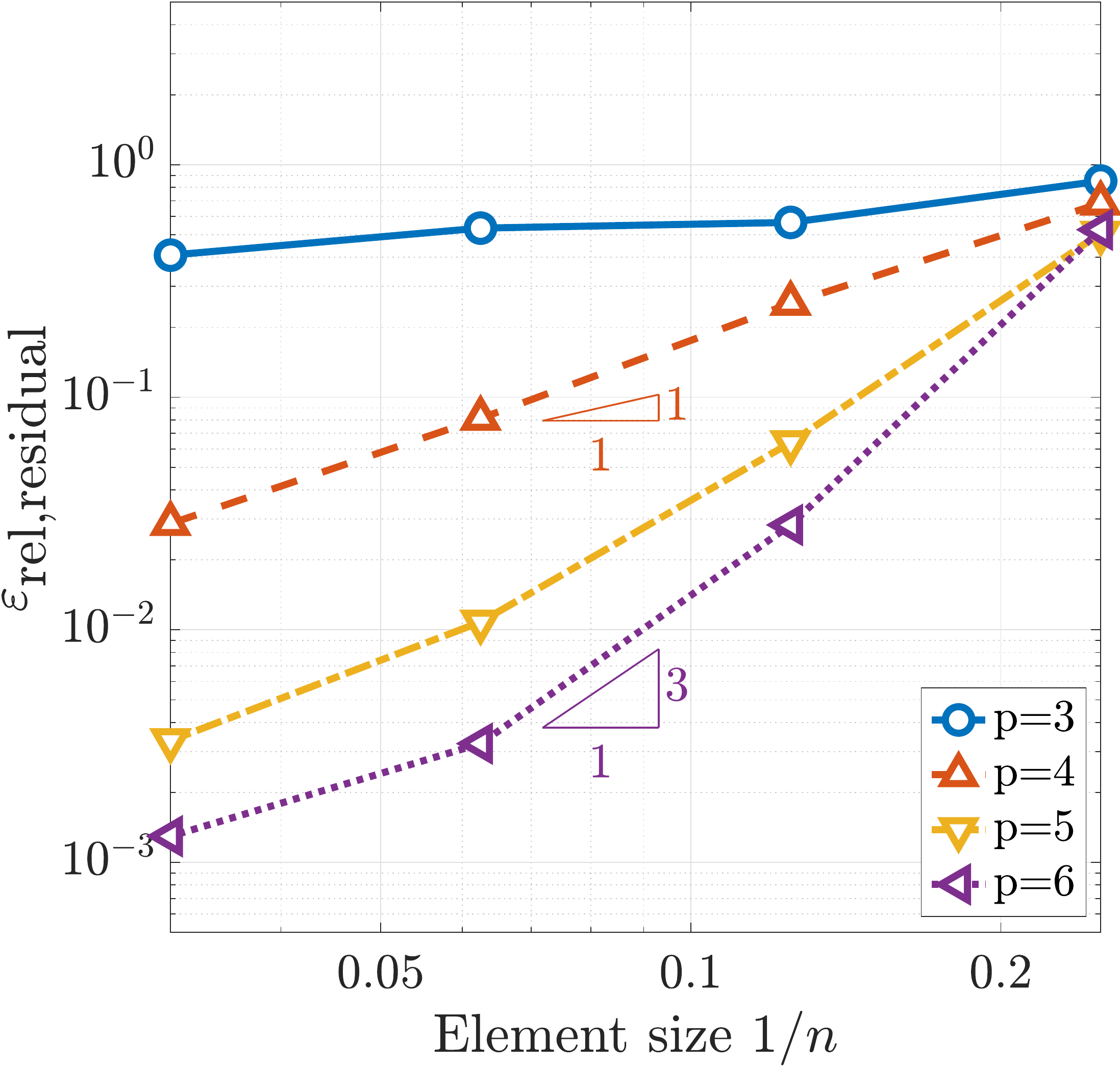}\label{fig:circsf}}
	\hfil
	\subfloat[energy error]{\includegraphics[width=\mw\textwidth]{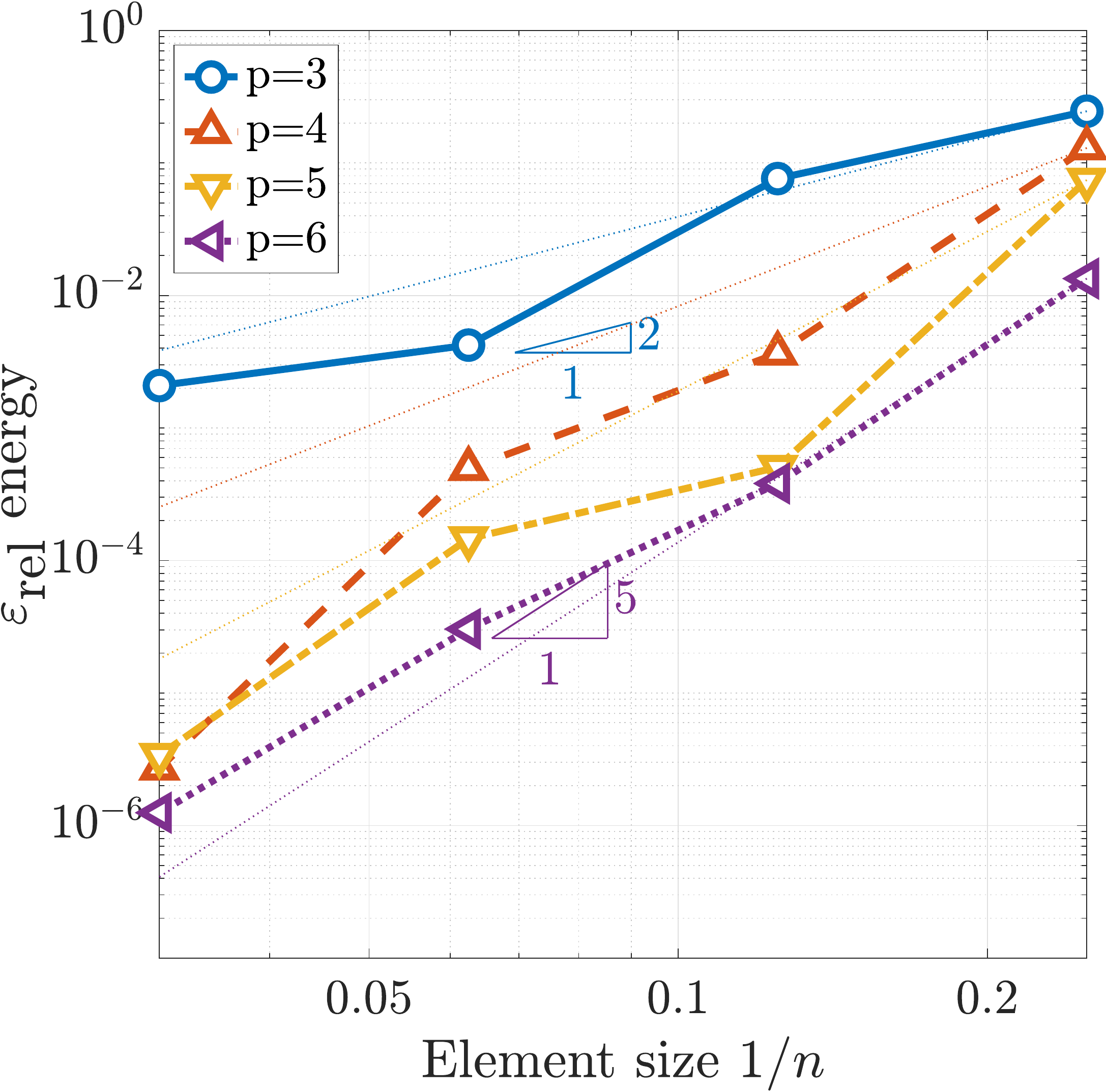}\label{fig:circeng}}
	\hfil
	\subfloat[condition number]{\includegraphics[width=\mw\textwidth]{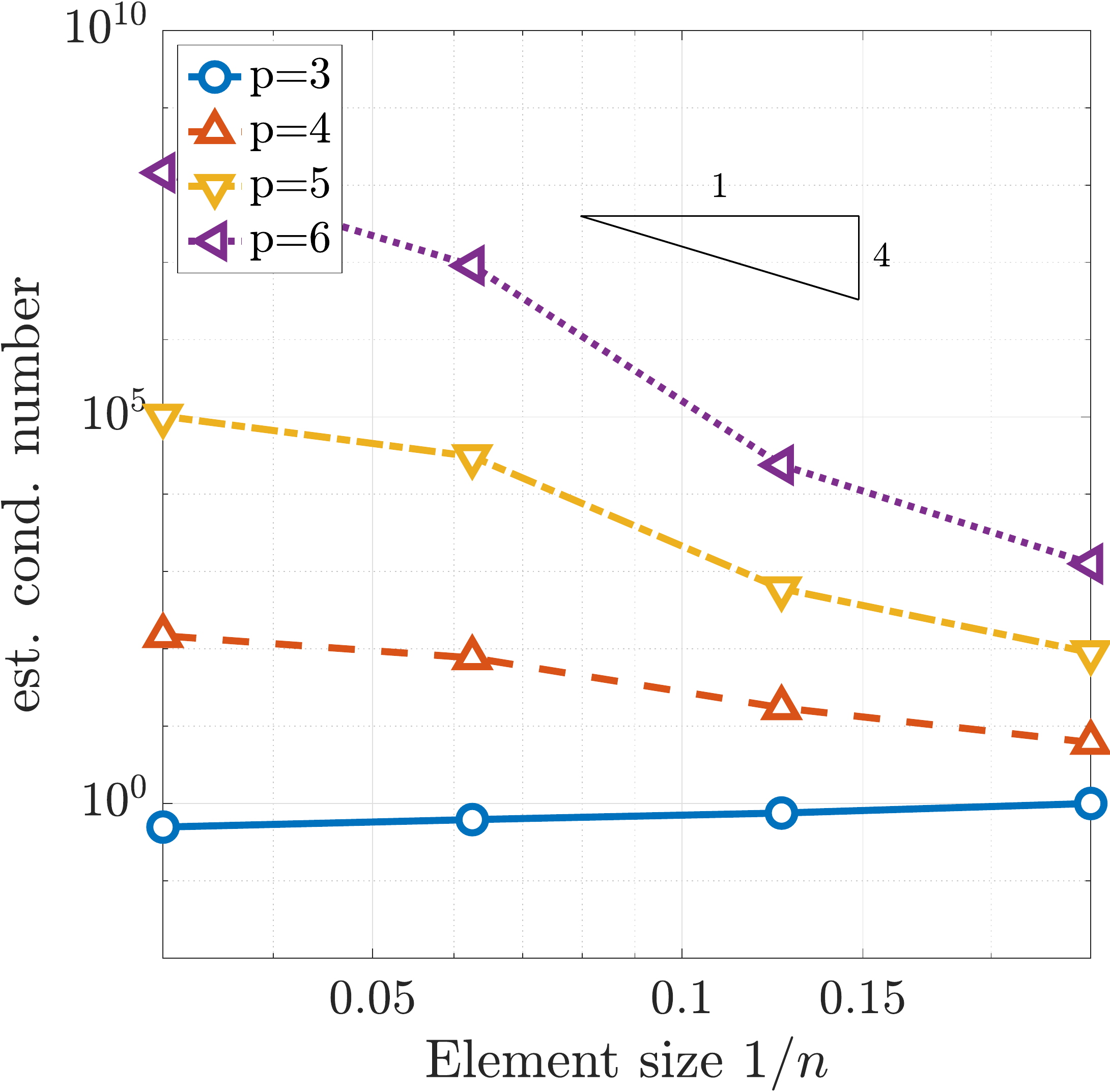}\label{fig:circond}}
	\caption{(a) Residual error $\varepsilon_{\t{rel,residual}}$, (b) relative energy error, the reference value is $\SI{5.85719e+01}{}$,  (c) normalized, estimated condition numbers, the reference value is
		$\SI{4.1539e+04}{}$ , which is the condition number at $n=4,\,p=3$.}
	\label{fig:circconv}
\end{figure}

In the convergence analyses, the element scale factor $n$ is varied as $\lbrace 4, 8, 16,    32\rbrace$ and the threshold $\alpha = 0.4$ for all orders and refinement levels. In \cref{fig:circsf}, the residual error is plotted. As expected, the third order $(p=3)$ does not convergence in the residual errors due to $\mathcal{O}(p-3) = \mathcal{O}(0)$. For the order $p\ge 4$ the optimal orders of convergence are achieved. In \cref{fig:circeng}, the relative error in the stored elastic energy is plotted. For the computation of the stored elastic energy, second order derivatives occur in the moment tensor and, therefore, the theoretical optimal order of convergence is $\mathcal{O}(p-1)$. It can be seen that optimal convergence rates are achieved for all orders.
In \cref{fig:circond}, the estimated condition numbers of the stiffness matrices in the convergence analyses are plotted and it can bee seen that the behaviour is similar to the second test case which is as expected for a fourth-order PDE. Analogously as above, the threshold $\alpha=0.4$ for $p=3$ may be slightly reduced in order to decrease the condition number.

\section{Conclusion}
\label{sec:conclusion}

\ifdefined\TodoListsOn
\begin{itemize}
    \item Higher-order consistent shell formulation
    \item Realization of higher-order convergence rates
    \item Emphasized various aspects that need to be properly addressed to achieve higher-order
    \item $\dots$     
    {\mycomment{\item Outlook: Criteria for threshold $\alpha$ which includes the order $p$}}
\end{itemize}
\fi

The use of higher-order basis functions for shell models is no novelty, especially since the introduction of isogeometric analysis (IGA). 
So it may come as a surprise that achieving higher-order convergence rates is still the exception.  
This circumstance is rooted in the fact that the related requirements on the geometry and the analysis are high. 
Already minor discrepancies in the discretization, application of boundary conditions, or the set-up of the system of equations diminish the accuracy of the results significantly. 

We present a Kirchhoff-Love (KL) shell formulation that treats \emph{all} these numerical aspects in a consistent higher-order fashion. 
From a geometric point of view, a B-spline tensor-product surface defines the geometry. 
It is at least $C^1$ to fulfil the requirements of the KL theory.  
Furthermore, the trimming concept allows the specification of arbitrary shapes over this single-patch representation.
Multi-patch models are not considered because the solution across different patches is non-smooth in general, and thus, these models are inherently low-order.
It is worth noting that this argument also holds for corners in the boundary of a shell for most cases.
Using a trimmed single-patch representation provides a remedy to this problem.
Regarding the analysis itself, the non-symmetric Nitsche method enforces the boundary conditions. 
The proposed definition of the Nitsche terms circumvents corner forces and shifts the tangent derivative on the drilling moment, which enables a more compact and less cumbersome implementation.
The system of equations is set up by a conventional Galerkin approach.
However, the treatment of cut elements, which arise due to trimming, requires special attention with respect to conditioning and integration.
We employ the extended B-spline concept and a reparameterization based on higher-order Lagrange elements to address these points. 
Combining these ingredients listed so far enables the consistent high-order accurate numerical treatment of \KL~shells.
Finally, our framework is completed by formulating the KL shell in the frame of tangential differential calculus (TDC) and using level-set functions as trimming curves. 
These two particular choices are, however, not necessary for the achievement of higher-order accuracy.

Numerical experiments demonstrate that the approach proposed yields optimal convergence rates. 
The Scordelis-Lo roof problem is investigated as a classic shell benchmark. 
Although the expected convergence behavior is shown, this example -- just like other traditional shell benchmark tests -- does not provide insight if higher-order accuracy has been achieved.
On the other hand, the results of a simply supported flat shell subjected to a manufactured solution show optimal higher-order convergence rates in the relative $L_2$-norm. 
This optimal behavior is not restricted to flat geometries, as confirmed by the problem of a general clamped shell with inhomogeneous boundary conditions.
Note that for such a complex problem, the concept of manufactured solutions is not practical. 
Hence, we employ the residual error, i.e., the relative $L_2$-error of the equilibrium in strong form.
This measure requires the evaluation of fourth-order surface derivatives, which is admittedly a challenge implementation-wise. 
Nevertheless, it allows a reliable assessment of higher-order accuracy.  

The numerical experiments also confirm that the extended B-spline concept controls the condition number of the system matrices.
For coarse high-order discretizations, however, the involved extension operation may affect the Nitsche method.
A possible solution may be the application of extended truncated hierarchical B-splines or another strategy for the conditioning. 
In general, the substitution of the techniques employed for each numerical aspect by an alternative scheme that allows a consistent higher-order treatment is a future topic worth exploring.

\appendix

\section{Computation of extrapolation weights}
\label{app:extrapolationWeights}

\ifdefined\TodoListsOn
\begin{itemize}
    \item[\checkmark] Details to efficient computation of the polynomial coefficients -- Benjamin
    \item[\checkmark] Adding of the algorithms -- Benjamin
    \item[\checkmark] Double check algorithms -- Benjamin 
\end{itemize}
\fi

The main task in the computation of the non-trivial extrapolation weights \labelcref{eq:deBoorFix_explicit} is the determination of the polynomial coefficients  $\poly$ and $\beta$. 
The latter corresponds to the Newton polynomial $\Nbasis_{\indexB,\pu} =  \sum^{\pu}_{\indexC=0} \beta_\indexC \: \uu^{\indexC}$ and can be explicitly expressed as
\begin{align}
	\label{eq:Nbasis_explicit_beta}
	 \beta_\indexC & =   (-1)^{\pu-\indexC}
	 \sum_{\indexG=1}^{ \totalG }  \prod_{\indexD \in \NSet_{\pu-\indexC,\indexG}} \uu_\indexD 
	 && \with &&
	\totalG = \binom{\pu}{\pu-\indexC}
\end{align}
where $\binom{\pu}{\indexD}$ is the binomial coefficient defined as
\begin{align}
	\label{eq:binomialcoefficient}
	\binom{\pu}{\indexD} \coloneqq \frac{\pu!}{(\pu-\indexD)! \:\indexD!} \: ,
\end{align}
and the sum over $\NSet_{\indexD,\indexG}$ denotes all $\indexD$-combinations with repetition
of the knots appearing in the definition~\labelcref{eq:deBoorFixNewtonBasis} of $\Nbasis_{\indexB,\pu}$, i.e., $\uu_{\indexB+1},\dots,\uu_{\indexB + \pu}$.

The coefficients  $\poly_\indexC$ of the segments $\BsplineSeg^{\indexSpan}_{\indexA}$ of $\Bspline_{\indexA,\pu}$ can be obtained by Taylor expansion
\begin{align}
	\label{eq:Bspline_Taylor}
	\BsplineSeg^{\indexSpan}_{\indexA}(\uu)
	& = \sum^{\pu}_{\indexC=0} \frac{\Bspline_{\indexA,\pu}^{(\indexC)}\left(\taylorpoint\,\right)}{\indexC!}\:\left(\uu-\taylorpoint\,\right)^{\indexC}
	= \sum^{\pu}_{\indexC=0} \alpha^*_\indexC\:\left(\uu-\taylorpoint\,\right)^{\indexC} ,  
	&& \taylorpoint \in \left[\uu_{\indexSpan}, \uu_{\indexSpan+1}\right)
\end{align}
and subsequent conversion to 
power basis form 
\begin{align}
	\label{eq:Bspline_PowerBasis}
	\BsplineSeg^{\indexSpan}_{\indexA}(\uu) &=
	\sum^{\pu}_{\indexC=0} \poly_\indexC \: \uu^\indexC 
	&& \with && 
	\poly_\indexC = \sum_{\indexD=\indexC}^{\pu} \binom{\indexD}{\indexC} \alpha^*_\indexD \left(-\taylorpoint\,\right)^{\indexD-\indexC}.
\end{align}

Alternatively, \cref{code:MatrixRepBsplineSegment1D} provides a derivation of the $\poly_\indexC$, which avoids the derivatives in \cref{eq:Bspline_Taylor}. It employs a MATLAB based syntax, and is inspired by the conversion of a B-spline to a matrix representation presented in \cite{qin2000a_BM}.
\begin{lstlisting}[language=Matlab,morekeywords={ones,nchoosek},caption={Computation of the polynomial coefficient matrix $\BsplineMatrix(\indexC,\indexA)=\poly^\indexA_\indexC$.},label=code:MatrixRepBsplineSegment1D]
function $\BsplineMatrixCode$ = MatrixRepresentationOfBsplineSegment($\indexSpan,\,\pu,\,\KV$)
% Input: 
%    $\color{gray}\indexSpan$... knot span
%    $\color{gray}\pu$... polynomial degree of the spline space
%    $\color{gray}\KV$... knot vector correspond to $\color{gray}\uu$
% Output: 
%    $\color{gray}\BsplineMatrixCode$... all polynomial coeffients $\color{gray}\poly^\indexA_\indexC$ related to $\color{gray}r$

$\BsplineMatrixCode$ = ones(1,1);
for i = 1 : $\pu$

    k = 1;
    $\textnormal{L}$ = zeros(i,i+1);
    $\textnormal{R}$ = zeros(i,i+1);
    for j = $\indexSpan$-i+1 : $\indexSpan$
        den = $\KV$(j+1) - $\KV$(j);
        $\textnormal{L}$(k,k)   = 1. - ( $\KV$($\indexSpan$)-$\KV$(j) ) / den;
        $\textnormal{L}$(k,k+1)= ( $\KV$($\indexSpan$)-$\KV$(j) ) / den;
        $\textnormal{R}$(k,k)   = -( $\KV$($\indexSpan$+1)-$\KV$($\indexSpan$) ) / den;
        $\textnormal{R}$(k,k+1) =  ( $\KV$($\indexSpan$+1)-$\KV$($\indexSpan$) ) / den;
        k = k + 1;  
    end

    $\textnormal{A}$i = $\BsplineMatrixCode$;
    $\BsplineMatrixCode$ = zeros(i+1,i+1);
    $\BsplineMatrixCode$(1:(*end*)-1,:) = $\BsplineMatrixCode$(1:(*end*)-1,:) + $\textnormal{A}$i * $\textnormal{L}$;
    $\BsplineMatrixCode$(2:(*end*),:) = $\BsplineMatrixCode$(2:(*end*),:) + $\textnormal{A}$i * $\textnormal{R}$;
end
\end{lstlisting}
%
%
The algorithm computes the coefficients $\poly^\indexA_\indexC$ for all non-zero B-splines $\Bspline_{\indexA,\pu}$ of the knot span and stores them in a matrix $\BsplineMatrix(\indexC,\indexA)=\poly^\indexA_\indexC$. Using $\BsplineMatrix$, \cref{eq:Bspline_PowerBasis} reads
\begin{align}
    \BsplineSeg^{\indexSpan}_{\indexA}(r) &= (1,~r,~\dots,~r^p) \BsplineMatrix(:,\indexA) && \with&& r = \frac{\uu - \uu_\indexSpan}{\uu_{\indexSpan+1} - \uu_\indexSpan}.
\end{align}
Note that the intrinsic coordinate is now $r\in[0,1)$ for each knot span. Hence, the entries of $\BsplineMatrix$ provide the $\poly^\indexA_\indexC$ sought only if the knot span $\indexSpan$ is defined from $0$ to $1$. This condition can be easily achieved by a linear transformation of the knots such that $r_\indexSpan=0$ and $r_{\indexSpan+1}=1$. When these transformed knots $r_\indexA$ are also employed for the Newton polynomials $\Nbasis_{\indexB,\pu}$, there is no alteration of the resulting extrapolation weights.

\Cref{code:codeDeBoorFixFunctional} computes the non-trivial extrapolation weights \labelcref{eq:deBoorFix_explicit} for the stable B-splines of a particular knot span $\indexSpan$.
We apply \cref{code:codeDeBoorFixFunctional} for each knot span that provides extensions. 
Note that $\myVec{r}_j$ in line 18 refer to the vector of all transformed knots $r_\indexA$ of $\Nbasis_{\indexB,\pu}$. The resulting matrix $\myMat{E}^\indexSpan$ contains the non-trivial extrapolation weights, where the stable B-spline of the knot span $\indexSpan$ and the associated degenerate ones given by $\degSet^{\indexSpan}$  correspond to the matrix's rows and columns, respectively. Finally, assembling all $\myMat{E}^\indexSpan$ according to the global degrees of freedom, togehter with the trivial extrapolations weights \labelcref{eq:exWeightsOne} and \labelcref{eq:exWeightsZero}, yields the overall extension matrix $\myMat{E}$.
\begin{lstlisting}[language=Matlab,morekeywords={ones,nchoosek},caption={Extrapolation weights for a particular knot span},label=code:codeDeBoorFixFunctional]
function $\myMat{E}^\indexSpan$ = ComputeExtrapolationWeights($\indexSpan,\,\degSet^{\indexSpan},\,\pu,\,\KV$)
% Input: 
%    $\color{gray}\indexSpan$... knot span providing the extensions
%    $\color{gray}\degSet^{\indexSpan}$... index set of all degenerate B-splines related to $\color{gray}\indexSpan$ 
%    $\color{gray}\pu$... polynomial degree of the spline space
%    $\color{gray}\KV$... knot vector corresponding to $\color{gray}\uu$
% Output: 
%    $\color{gray}\myMat{E}^\indexSpan$... extrapolation weights for the stable B-splines of $\color{gray}\indexSpan$  

% polynomial coefficients of the extensions 
$\BsplineMatrix$ = MatrixRepresentationOfBsplineSegment($\indexSpan,\,\pu,\,\KV$)
% knot values mapped to intrinsic coordinate r
$\KV_r$ = ($\KV$-$\KV$($\indexSpan$)) ./ ($\KV$($\indexSpan$+1)-$\KV$($\indexSpan$));

% coefficients of the Newton polynomial based on $\color{gray}\KV_r$
$\textnormal{B}$ = zeros($\pu$+1,length($\degSet^{\indexSpan}$));
for n = 1 : length($\degSet^{\indexSpan}$)
    $\myVec{r}_j$ = $\KV_r$( $\degSet^{\indexSpan}$(n)+1 : $\degSet^{\indexSpan}$(n)+$\pu$ );
    for k = 0 : $\pu$-1 

        L = nchoosek($\pu$,$\pu$-k);
        if L == 1 
            $\NSet$ = $\myVec{r}_j$;
        else
            $\NSet$ = nchoosek($\myVec{r}_j$,$\pu$-k);
        end

        tmp = 0;
        for l = 1 : L            
            tmp = tmp + prod($\NSet$(l,:));         
        end
        $\textnormal{B}$(k+1,n) = (-1)^($\pu$-k) * tmp; 
        
    end
end

% extrapolation weights 
$\myMat{E}^\indexSpan$ = zeros(size($\BsplineMatrix$,2),length($\degSet^{\indexSpan}$));
kFactorial = [1 cumprod(1:size($\BsplineMatrix$,1)-1)]; % pre-computation of k!
for j = 1 : size($\myMat{E}^\indexSpan$,2)
    for i = 1 : size($\myMat{E}^\indexSpan$,1)
         for k = 0 : $\pu$
             $\myMat{E}^\indexSpan$(i,j) = $\myMat{E}^\indexSpan$(i,j) + (-1)^k * ...
                 kFactorial((*end*)-k) * $\textnormal{B}$((*end*)-k,j) * ...
                 kFactorial(k+1) * $\BsplineMatrix$(k+1,i);
         end
    end
end
$\myMat{E}^\indexSpan$ = $\myMat{E}^\indexSpan$ ./ kFactorial((*end*));
\end{lstlisting}
%

\bibliographystyle{schanz_mod}
\addcontentsline{toc}{section}{\refname}
\bibliography{\pathToBibFile}

\end{document}